	\documentclass[referee]{aa}  

	\usepackage{lscape,graphicx}
	\usepackage{psfig}  
	\usepackage{natbib}
	 \bibpunct{(}{)}{;}{a}{}{,} 
	\usepackage{aas_macros}

	\def\ww{luminosity-weighted }

\begin{document}


	   \title{Galaxy Evolution in Local Group Analogs. I. \\
	   A {\it GALEX} study of nearby groups \\
	dominated by late-type galaxies\thanks{Based on  {\it GALEX}  observations (GI2-121 PI L.M. Buson)}}


	   \author{A.~Marino,
	   \inst{1} 
	   L.~Bianchi,
	   \inst{1}
	   R.~Rampazzo,\inst{2}
	   L.M.~Buson,
	   \inst{2}
	    \and 
	   D.~Bettoni
	   \inst{2} 
	    }

	   \offprints{A. Marino}

	   \institute{
       Dept. of Physics and Astronomy, Johns Hopkins University, 3400 North Charles 
	   Street, Baltimore, MD 21218  USA\\
	   \email{bianchi@pha.jhu.edu, amarino@pha.jhu.edu}
	   \and
	    INAF Osservatorio Astronomico di Padova, vicolo dell'Osservatorio~5, I-35122  Padova, Italy\\
	   \email{daniela.bettoni@oapd.inaf.it, lucio.buson@oapd.inaf.it, roberto.rampazzo@oapd.inaf.it}
	    }

	   \date{Received / accepted}

	   \titlerunning{A {\it GALEX} view of nearby, late-type galaxy dominated groups}
	   \authorrunning{Marino et al.}

	\abstract
	 {Understanding the astrophysical processes acting within galaxy groups and 
	 their  effects on the evolution of the galaxy population is one of the crucial
	  topic of  modern cosmology, as almost 60\% of galaxies in the 
	  Local Universe are found in groups.}
	  {We aim at learning about galaxy evolution within  nearby groups dominated
	  by late-type galaxies, specifically by studying  their  ultraviolet-emitting stellar population.}
	  {We imaged in the far (FUV, $\lambda_{eff}$ = 1539 \AA) and near  ultraviolet
	 (NUV, $\lambda_{eff}$ = 2316 \AA)   
	    with the Galaxy Evolution Explorer ({\it GALEX}) 
	    three nearby groups, namely LGG~93, LGG~127 and LGG~ 225. 
	    We obtained the UV galaxy surface photometry and, 
	    for LGG~225, the only group covered by the SDSS,
	    the photometry in  u, g, r, i, z bands.  We discuss galaxy morphologies
	    looking for  interaction signatures
	   and  we analyze the spectral energy distribution 
	   of  galaxies  to infer their luminosity--weighted ages.
	    The UV and optical photometry was also used to perform a 
	    luminosity-weighted kinematical and dynamical analysis of each group and to evaluate the stellar mass. 
	    }
	   {A few member galaxies in LGG~225  show a distorted UV morphology 
	   due to ongoing interactions.  (FUV-NUV) colors  suggest  that spirals in LGG~93 and 
	   LGG~225   host stellar populations  in their outskirts 
	   younger  than that of M31 and M33 in the Local Group or with less extinction.  
	   The irregular interacting galaxy NGC~3447A
	has a significantly younger stellar population (few Myr old) than the average 
	of the other irregular galaxies in LGG~225 suggesting that the encounter triggered
	 star formation. 
	 The early-type members of LGG~225, NGC 3457 and  NGC~3522, have masses of the order of a 
	few 10$^9$ M$_\odot$, comparable to the Local Group ellipticals. For the most massive spiral
	in LGG 225, we estimate a stellar  mass of $\approx$ 4 $\times$ 10$^{10}$ M$_\odot$, comparable to M33 in the 
	Local Group. Ages of stellar populations range from a few to $\approx$ 7 Gyr for the galaxies in LGG 225.
	 The kinematical and dynamical analysis indicates that  
	   LGG~127 and LGG~225  are in a pre-virial collapse phase, i.e. still undergoing  
	dynamical relaxation, while  LGG~93  is likely virialized.
	Both the photometric and the dynamical analysis suggest that 
	  LGG~225 group  is in a more active evolution phase than 
	   both LGG~93 and LGG~127.}


	   \maketitle

	\section{Introduction}

	\noindent

	Although  the majority of galaxies ($\sim$ 60\%)
	in the local universe resides in groups 
	\citep[e.g.][] {Huchra82, Tully87, Giuricin00, Ramella02, Eke04, Tago08}, 
	our knowledge of galaxy  evolution in low density environments, of which groups are
	the defining galaxy association, is still scanty. In particular,
	whether a link  between the elliptical-galaxy dominated and 
	late-type galaxy dominated groups exists, i.e. whether they represent two evolutionary stages
	in a hierarchical cosmological framework, is still 
	unclear. 

	 Some important physical  parameters mark the  difference between 
	 the possible evolution within  groups and clusters. Among them,
	 the galaxy velocity dispersion in groups is comparable 
	to the inner velocity dispersion of  individual galaxies. Processes 
	such as galaxy merging, and accretion e.g. of HI-rich high-velocity 
	clouds \citep{Zablu98, Blitz99}, are much more effective in groups 
	than in clusters. Conversely, mechanisms regulating  galaxy
	evolution in the cluster environment, such as ram-pressure stripping 
	\citep[see e.g.][]{Vollmer02} and galaxy harassment 
	\citep{Moore99}, are less relevant in groups. In the group environment, 
	galaxy encounters may then  completely reshape the member's morphology 
	and trigger secondary star formation episodes even in early-type galaxies 
	 \citep[see e.g.][and references therein]{Annibali07} .

	\begin{table*}
	\caption{Characteristics of the LGG 93, LGG 127 and LGG 225 groups$^a$.}	 
	 \scriptsize
	\begin{tabular}{lllllllllll}
	\hline\hline
	   Group  & RA & Dec.&  Morph.& RC3 & E(B-V)$^d$& Incl. & Semi-major and & P.A.$^c$ & Mean Hel.& B$_T$ \\
	   Galaxies & (J2000) & (J2000) & type & type & &  & minor axis$^b$  &  &  Vel  & \\
		   & (h:m:s) & (d:m:s)   &  &  &  & [deg] &[arcmin] & [deg] &  [km/s] & (AB mag)\\
				  
	\hline
	LGG 93   & & & & & & & & & &  \\
	\hline 
		   NGC 1249  & 03:10:01.2 & -53:20:09	 & SBc& 6.0 $\pm$ 0.3 & 0.017   &69.0& 2.45 ~ 1.15&86  & 1072 $\pm$ 14 &  12.03	$\pm$ 0.19 \\ 
		   NGC 1311  & 03:20:06.9 & -52:11:08	 & SBm& 8.8 $\pm$ 0.9 & 0.021   &90.0& 1.50 ~ 0.40&40   & 571  $\pm$ 5 &   13.24   $\pm$ 0.20\\ 
		   IC 1933   & 03:25:39.9 & -52:47:08	 & Sc & 6.1 $\pm$ 0.7 & 0.017  &59.4 & 1.10 ~ 0.60&55   & 1063 $\pm$ 5 &	12.77	$\pm$ 0.13\\ 
		   IC 1954   & 03:31:31.4 & -51:54:17	 & SBb& 3.2 $\pm$ 0.8 & 0.016  &58.0 & 1.60 ~ 0.75&66   & 1062 $\pm$ 4  &  11.97   $\pm$ 0.08 \\ 
		   IC 1959   & 03:33:12.6 & -50:24:51	 & SBd& 8.4 $\pm$ 1.5 & 0.011  &90.0 & 1.40 ~ 0.35&147  & 641  $\pm$ 5 &   13.04   $\pm$ 0.13\\ 
																      
	\hline														 	      
	LGG 127    & & & & & & & & & &  \\
	\hline 
		    NGC 1744  & 04:59:57.8 & -26:01:20 & SBcd& 6.7 $\pm$ 1.4  & 0.041 & 69.9 & 4.05 ~ 2.20 & 168 & 740 $\pm$ 6  &  11.55  $\pm$ 0.19\\ 
		    NGC 1792  & 05:05:14.4 & -37:58:51 & SBbc& 4.0 $\pm$ 0.2  &0.023 & 62.8  & 2.60 ~ 1.30 & 137 & 1207 $\pm$ 4  &  10.60  $\pm$ 0.13 \\ 
		    NGC 1800  & 05:06:25.7 & -31:57:15 & Sd &  8.2 $\pm$ 3.9  & 0.014 & 47.0 & 1.00 ~ 0.55 &113  & 806 $\pm$ 5    &  12.96  $\pm$ 0.05\\ 
		    NGC 1808  & 05:07:42.3 & -37:30:47 & SABa &1.2 $\pm$ 0.5  & 0.030 & 83.9 & 3.25 ~ 1.95 & 133 & 1001 $\pm$ 5 &  10.66  $\pm$ 0.06\\ 
		    NGC 1827  & 05:10:04.6 & -36:57:37 & SABc &5.9 $\pm$ 0.5  & 0.028 & 84.9 & 1.50 ~ 0.35 & 120 & 1041 $\pm$ 7&   13.08  $\pm$ 0.20\\ 
		   ESO305-009 & 05:08:07.6 & -38:18:33 & SBd&  8.0 $\pm$ 0.4 & 0.027 & 53.0  & 1.75 ~ 1.40 & 63  & 1022 $\pm$ 5&  13.23  $\pm$ 0.77 \\ 
		   ESO305-017 & 05:15:00.6 & -41:23:33 & IB &  9.9 $\pm$ 0.5 & 0.030 & 86.0  & 0.90 ~ 0.35 & 64  & 1071  $\pm$ 11 &  14.51  $\pm$ 0.23\\ 
		   ESO362-011 & 05:16:38.8 & -37:06:09 & Sbc&  4.2 $\pm$ 0.6 & 0.048 & 90.0  & 2.25 ~ 0.35 & 76  &   1345 $\pm$ 6 &	  12.85  $\pm$ 0.23\\ 
		   ESO362-019 & 05:21:04.2 & -36:57:25 & SBm & 8.9 $\pm$ 0.6 & 0.043 & 90.0  & 1.10 ~ 0.35 & 3	&   1282 $\pm$  8 &  13.90  $\pm$ 0.28 \\ 

	\hline
	LGG 225   & & & & & & & & & &   \\
	\hline
		   NGC 3370  &10:47:04.0  & +17:16:25  &Sc  & 5.1 $\pm$ 1.1 & 0.031 & 56.2 & 1.60 ~ 0.90 & 148  &1281 $\pm$ 3&  12.03	$\pm$ 0.27 \\ 
		   NGC 3443  &10:53:00.1  & +17:34:25  &Scd & 6.6 $\pm$ 0.8 & 0.036 & 61.2 & 1.40 ~  0.70& 145   &1132 $\pm$ 8 &  14.62 $\pm$ 0.44 \\ 
		   NGC 3447  &10:53:23.99 & +16:46:21  &Sm  & 8.8 $\pm$ 0.7 & 0.030 & 64.1 & 1.85 ~  1.05 & 26  &1069 $\pm$3 & 14.28 $\pm$ 0.68 \\ 
		   NGC 3447A &10:53:29.69 & +16:47:9.9 &IB  & 9.9 $\pm$ 0.4 & 0.030 & 69.3 & 0.88  ~ 0.55& 106   &1094 $\pm$ 7 & 15.97 $\pm$ 0.50\\ 
		   NGC 3454  &10:54:29.4  & +17:20:38  &SBc & 5.5 $\pm$ 0.9 & 0.034 & 83.7 & 1.40  ~  0.18& 116   &1114 $\pm$ 10&  13.53   $\pm$ 0.08\\ 
		   NGC 3455  &10:54:31.1  & +17:17:05  &SABb& 3.1 $\pm$ 0.8 &0.033 &  53.6 & 1.25  ~  0.75& 80   &1107 $\pm$ 5&   14.15   $\pm$ 0.25\\ 
		   NGC 3457  &10:54:48.6  & +17:37:16  &E   & -5.0$\pm$ 1.2 & 0.031 & 23.7 & 0.45  ~  0.45	 &&1158 $\pm$ 6&   12.77   $\pm$ 0.41 \\ 
		   NGC 3501  &11:02:47.3  & +17:59:22  &Sc  & 5.9 $\pm$ 0.5 & 0.023 & 90.0 & 2.38  ~  0.27& 27    &1131 $\pm$ 7&   13.41   $\pm$ 0.22\\ 
		   NGC 3507  &11:03:25.4  & +18:08:07	&SBb & 3.1 $\pm$ 0.4 & 0.024 & 31.9 & 1.70  ~  1.45&110     &975  $\pm$ 9 &  11.93   $\pm$ 0.51\\ 
		   NGC 3522  &11:06:40.4  & +20:05:08	&E   & -4.9$\pm$ 0.4  & 0.023 &90.0 & 0.60  ~  0.35 & 117  &1225 $\pm$ 11 & 13.72 $\pm$ 0.41 \\ 
		   UGC 6022  &10:54:15.38 & +17:48:34.4  &I   & 9.9 $\pm$ 0.5 & 0.032 & 63.9 & 0.60  ~  0.35  & 10   & 973 $\pm$ 6 &		  \\
		   UGC 6035  &10:55:28.7  & +17:08:33	&IB  & 9.9 $\pm$ 0.4 & 0.031 & 57.2 & 0.70  ~  0.65 &177    &1072 $\pm$ 3  & 15.28   $\pm$ 0.50 \\ 
		   UGC 6083  &11:00:23.8  & +16:41:32	&Sbc & 4.1 $\pm$ 0.5 & 0.020 & 90.0 & 0.78  ~  0.09& 143   &938  $\pm$ 3 &  14.96  $\pm$ 0.32 \\ 
		   UGC 6112  &11:02:35.3  & +16:44:05	&Scd & 7.4 $\pm$ 0.8 & 0.020 & 90.0 & 1.25  ~  0.40&123    &1038 $\pm$ 6  & 14.56   $\pm$ 0.53 \\ 
		   UGC 6171  &11:07:10.0  & +18:34:10	&IB  & 9.9 $\pm$ 0.4 & 0.022 & 90.0 & 1.25  ~  0.30 &68    &1199 $\pm$ 9&   14.97   $\pm$ 0.39\\ 
	\hline     						     
	     \end{tabular} 					      
								     
	  $^a$data from \tt HYPERLEDA http://leda.univ-lyon1.fr \citep{Paturel03} 
	  except where otherwise noted.\\  
	  $^b$ Major and minor axes of the elliptical apertures used to compute FUV and NUV magnitudes.\\ 
	  $^c$ P.A. from RC3 catalog
	  \citep{devau91}.  UGC 6035's P.A. is from HYPERLEDA. \\ 
	  $^d$ taken from NED.
	   
	   \label{table1}

	\medskip
	 \end{table*}

	\begin{figure*}[!ht]
	\begin{tabular}{cc}
	\includegraphics[height=16.5cm]{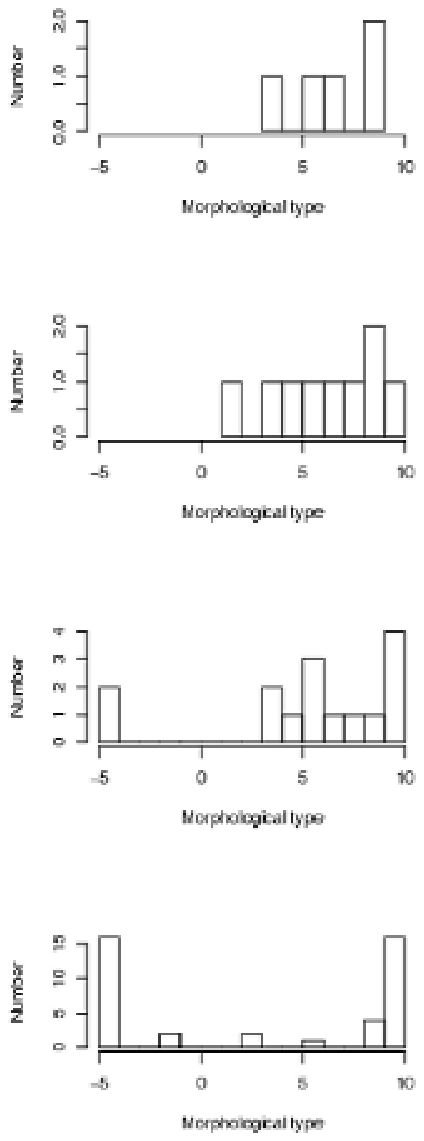} &
	\includegraphics[height=16.5cm]{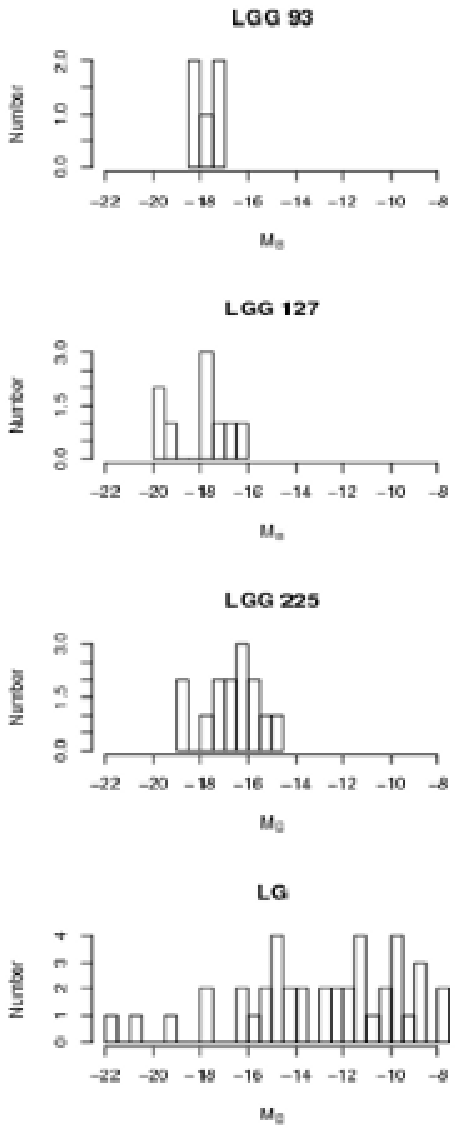}\\
	 \end{tabular}
	 \caption{From top to bottom: 
	morphological type distributions (left) and B absolute magnitudes of LGG~93, LGG~127, LGG 225 
	 and 
	of the LG from \citet{Pasetto07}.
	}
	  \label{fig0}
	 \end{figure*}

	\begin{table*}[!t]
	\caption{Journal of the {\it GALEX} observations}
	\scriptsize
	\begin{tabular}{lllcllll}
	 \hline\hline
	 \multicolumn{1}{l}{LGG}&
	\multicolumn{1}{l}{RA}&
	\multicolumn{1}{l}{Dec} &
	 \multicolumn{1}{c}{Observing} &
	  \multicolumn{1}{l}{FUV }&
	   \multicolumn{1}{l}{NUV } &
	\multicolumn{1}{l}{Observing } \\
	\multicolumn{1}{l}{} &
	\multicolumn{1}{l}{[deg]}&
	\multicolumn{1}{l}{[deg]} &
	\multicolumn{1}{c}{Date}&
	\multicolumn{1}{l}{Exp. Time [sec]} &
	\multicolumn{1}{l}{Exp. Time [sec]} &
	\multicolumn{1}{l}{program} \\
	\hline
	LGG 93 P1  & 47.5050  & -53.3358  & 2006-12-02 & 1172 & 1172 & GI1 047 \\
	LGG 93 P2  & 50.1375  & -52.63825 &  2006-10-10  & 1860 & 1860  & MIS \\
	LGG 93 P3  & 51.7627   & -52.87209 &  2006-10-10  & 1516 & 1517  & MIS \\
	LGG 93 P4  & 52.9283   &-51.9025   & 2004-12-09    & 1614   & 1614   & GI1 009 \\ 
	LGG 93 P5  &  53.3025   & -50.41417  & 2004-12-08  &  1513   & 1513  & GI1 047 \\
	\hline 
	LGG 127 P1  & 77.2175	 &-37.2449 &  2005-12-05 &2155 & 2155   & GI2 121   \\
	LGG 127 P2  & 74.9908 & -26.0222 & 2004-12-11 & 1905 & 3907  & GI1 047     \\
	LGG 127 P3  & 76.3163	& -37.9719 &2004-12-10  & 1957& 3954 & GI1 009 \\
	LGG 127  P4 &76.6115 & -31.9493 & 2003-11-27 & 1698 &1698 & NGA\\
	LGG 127  P5 & 77.0408 & -38.3053 & 2004-12-10 & 2835 & 2835. &GI1 009  \\
	LGG 127  P6 & 78.7604 & -41.3572 & 2004-11-12/11 &2719 & 4060 & GI1 009 \\ 
	LGG 127  P7  &79.1738 & -37.0953& 2004-11-12/2005-01-29 & 1490 & 8006 & GI1 009 \\ 
	LGG 127  P8 & 80.7694 & -37.1592 & 2005-10-31 & 116 & 116 & AIS \\
	\hline
	LGG 225 P1  & 163.3617 &16.77889 & 2006-03-22 & 4160  & 4651&  GI2 121   \\
	LGG 225 P2  & 163.7025&17.6212 & 2006-03-23 & 1609    &1609  & GI2 121   \\
	LGG 225 P3  & 161.9588  &  17.6503  & 2005-02-28 & 104 & 104 & AIS\\ 
	LGG 225 P4  & 165.0826 & 16.25137   & 2005-02-28 & 108 & 108 & AIS\\    
	LGG 225 P5 & 165.5296 &   17.20472 &  2005-02-28   & 109 & 109 & AIS \\  
	LGG 225 P6  &  165.9695 & 18.14557  & 2005-02-28 & 107 & 107 & AIS\\
	LGG 225 P7  & 166.6684 & 20.0865   & 2008-02-14 & 1706 & 1706 & GI3 046  \\  
	LGG 225 P8  & 167.0216 &   18.2238  & 2005-02-28 & 107 & 107 & AIS \\                                
					      
	\hline
	\end{tabular}
	\label{table2}
	\end{table*}
	
	 \normalsize
	 
	In the context of group evolution, the Local Group (LG hereafter) 
	offers clear examples of ongoing interactions and/or merging episodes. 
	Both observations and models suggest that the life in the LG is highly 
	dynamic. Although the Milky Way, M31 and M33 do not
	appear severely distorted by the mutual gravitational interaction, except for the outermost
	regions \citep{Mc09}, 
	the disruption and accretion of small galaxies by the
	spiral members seems very frequent. These accretions give rise to 
	gaseous and stellar tidal streams that continue to orbit the 
	accreting galaxy as fossil relics of the mass transfer activity.
	Ongoing accretion events have been detected around the
	Milky Way.  We quote some cases  discovered in the recent years: the Pal~5 
	stream \citep[e.g.][]{Odenkirchen01}, the Sagittarius stream
	\citep[e.g.][]{Ibata01}, the Monoceros stream \citep[e.g.][]{Yanny03},
	the orphan stream \citep[]{Belokurov06} and the anticenter stream
	\citep[]{Grillmair06}. The most prominent and the earliest
	discovered is the Magellanic stream 
	\citep{Bruns05} that dynamical models suggest to be 
	about 1.7 Gyr old \citep[e.g.][]{Nidever08}.

	In order to place the observed properties of the LG in the general 
	evolutionary framework of loose groups dominated by late-type galaxies, 
	we need to compare its basic parameters (e.g presence of tidal phenomena,
	star formation activity) with nearby, very similar systems, possibly so similar to 
	be considered basically analogs of the galaxy system we inhabit. 
	In this context, we started  a study of a sample 
	of late-type galaxy dominated groups   with {\it  GALEX}.  
	Our sample includes  three nearby groups at
	approximately the same distance, namely 
	LGG~93,  LGG~127 and LGG~225. LGG~93 and LGG~127   
	are completely dominated by spiral galaxies while 
	 LGG~225  contains two early-type galaxies.
	This study, through a detailed analysis of each member galaxy
	in the UV and, where possible, in the optical, combined with a
	luminosity-weighted dynamical study of each group as a whole, 
	 provides new elements  to infer the evolutionary
	state of the group in a hierarchical evolutionary scenario.
	{\it GALEX}'s wide field of view  allowed us to obtain 
	a snapshot of the entire groups, of tidal features and 
	to map the recent star formation (SF)
	in late-type galaxies.   

	The paper is arranged as follows. Section~2  describes the sample and its properties, 
	as compiled from the current literature. Sections~3 presents the UV observations and,
	for LGG 225, the SDSS observations.  
	The UV morphology and the optical and  UV  photometry  are presented in Section~4. 
	In Section~5 we use  
	synthetic galaxy populations models   
	to interpret   the energy distribution,  
	estimate ages and masses of group galaxies and  compare them to the groups dynamical
	properties.  
	Discussion and conclusions  are given in Section~6. 
	H$_0$=75 km~s$^{-1}$Mpc$^{-1}$ is used throughout the paper.

	\section{The sample}

	 Table~1 compiles the main characteristics of the galaxies 
	members of  LGG~93, LGG~127 and  LGG~225 analyzed in the present study.
	It includes for each galaxy the J2000 coordinates, the morphological type, 
	the foreground galactic extinction from NED, 
	the  inclination, the semi-major and semi minor axis length, the 
	position angle (P.A.), the heliocentric
	systemic velocity and  the B-band total apparent magnitude. 

	 LGG~93 and LGG~127 are located in the southern hemisphere. The three
	groups have a similar average recession velocity of $\sim$1000 km~s$^{-1}$. The galaxy group
	membership list is obtained from {\tt HYPERLEDA}
	on the ground of the redshift analysis performed by
	\citet{Huchra82} and \citet{Garcia93}. For LGG~225, we integrated the {\tt
	HYPERLEDA} member list with \citet{Giuricin00} catalog.
	 
	LGG~93 is composed of 5 identified members ($\Delta~V_{max}$=501 km~s$^{-1}$), all spirals
	of similar apparent magnitude ($\langle B_T \rangle $=12.61$\pm$0.58). Four of them are
	classified as barred spiral. LGG~127 is composed of 9 identified members
	($\Delta~V_{max}$=605 km~s$^{-1}$). The faintest member, ESO305-017, is the only irregular
	galaxy identified in this group composed of spirals, most of which of late morphological
	types. The spread in luminosity ($\langle B_T \rangle$=12.59$\pm$1.37) is larger than in LGG 93. 
	 LGG~225 is the richest group, composed of 15 identified members with a spread of 
	$\Delta~V_{max}$=343 km~s$^{-1}$.  
	Together with a large number of spirals  and five irregulars it contains also two 
	elliptical galaxies. The group is also characterized by a large spread in the 
	galaxy apparent magnitude ($\langle B_T \rangle
	$=14.01$\pm$1.19). 
         
	The morphological type  and absolute B magnitudes distributions of group members and,  
	for comparison, of our LG  are shown in Figure~\ref{fig0}.  
	For the LG members, we used the list of galaxies  by \citet{Pasetto07}.
	The list of our group members 
 does not include galaxies fainter than M$_B$= -15 which are the 
 larger fraction in the LG. Such faint galaxies are below the detection limits of our
 source catalogues \citep[see e.g.][]{Garcia93}. 
 Most of the LG galaxies with Type=-5 are dwarf spheroidal galaxies, clearly
 different in  magnitude (and mass) range from  early-type galaxies
 present in LGG~225. Our groups show an overabundance of intermediate 
 luminosity galaxies with respect to LG and a lack of bright (M$_B \leq$-20),
 M31 and MW-like, members. 	    
	    
        In the recent group compilations,   
        of \citet{Tago08}, based on SDSS data
       release 5, and  \citet{Eke04}  from the Two-degree Field Galaxy Redshift Survey,
       the majority (81\%, and 76\%, respectively) of groups have less than 4 members.  
       The percentage of groups with  4 to 10 members  drops to $\sim$ 17\% and 21\%  
        respectively, and to $\sim$ 2\% and 3\%, for groups with 11 to 20 members.
        In this context, although the above group compilations suffer of some biases, the 
        most important of which is a redshift-dependent bias underlined e.g. by Plionis et al. (2006), 
        LGG~225, dominated by spirals with at least 15 members, may be considered
        among the more rare and richest groups.

	\begin{figure}[]
	\centerline{\psfig{figure=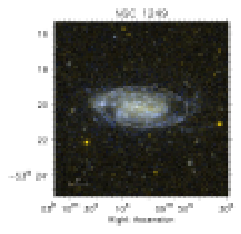,width=4.1cm} \psfig{figure=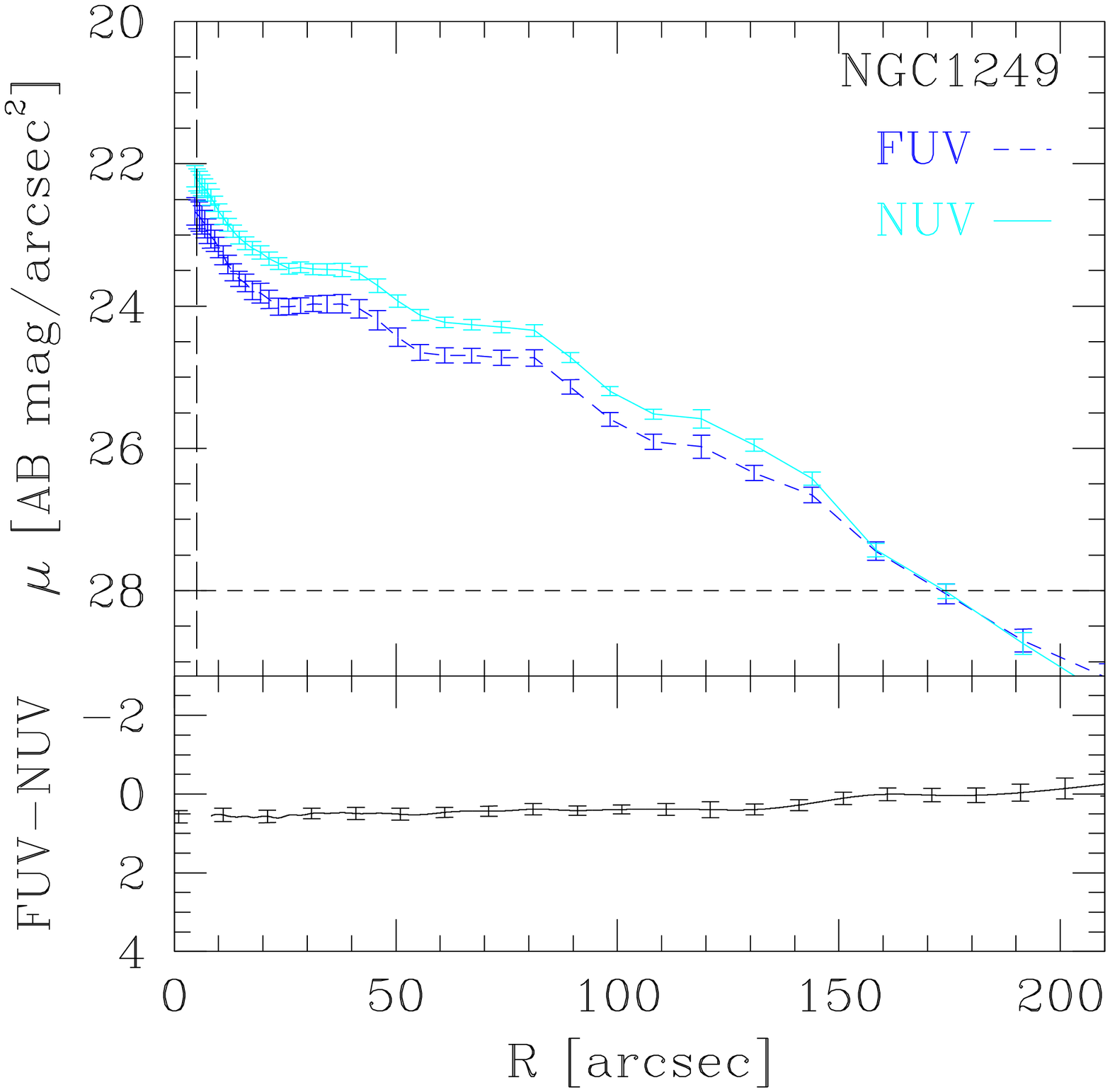,width=4cm}}
	\centerline{\psfig{figure=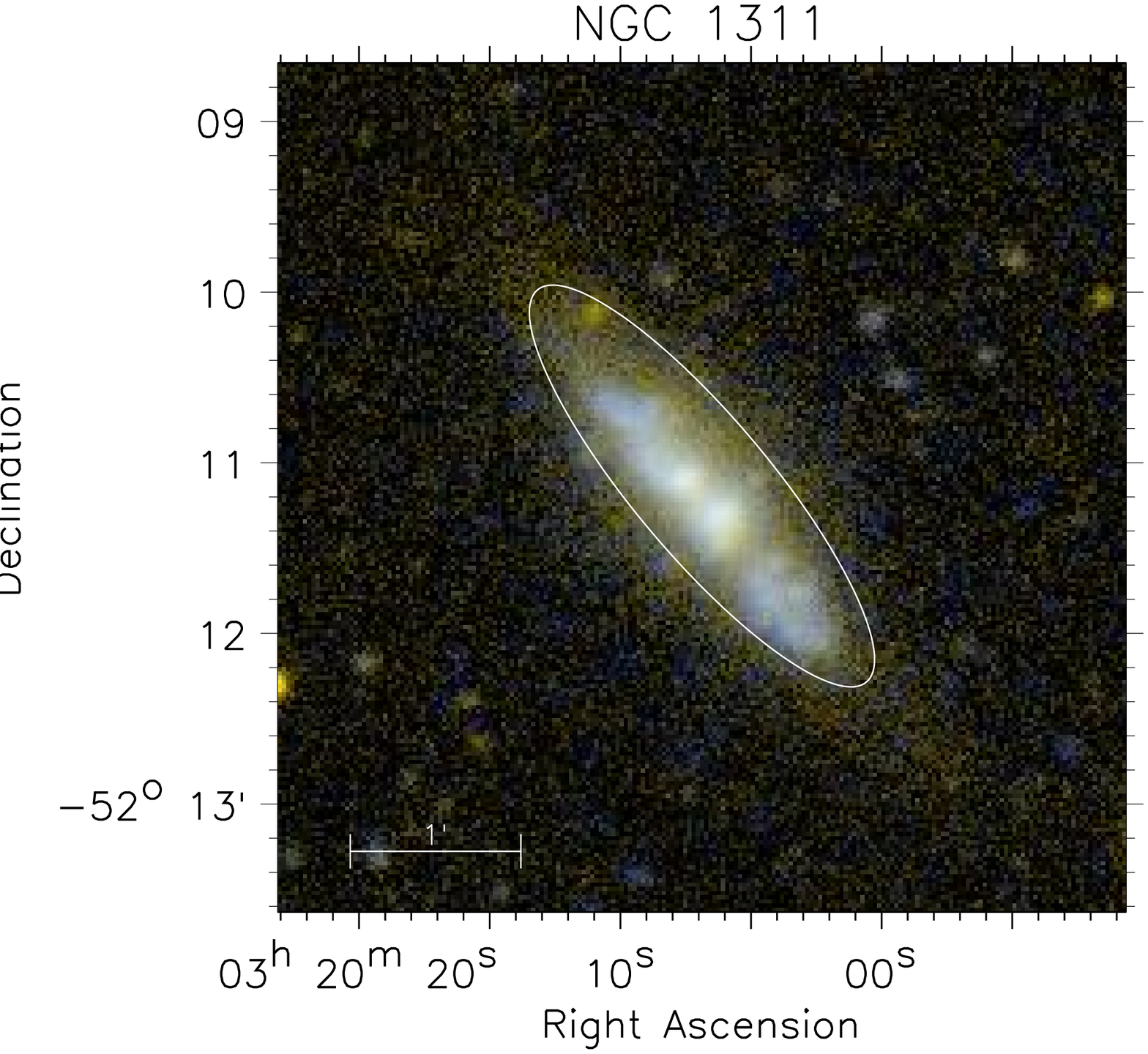,width=4.1cm} \psfig{figure=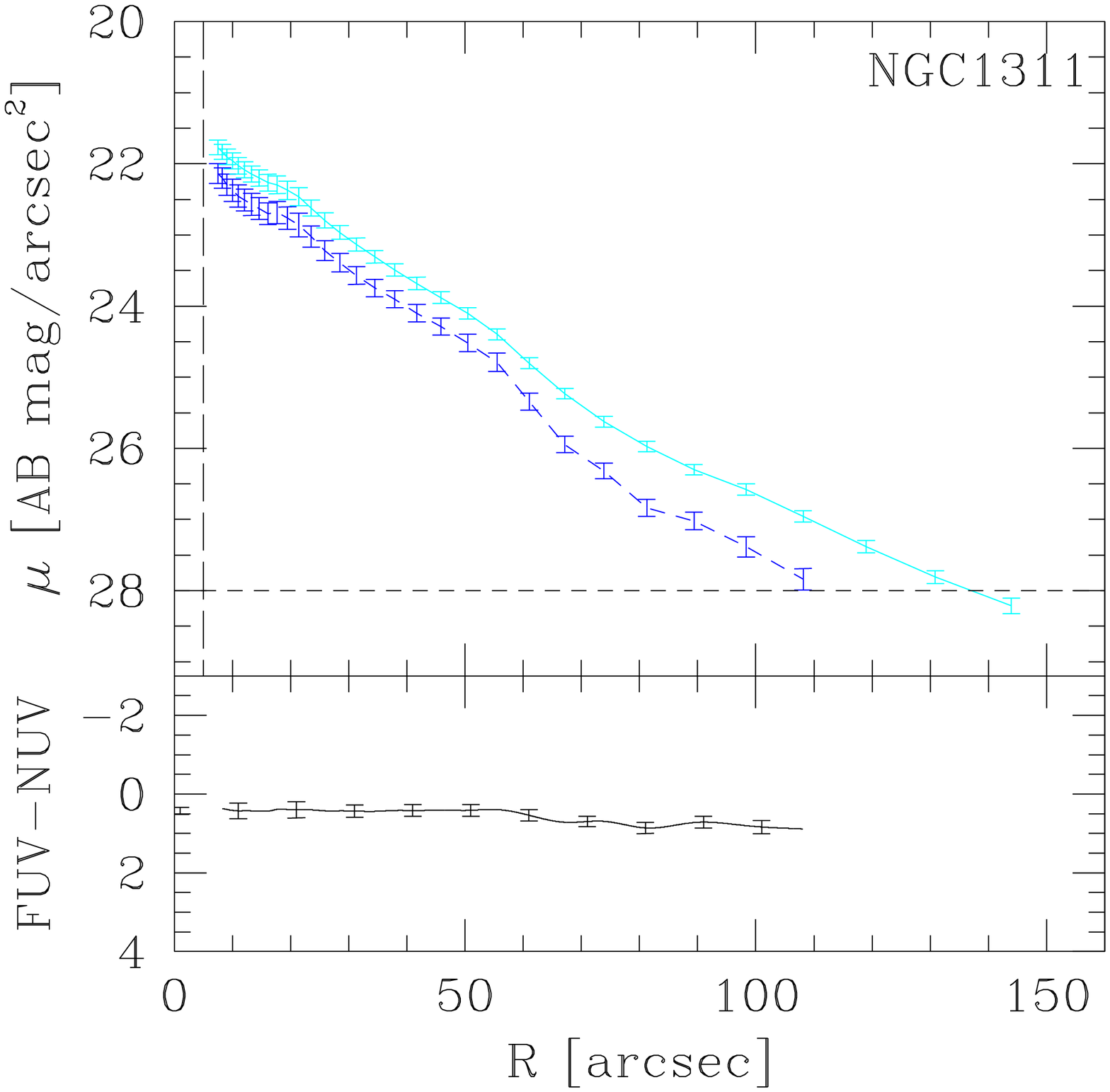,width=4cm}}
	\centerline{\psfig{figure=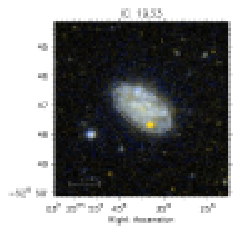,width=4.1cm} \psfig{figure=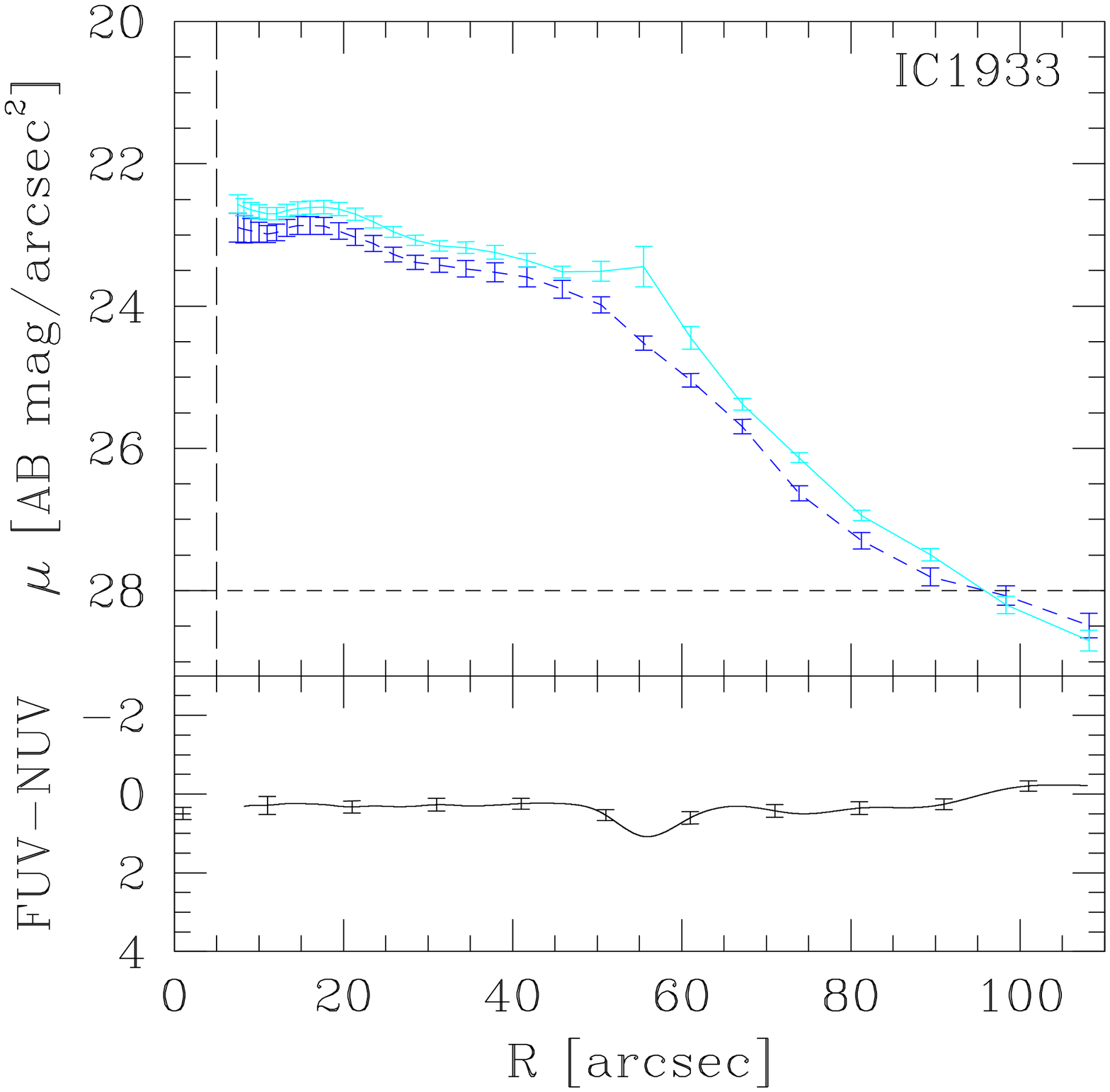,width=4cm}}
	\centerline{\psfig{figure=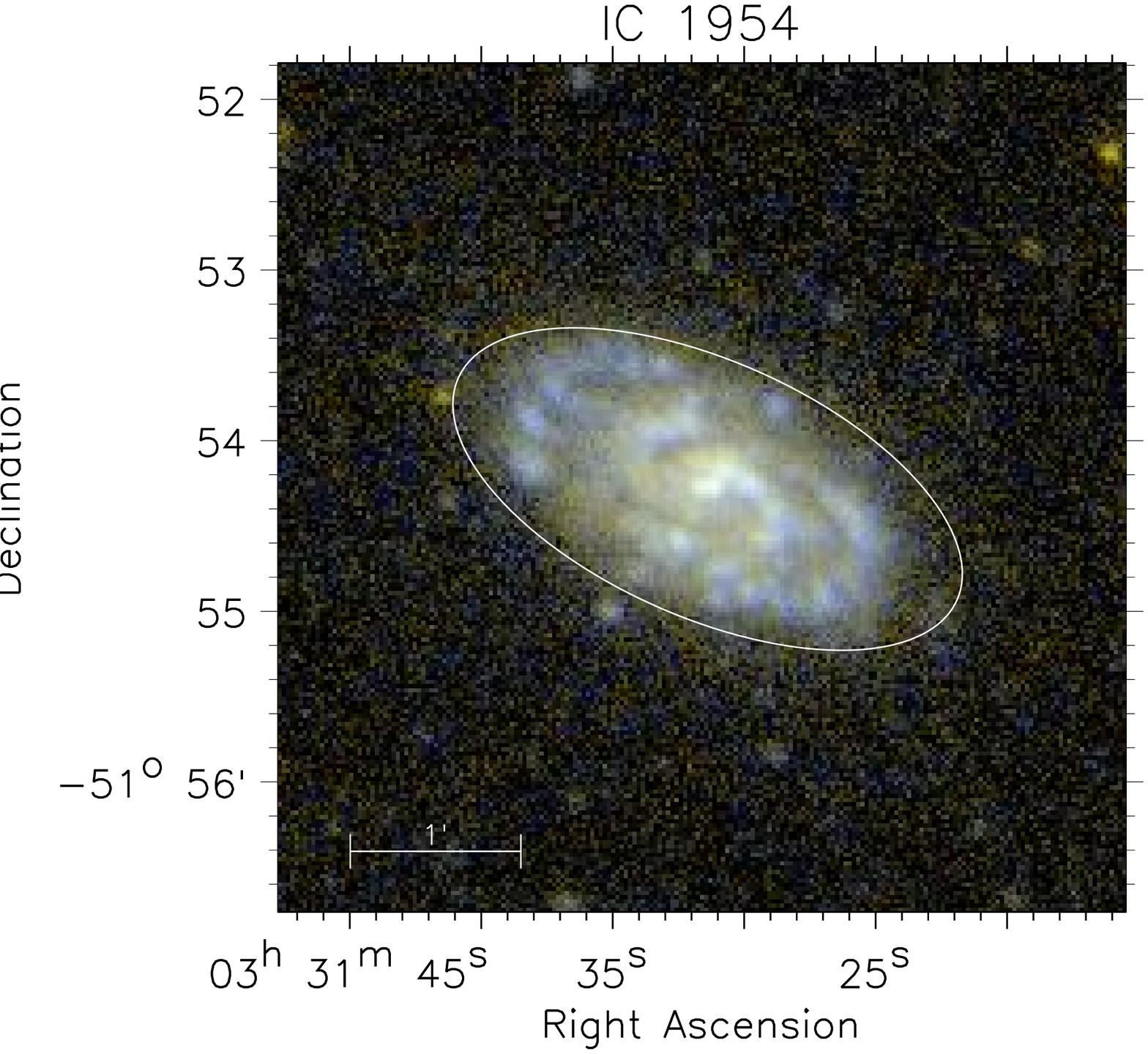,width=4.1cm} \psfig{figure=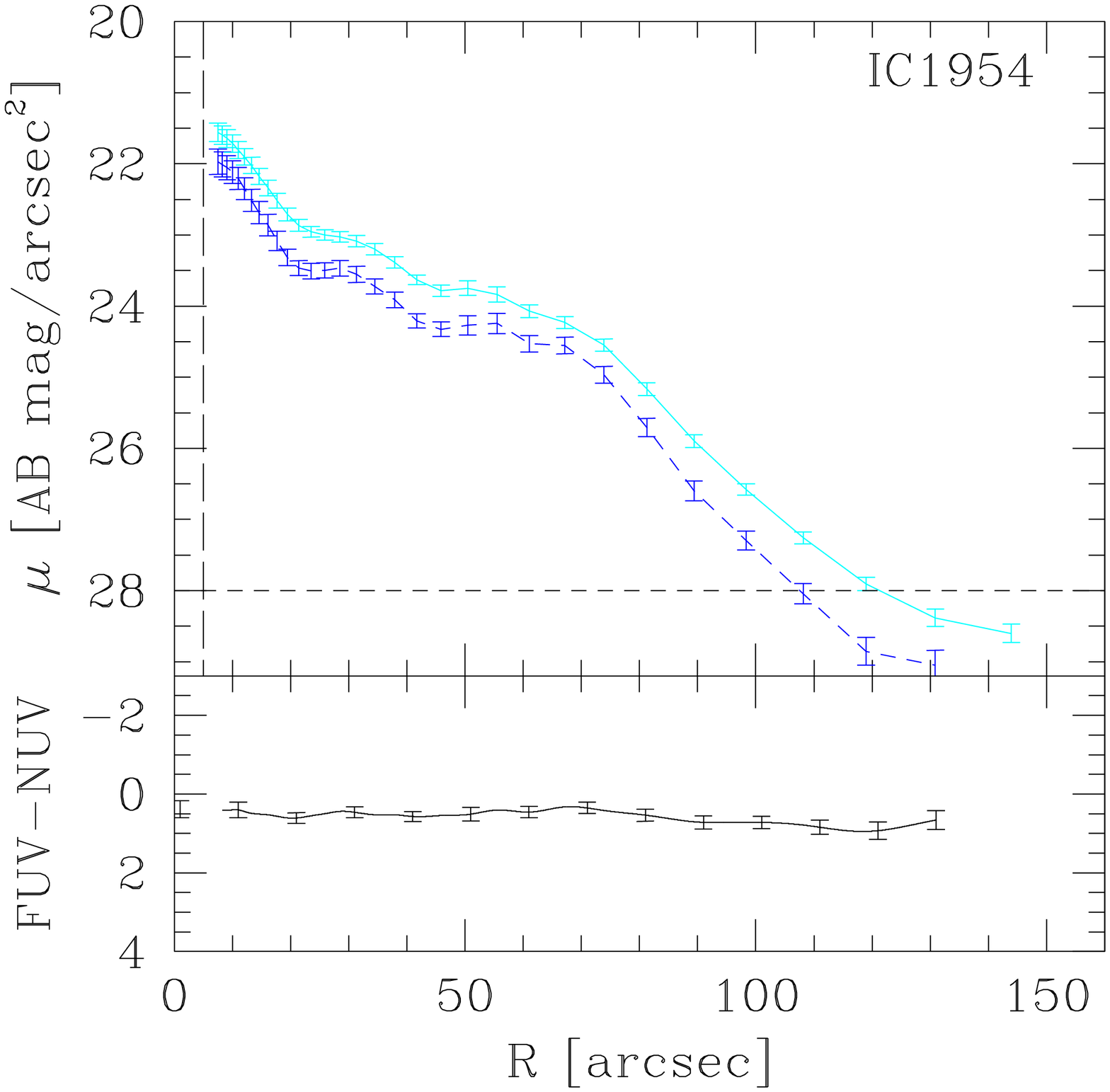,width=4cm}}
	\centerline{\psfig{figure=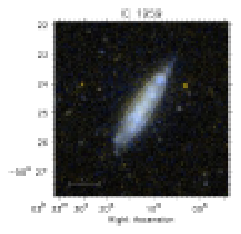,width=4.1cm} \psfig{figure=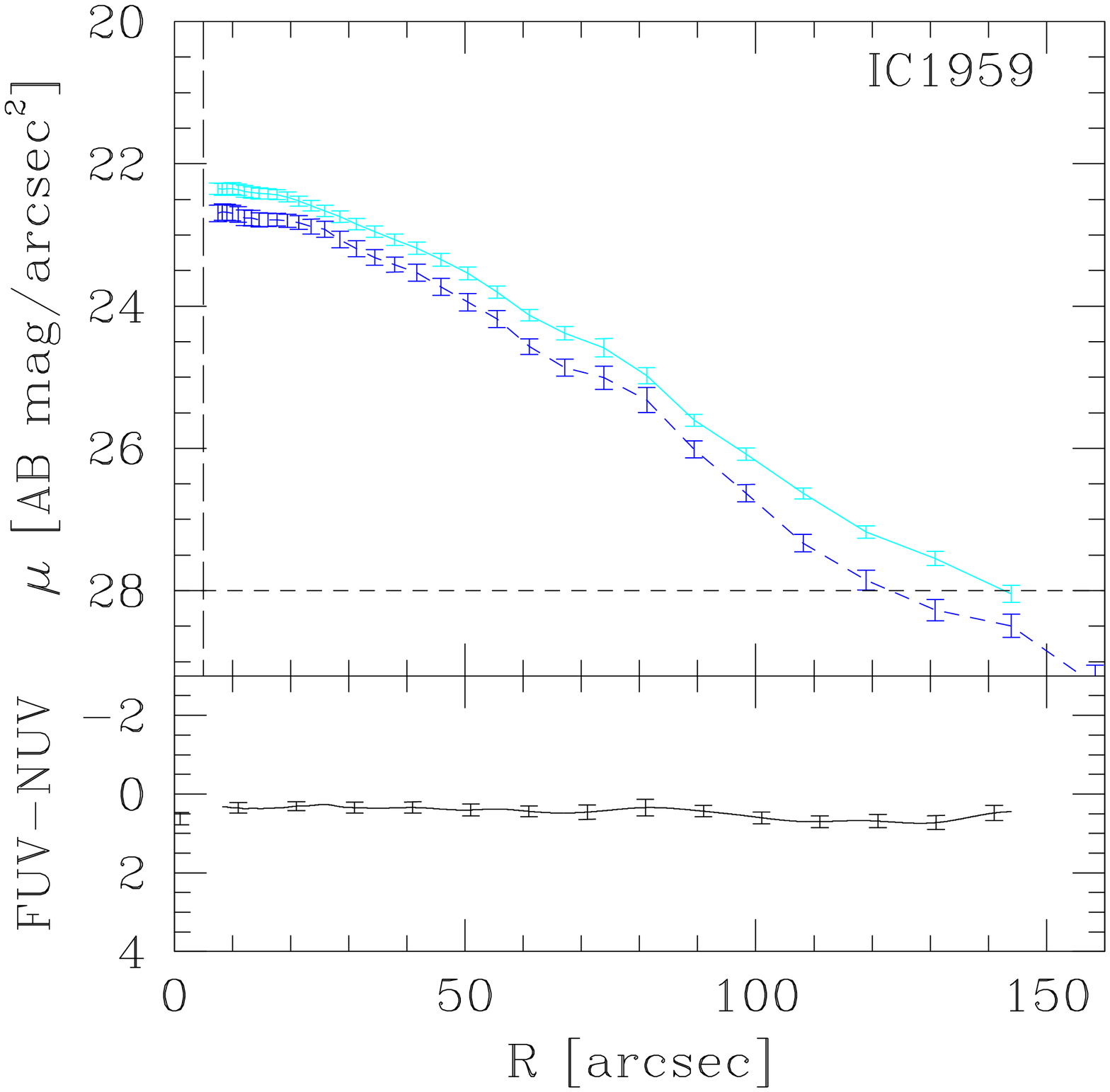,width=4cm}}
	\caption{Left: {\it GALEX} FUV  and NUV composite images of 
	LGG 93 members. Ellipses mark the regions where 
	we measure the integrated magnitude (see text). Right: 
	{\it GALEX} surface brightness profiles (top panel) with the vertical 
	dashed line at 5 arcsec showing the  approximate FWHM of the
	{\it GALEX} point spread function and the horizontal line the nominal UV surface brightness. (FUV-NUV)  
	color profile (bottom panel) vs. the galactocentric distance along  the semi-major axis of the fitted ellipse are also shown.}  
	\label{93}
	\end{figure}

	\begin{figure}[]
	\centerline{\psfig{figure=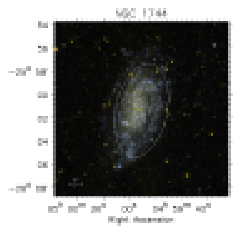,width=4.1cm} \psfig{figure=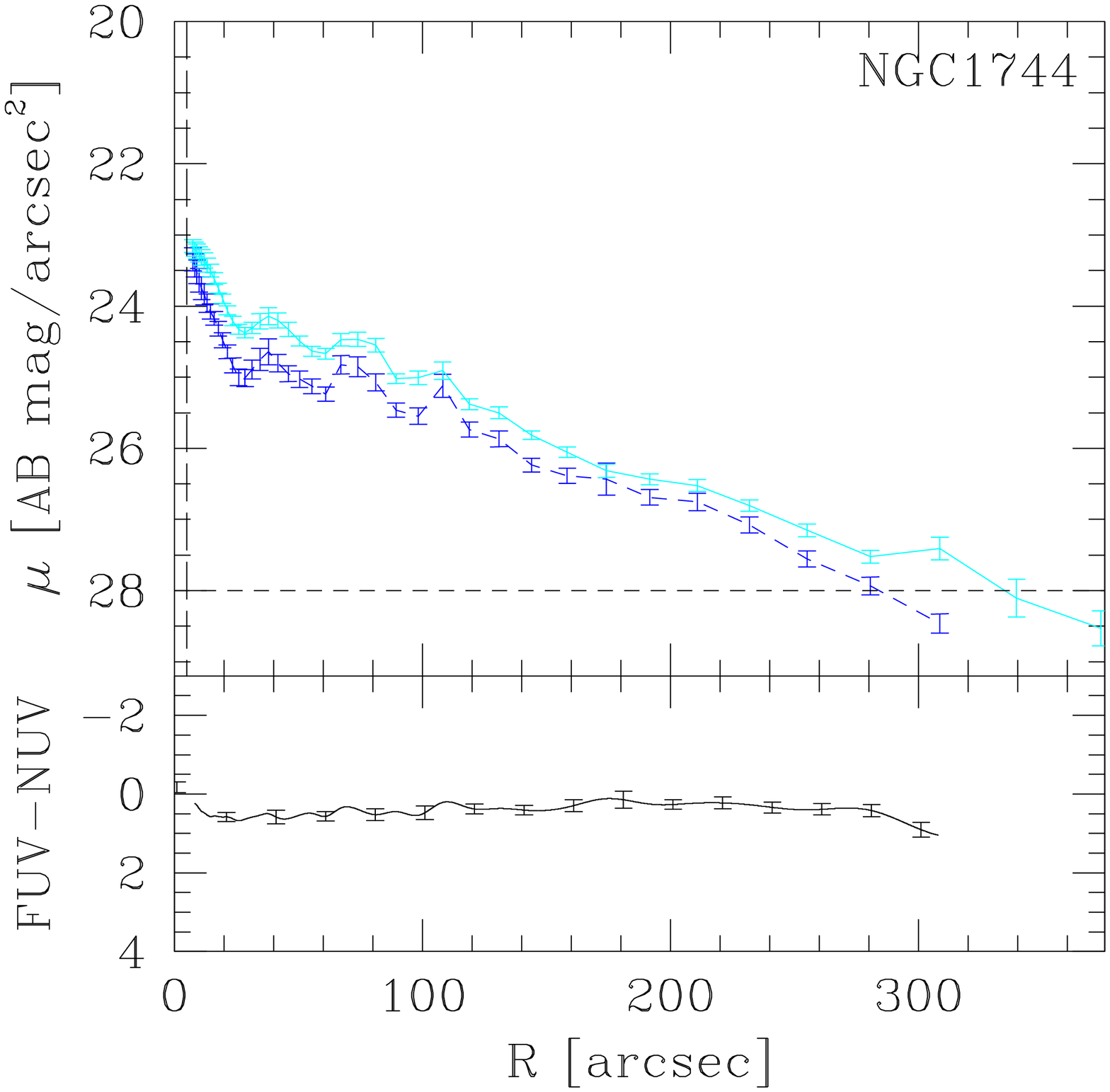,width=4cm}}
	\centerline{\psfig{figure=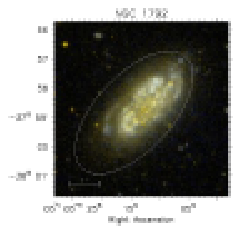,width=4.1cm} \psfig{figure=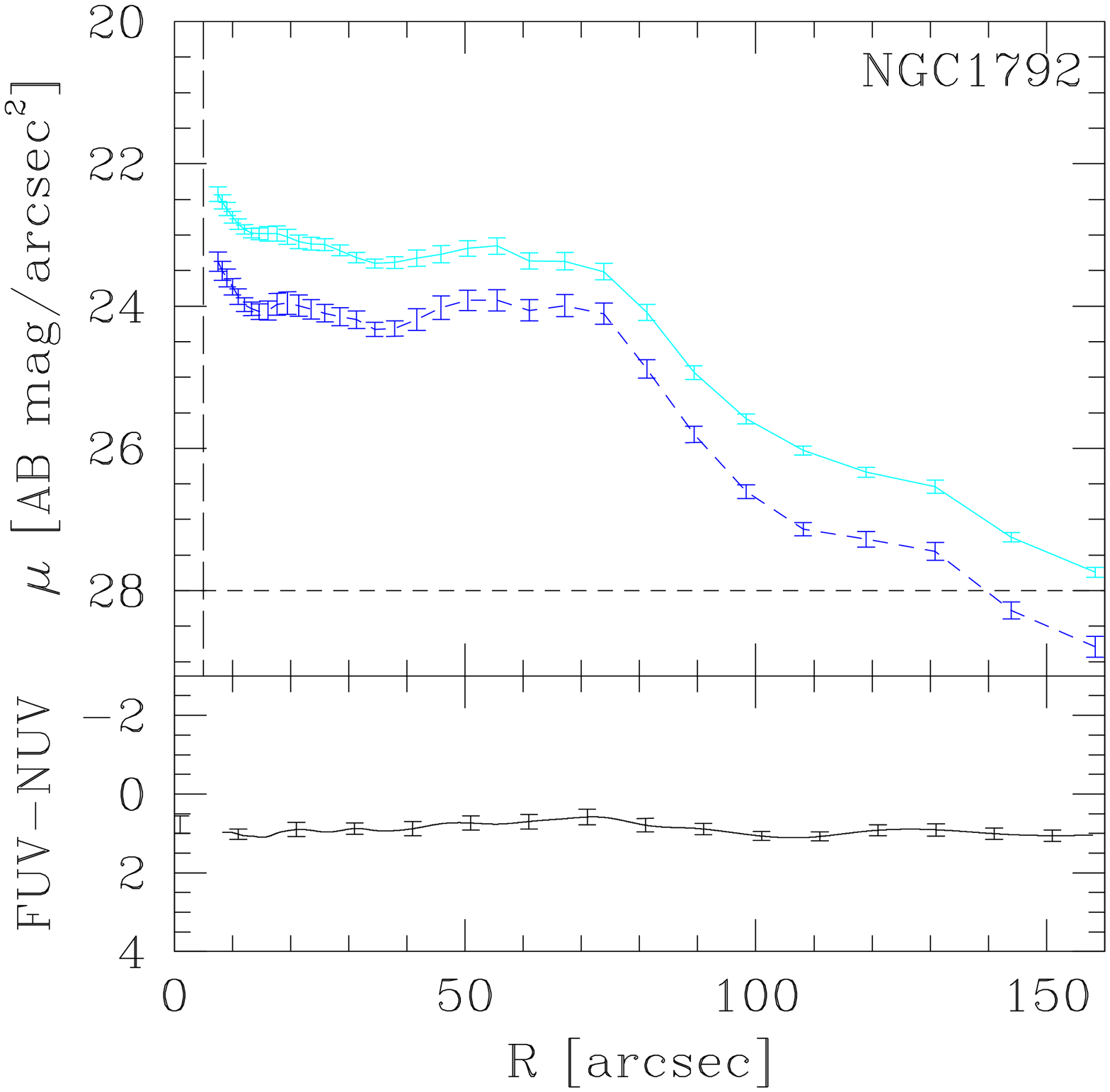,width=4cm}}
	\centerline{\psfig{figure=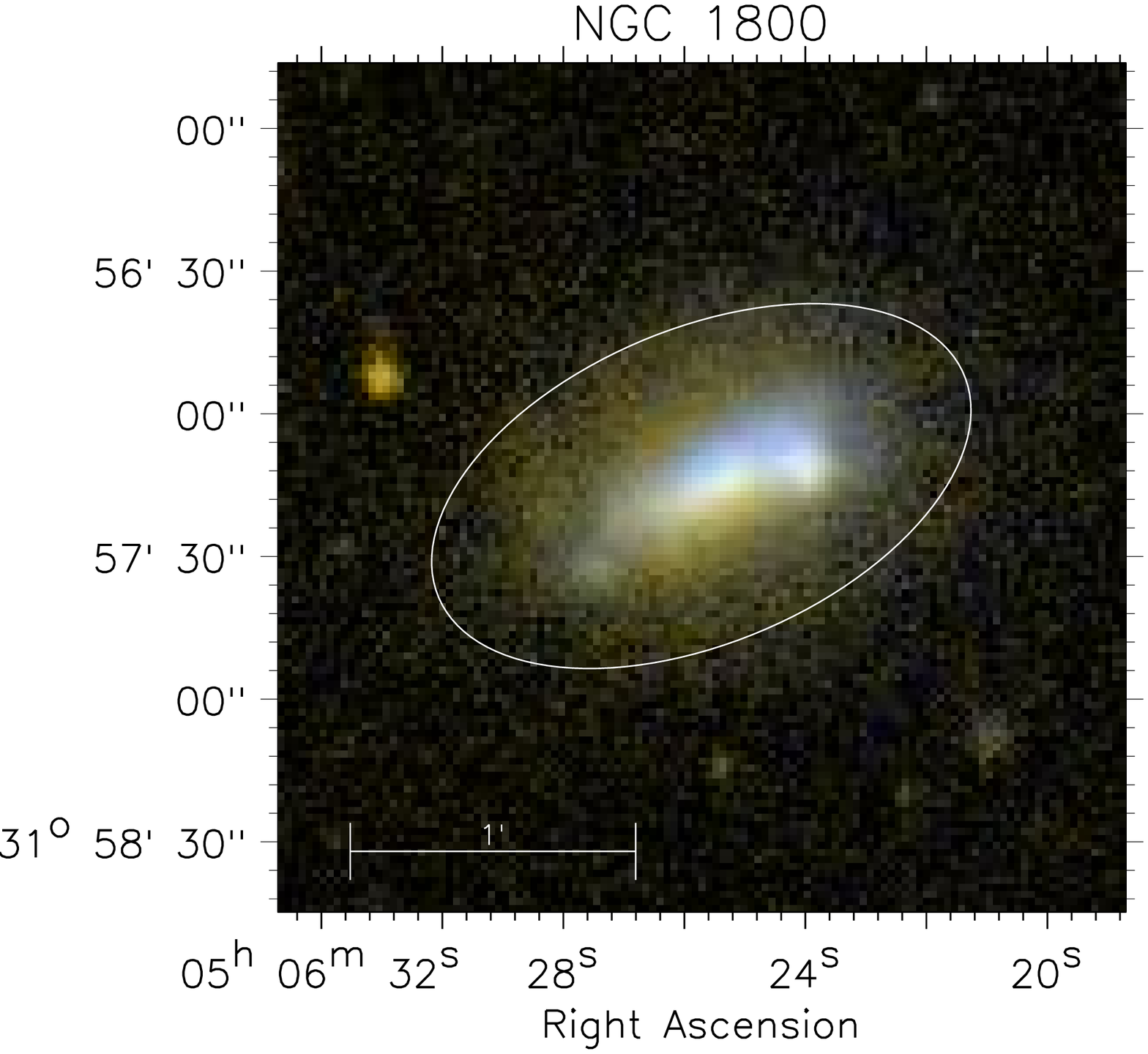,width=4.1cm} \psfig{figure=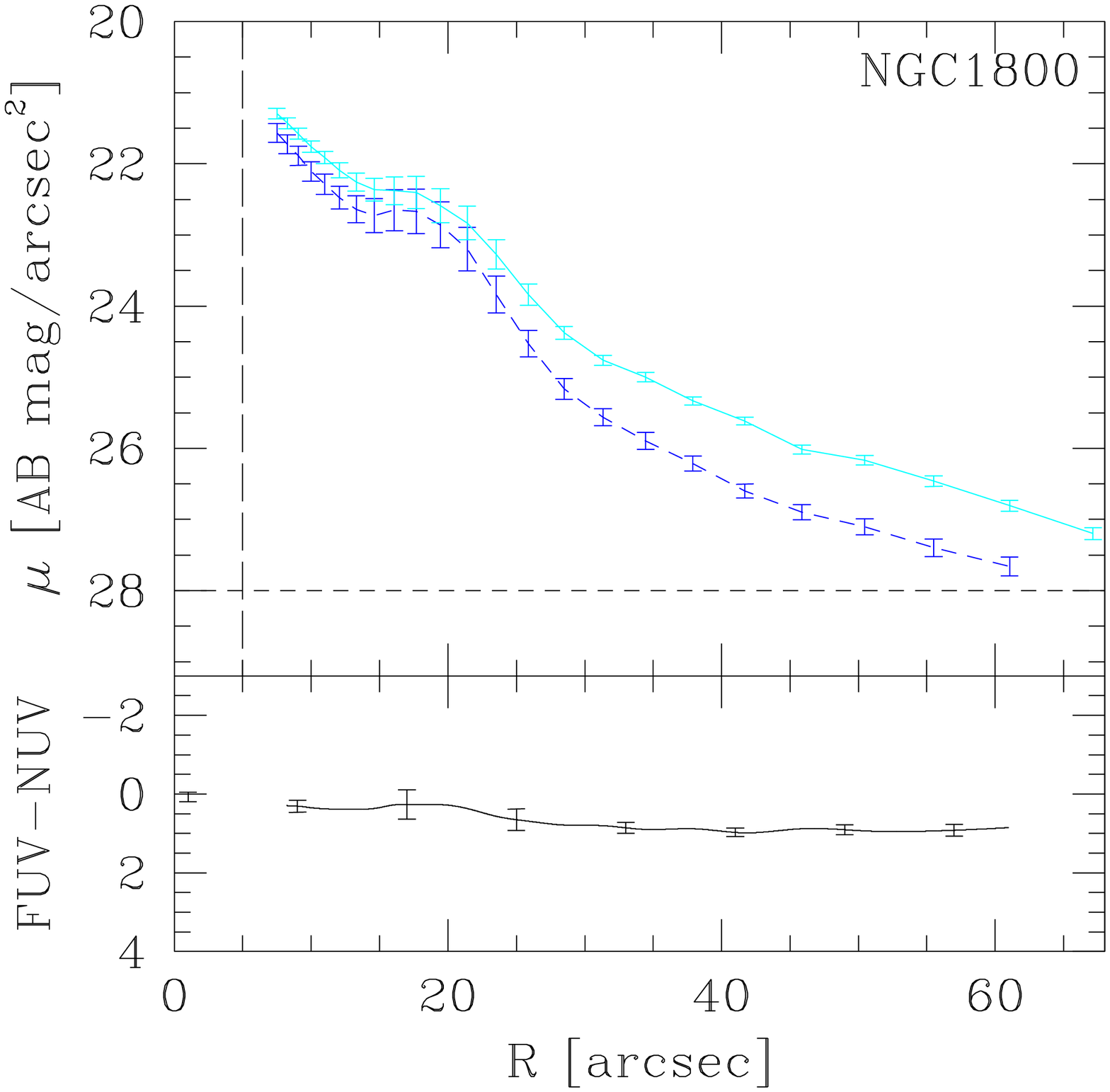,width=4cm}}
	\centerline{\psfig{figure=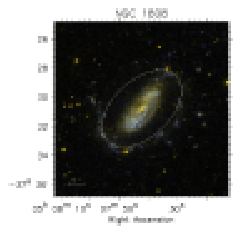,width=4.1cm} \psfig{figure=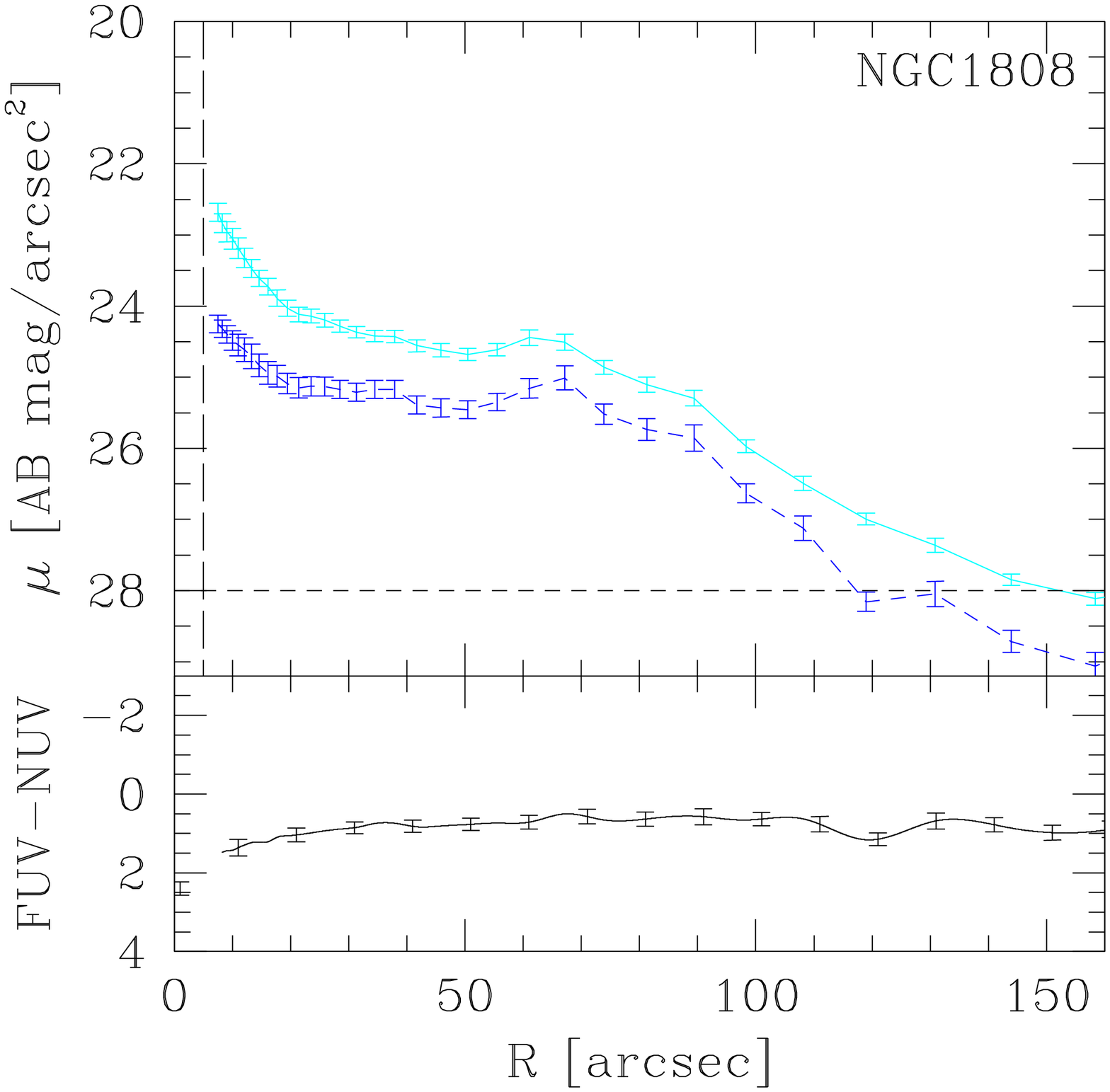,width=4cm}}
	\centerline{\psfig{figure=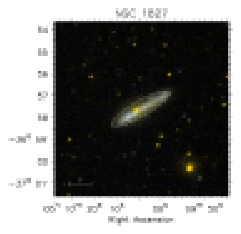,width=4.1cm} \psfig{figure=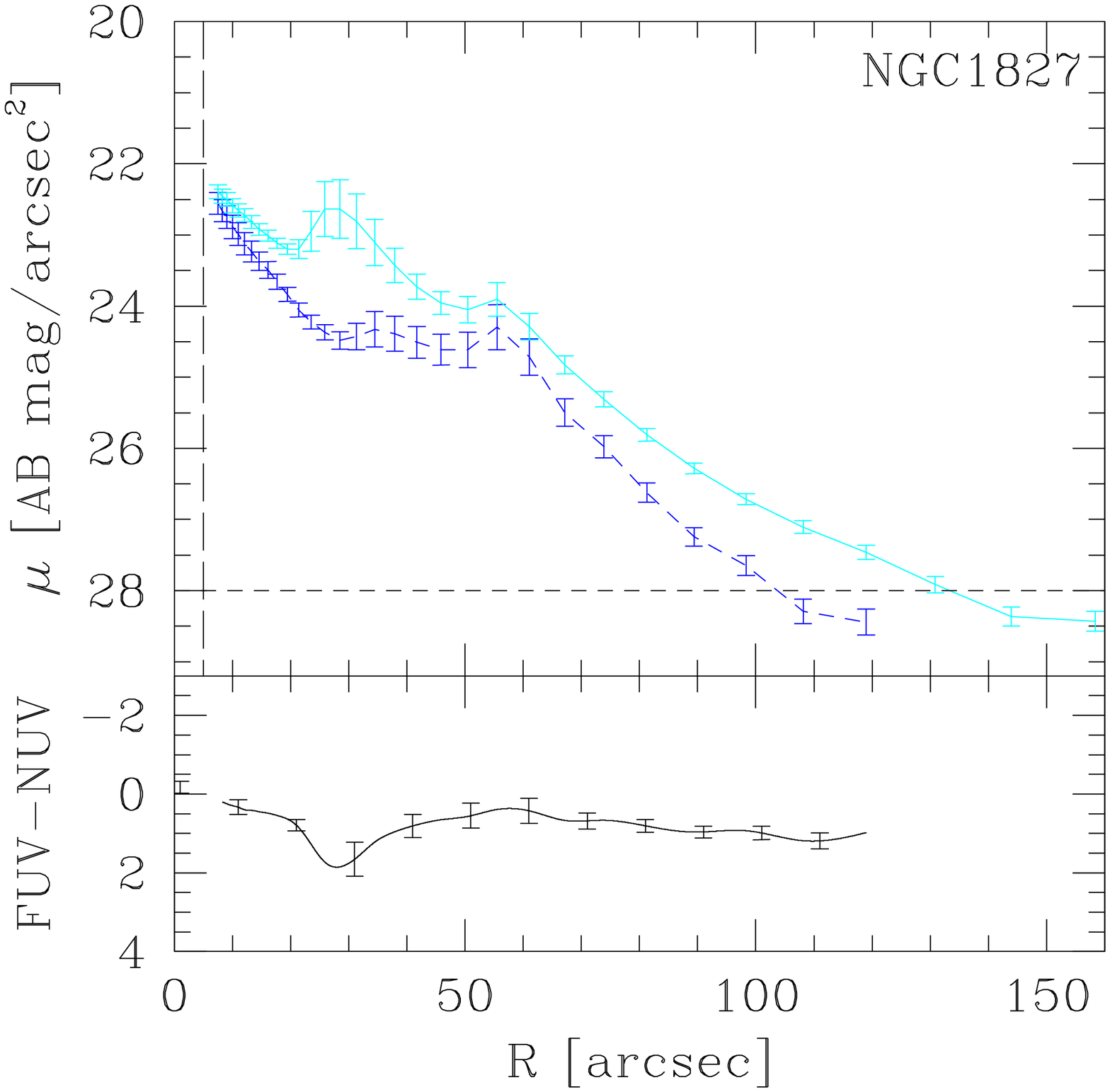,width=4cm}}
	 \caption{As in Fig. \ref{93} for LGG 127 members.} 
	\label{127}
	\end{figure}

	\begin{figure}[]
	\centerline{\psfig{figure=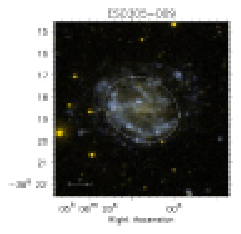,width=4.1cm}  \psfig{figure=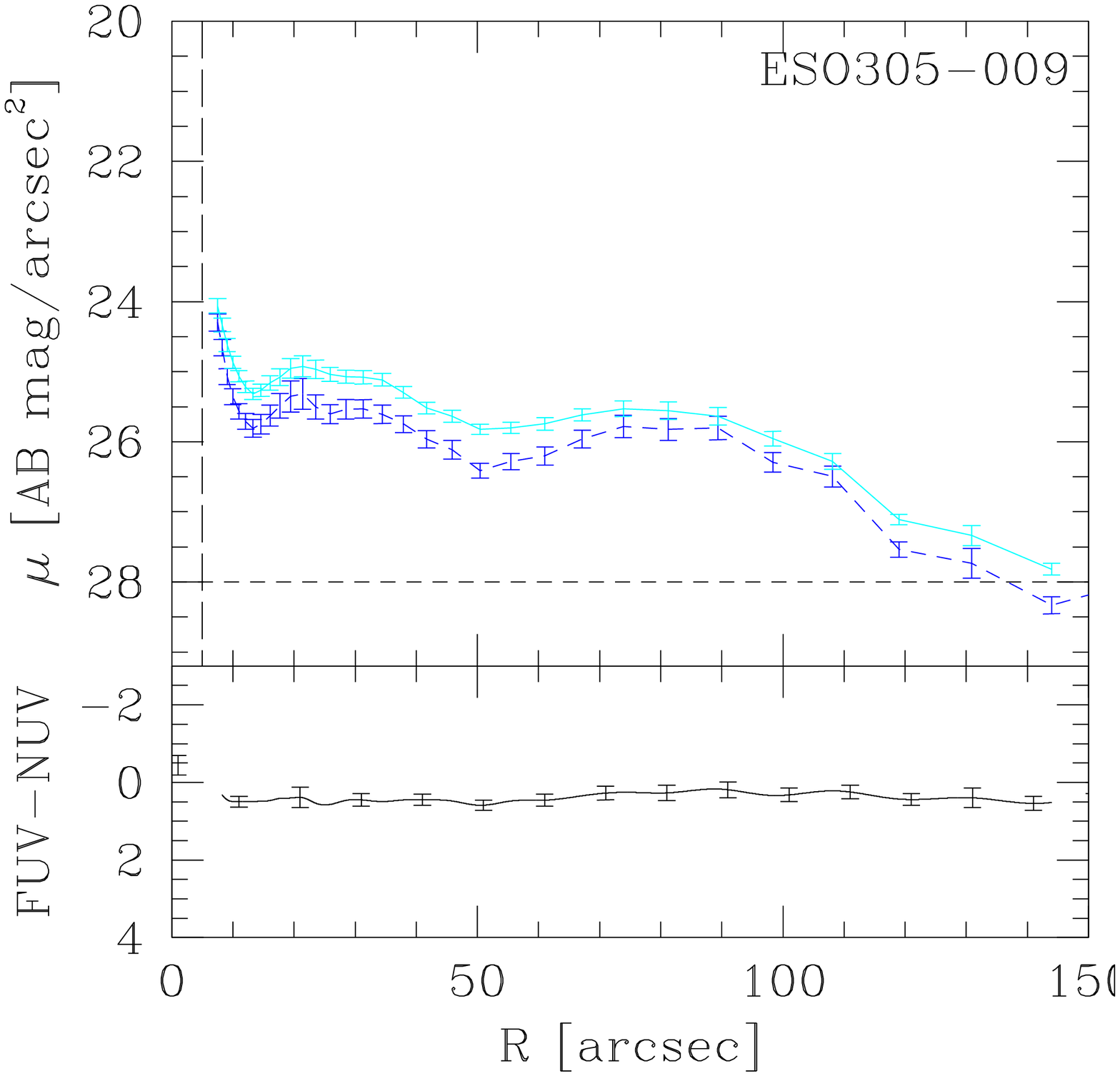,width=4cm}}
	\centerline{\psfig{figure=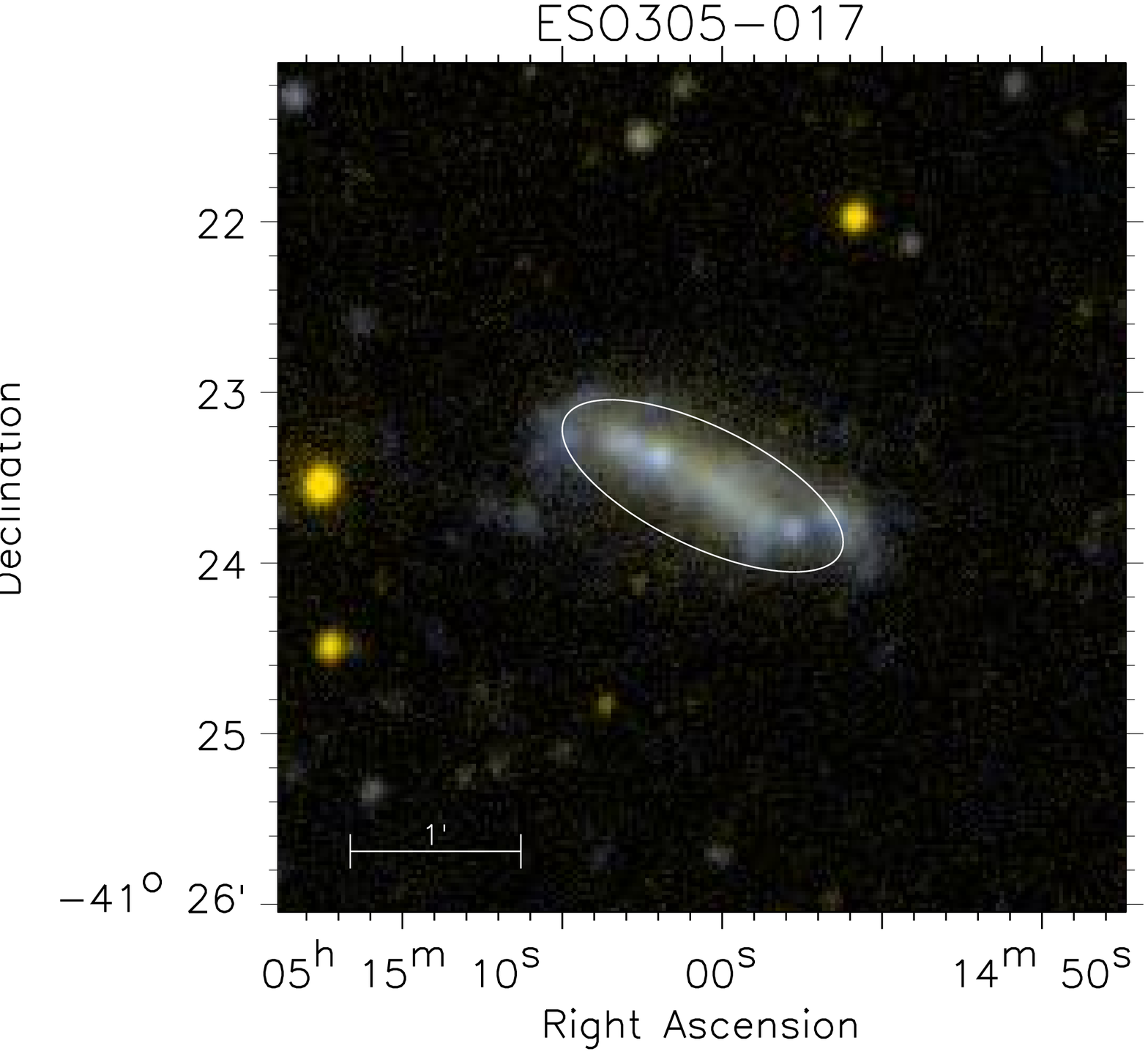,width=4.1cm}  \psfig{figure=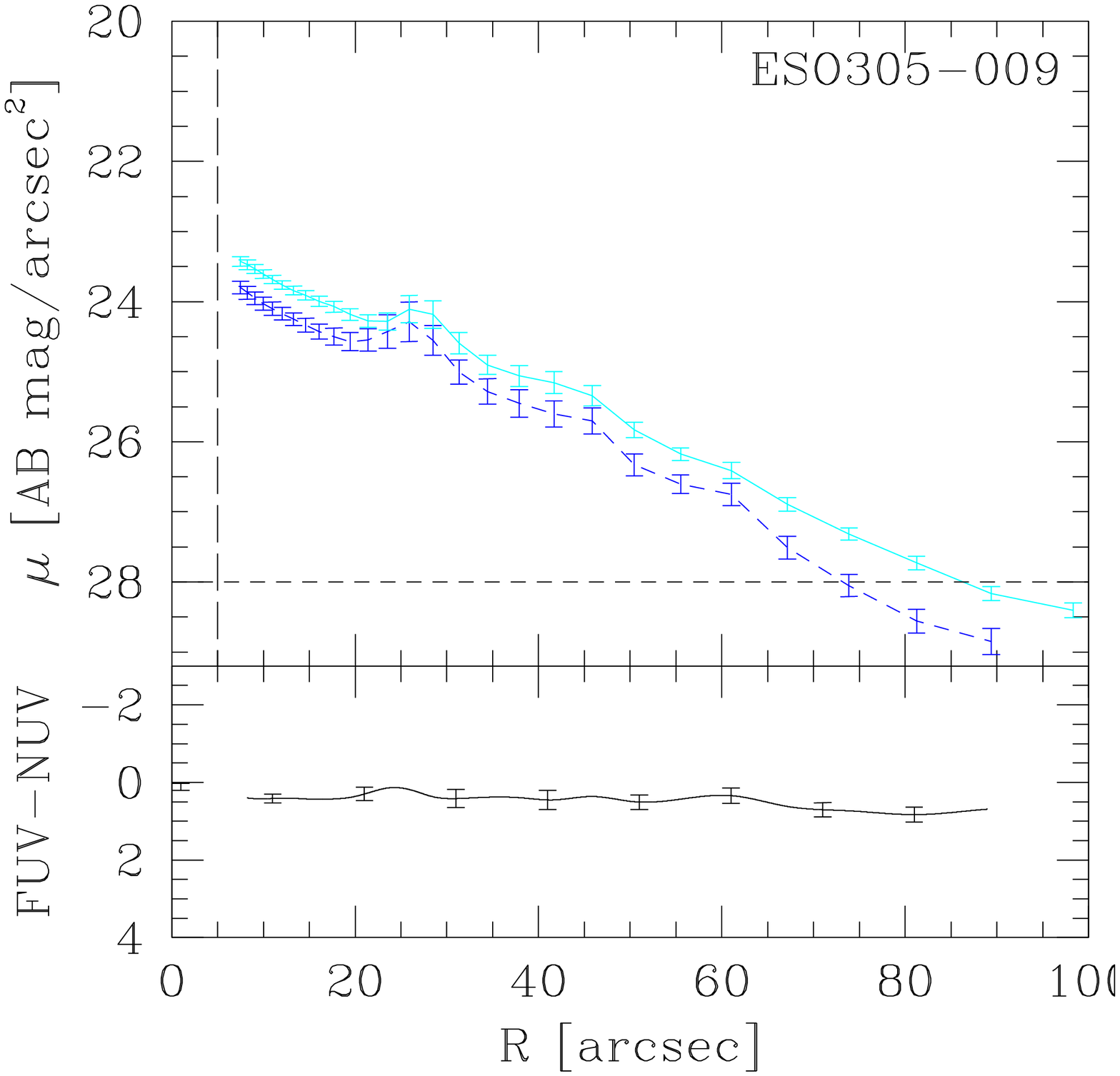,width=4cm}}
	\centerline{\psfig{figure=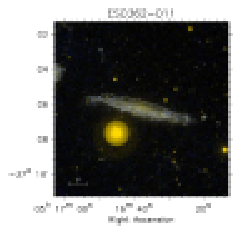,width=4.1cm} \psfig{figure=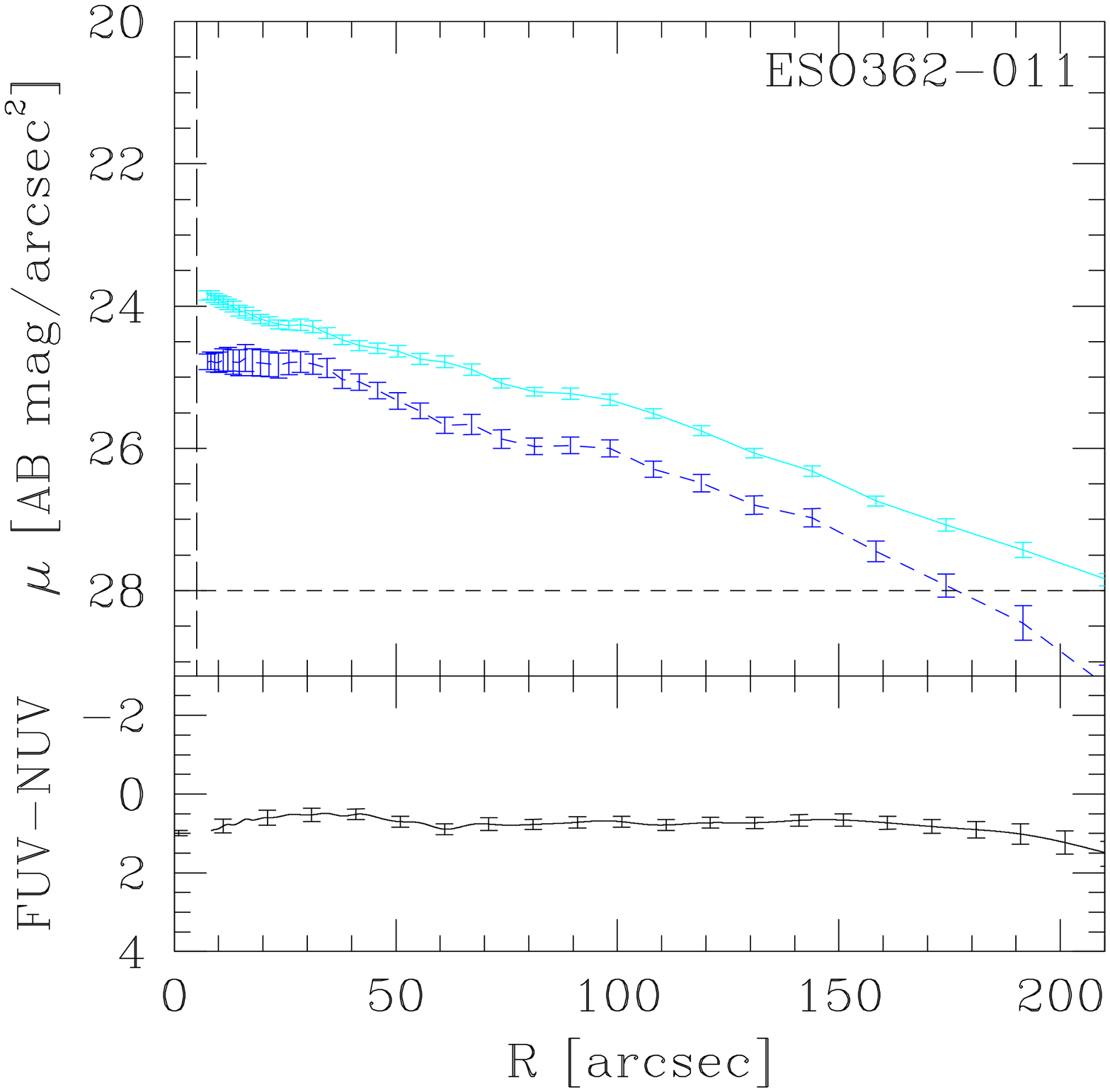,width=4cm}}	
	\centerline{\psfig{figure=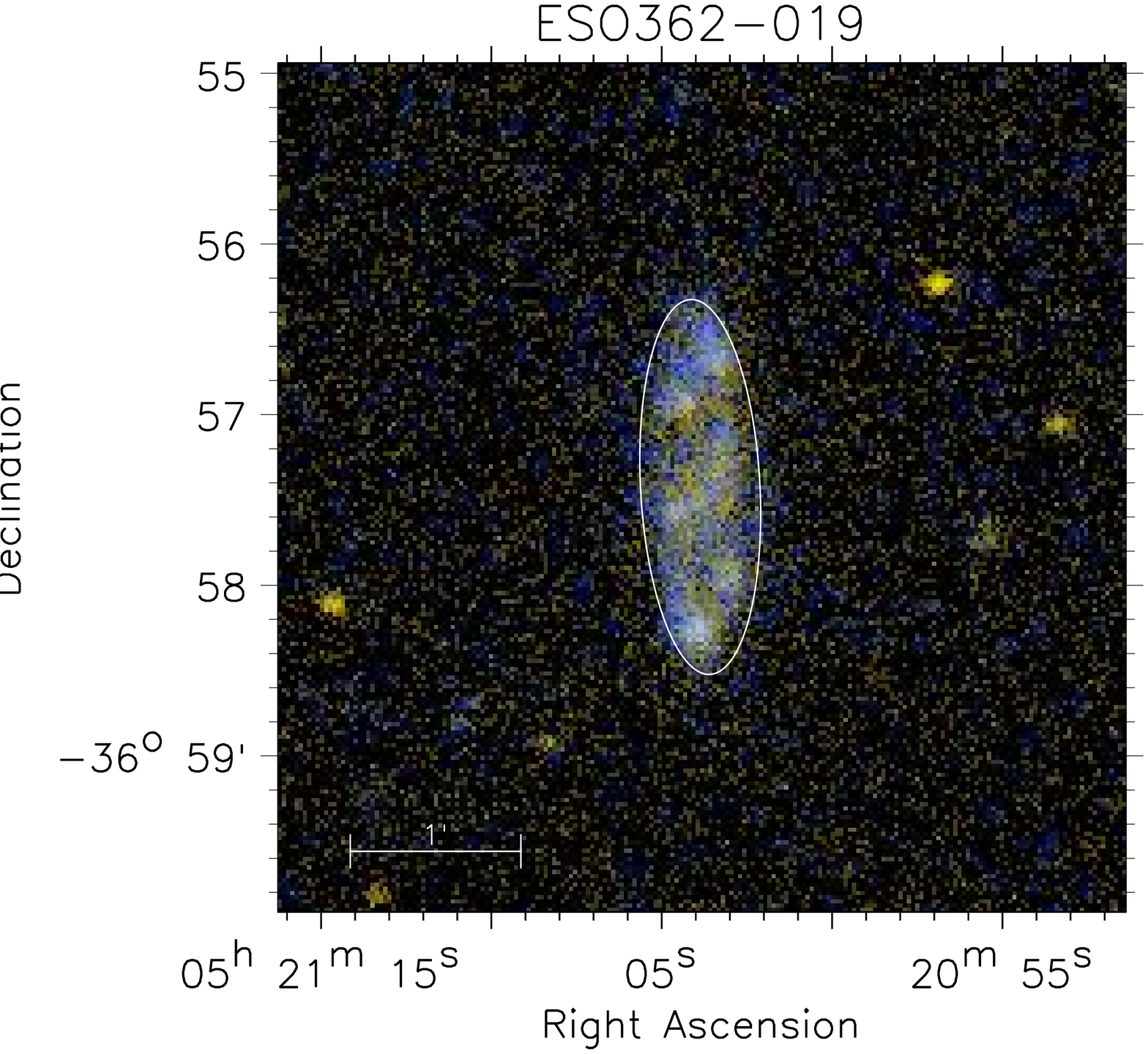,width=4.1cm}  \psfig{figure=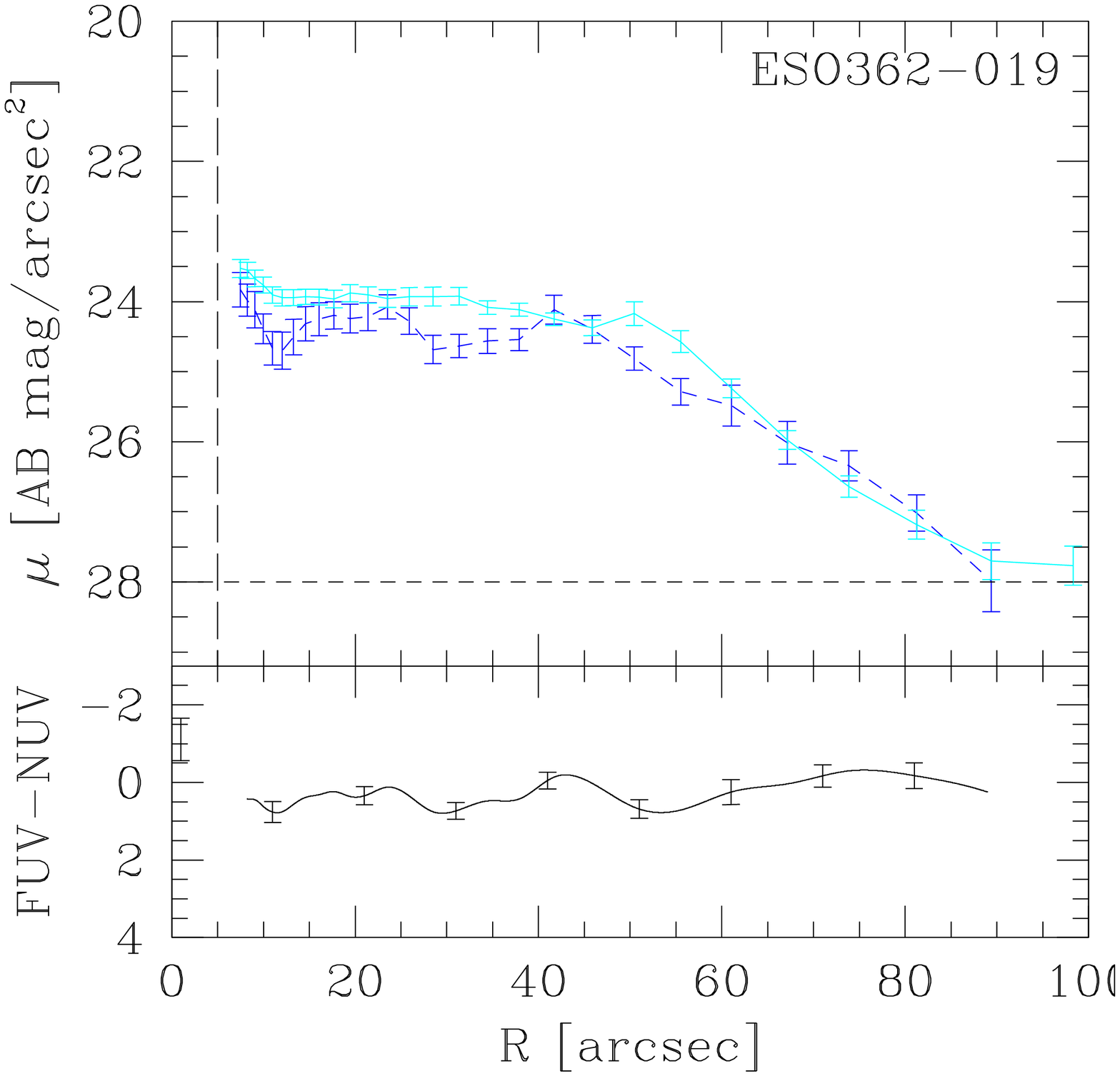,width=4cm}}
	 \addtocounter{figure}{-1}
	 \caption{continued.}
	\end{figure}

	\section{Observations}

	The UV imaging was obtained with  GALEX  \citep{Martin05, 
	Morrissey07} in two ultraviolet bands, FUV (1344 -- 1786 \AA) ~ and NUV (1771 -- 2831
	\AA). The instrument 
	has a very wide field of view (1.25 degrees diameter) and a spatial resolution    
	$\approx$4\farcs2  and 5\farcs3 FWHM in FUV and NUV
	 respectively, sampled with 1\farcs 5$\times$1\farcs 5 pixels \citep{Morrissey07}.  
	 
	We observed   
	our sample groups as 
	part of {\it GALEX} GI2 program 121, P.I. L.M. Buson. 
	In order to analyze all members of the three groups we also searched for
	other observations in the public {\it GALEX} archive (see Table~2).  
	The exposure times of our  sample range from $\sim$ 100 sec for All Sky Survey (AIS) archival data  
	(limiting AB magnitude in FUV/NUV of $\sim$ 19.9/20.8) to $\sim$ 8 ksec.
	 AIS archival data used to complete the sample have exposure 
	times $\sim$ 20 times shorter than our GI data, therefore a $\sim$ 3.2 mag brighter limit.

	We used FUV and NUV   background-subtracted intensity 
	images from the {\it GALEX} pipeline to compute integrated photometry 
	of the galaxies and light profiles, as described
	in the next Section. Background counts were estimated from the sky 
	 background image and  high resolution relative response map 
	provided by the {\it GALEX} pipeline (Section 4.2). 

	In addition, we used optical     
	 Sloan Digital Sky Survey (SDSS) archival data \citep{Ade08} in
	the u [2980-4130 \AA], g [3630-5830 \AA], r [5380-7230 \AA], i
	[6430-8630 \AA], z [7730-11230 \AA] bands available for LGG 225, and B magnitudes
	from Table 1 for the other two groups.

	\begin{landscape}
	\begin{figure*}[!h]
	\begin{tabular}{ccccccc}
	\hspace{-0.8cm}
	 \includegraphics[width=4.cm]{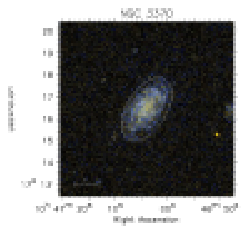}&
	 \includegraphics[width=4.cm]{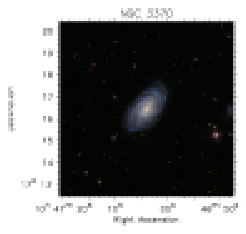}&
	 \includegraphics[width=3.5cm]{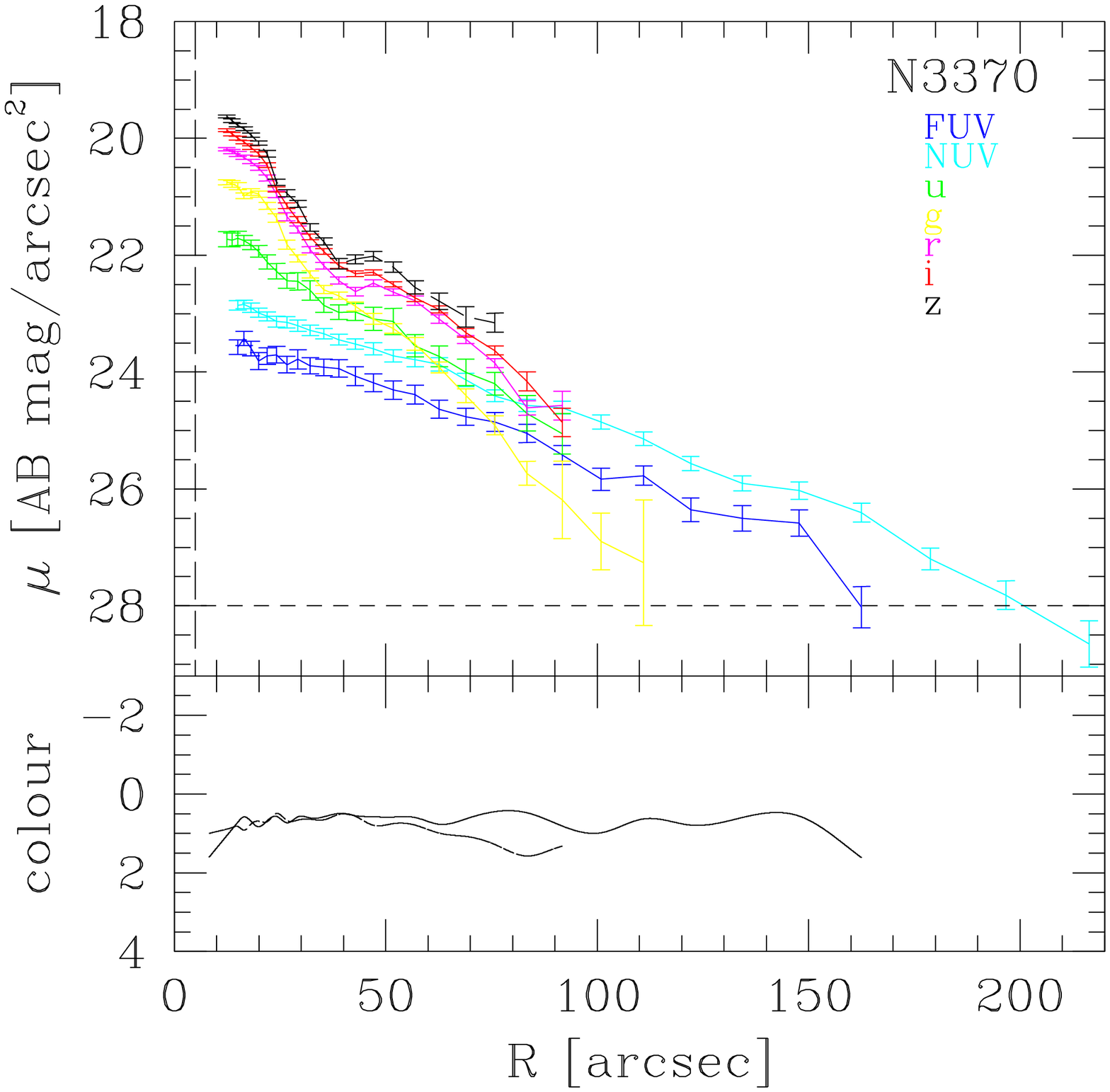}& 
	 \includegraphics[width=4.cm]{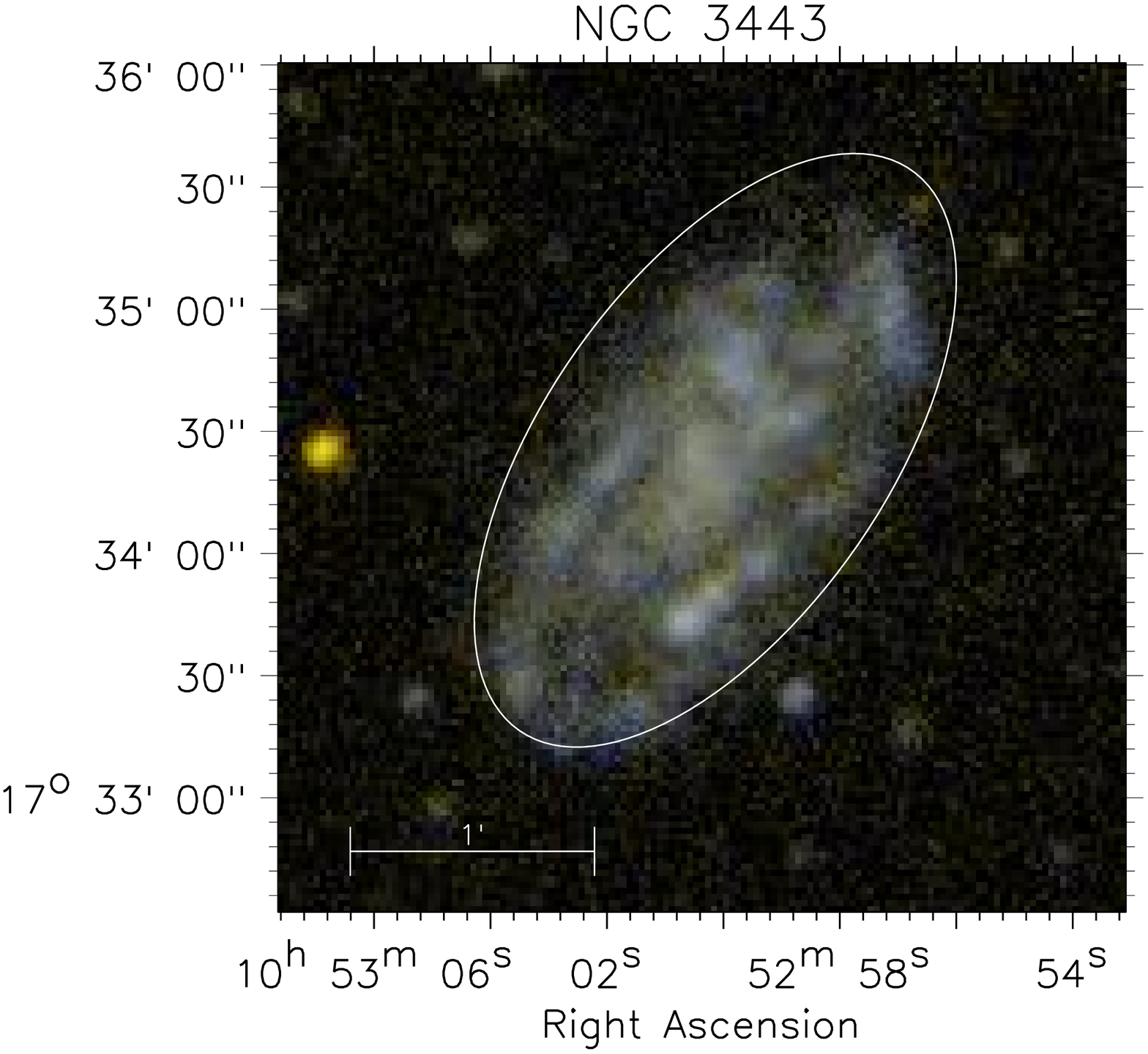} &
	 \includegraphics[width=4.cm]{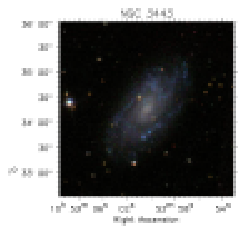} &
	 \includegraphics[width=3.5cm]{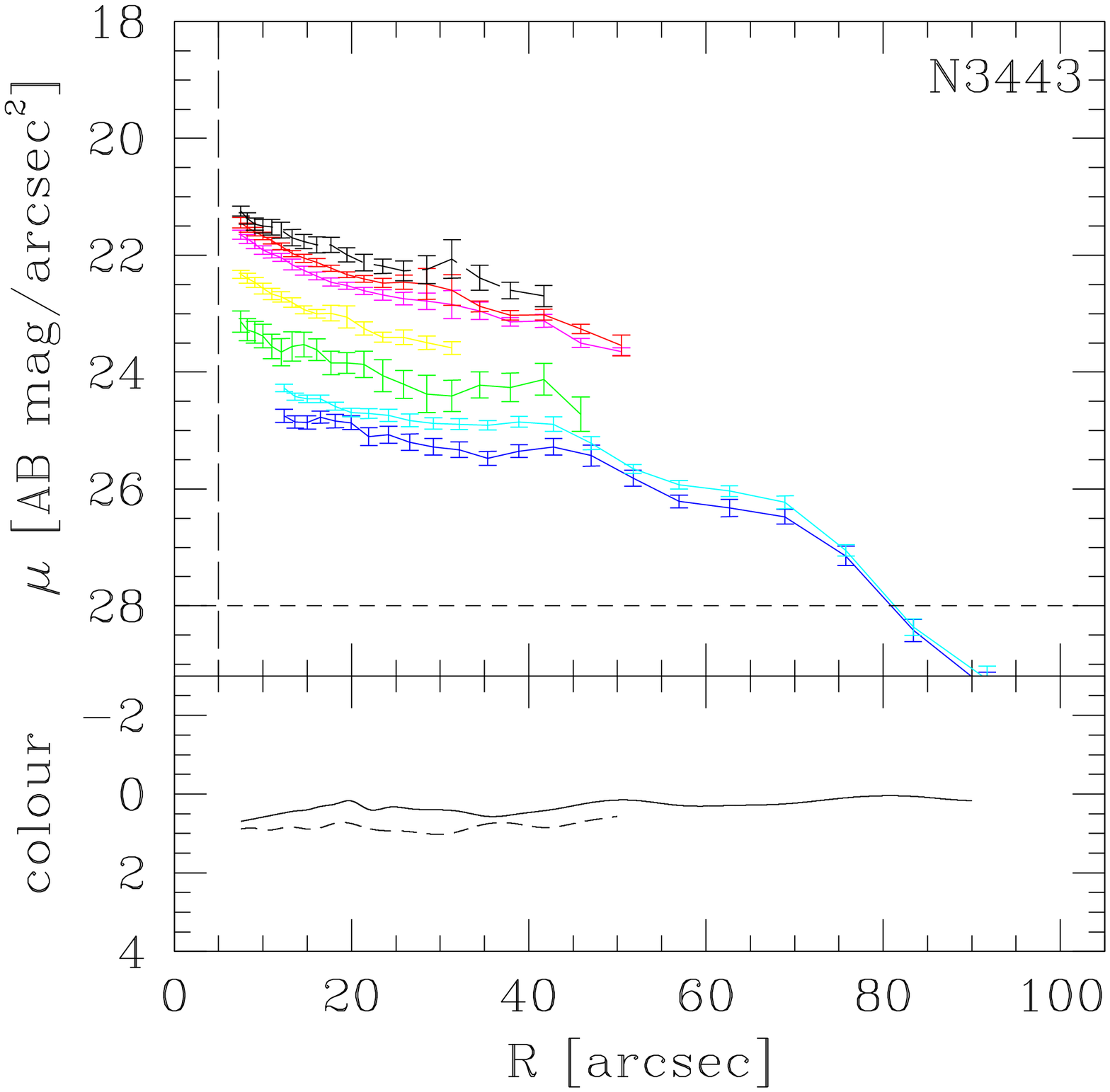}&\\
	 \hspace{-0.8cm}
	 \includegraphics[width=4.cm]{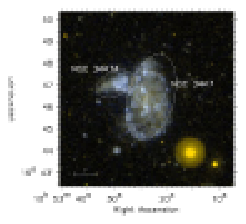}  &
	 \includegraphics[width=4.cm]{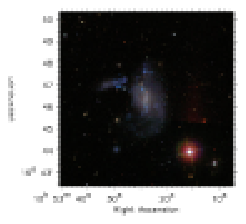}& 
	 \includegraphics[width=3.5cm]{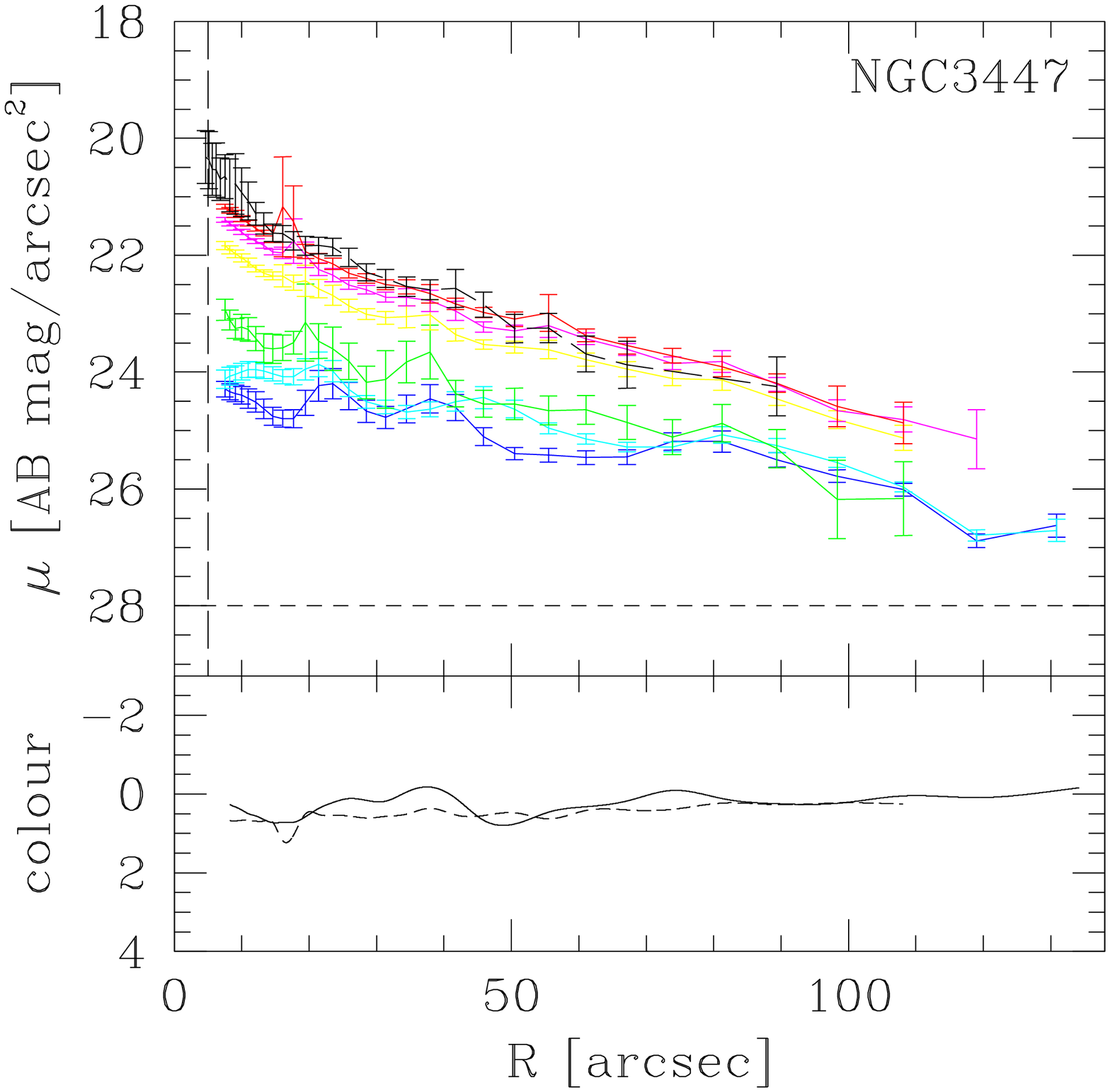}&
	 \includegraphics[width=4.cm]{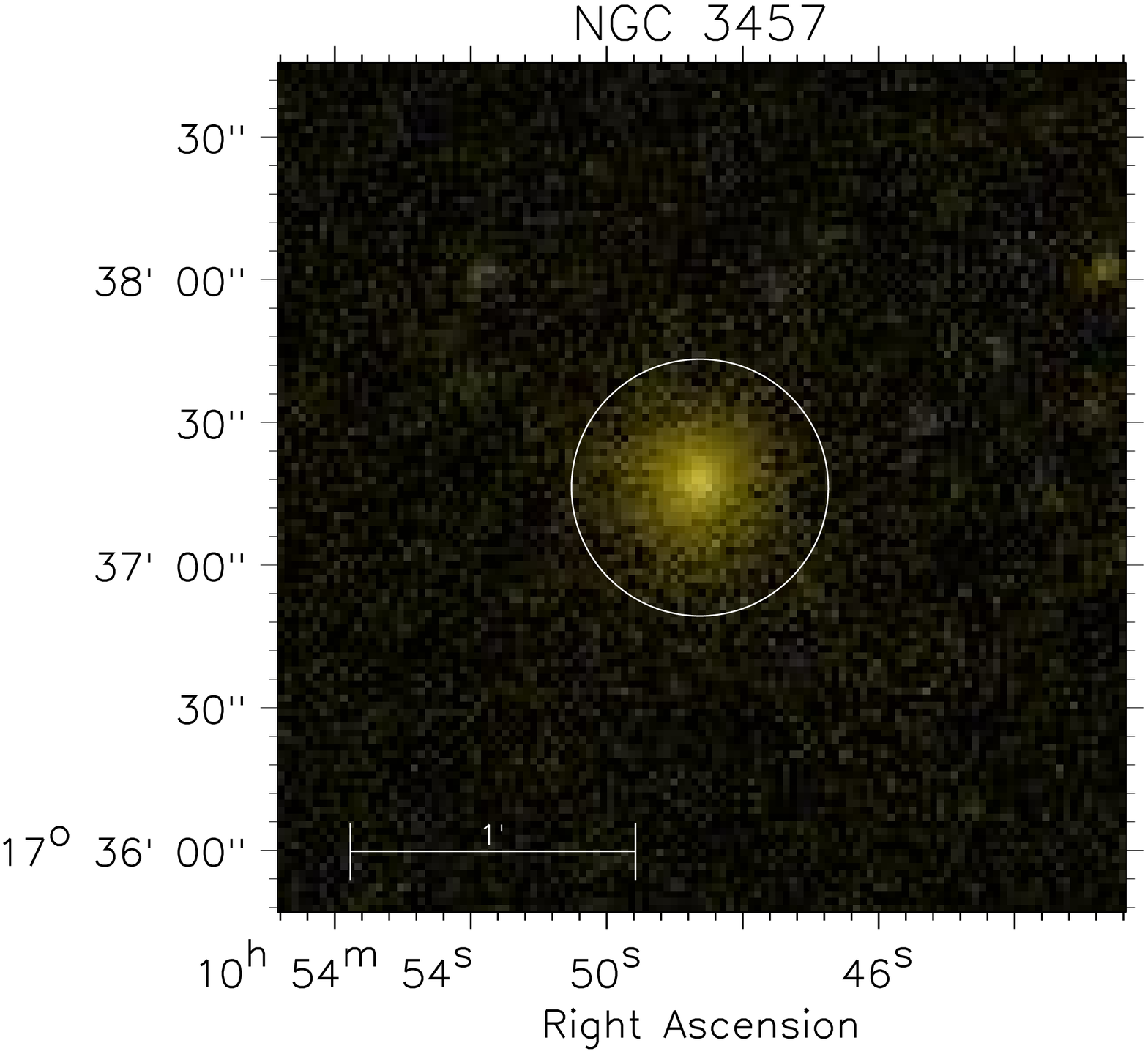}&
	 \includegraphics[width=4.cm]{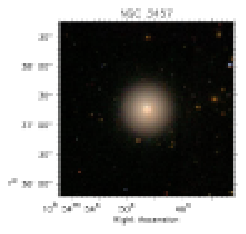}&
	 \includegraphics[width=3.5cm]{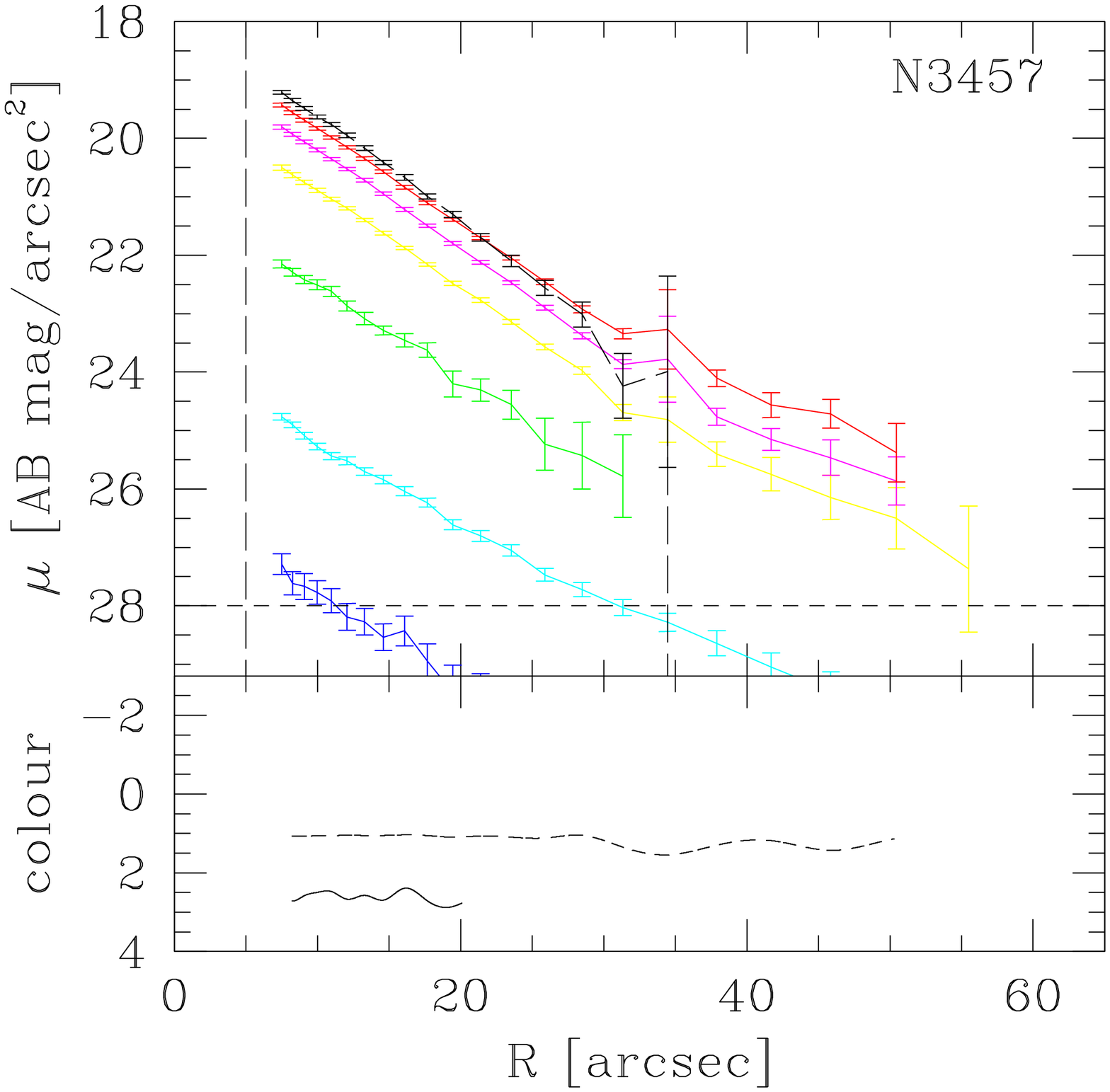}&\\
	 \hspace{-0.8cm}
	 \includegraphics[width=4.cm]{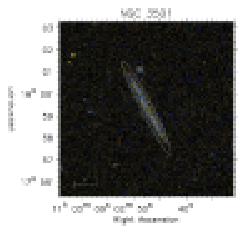}  &
	 \includegraphics[width=4.cm]{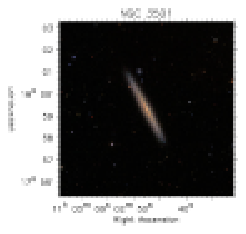}& 
	 \includegraphics[width=3.5cm]{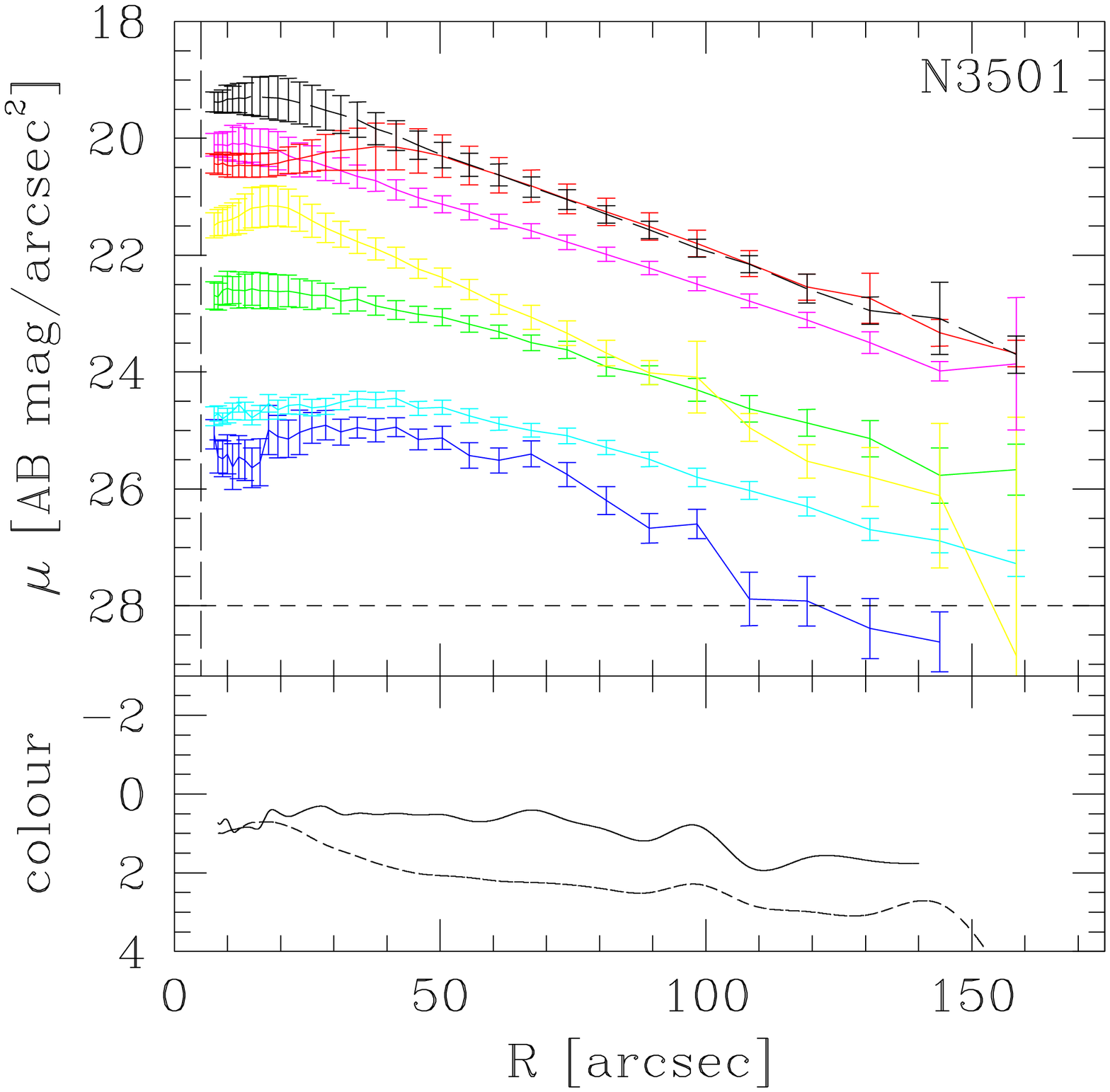}&
	 \includegraphics[width=4.cm]{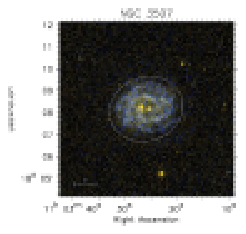}&
	 \includegraphics[width=4.cm]{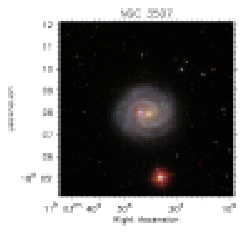}&
	 \includegraphics[width=3.5cm]{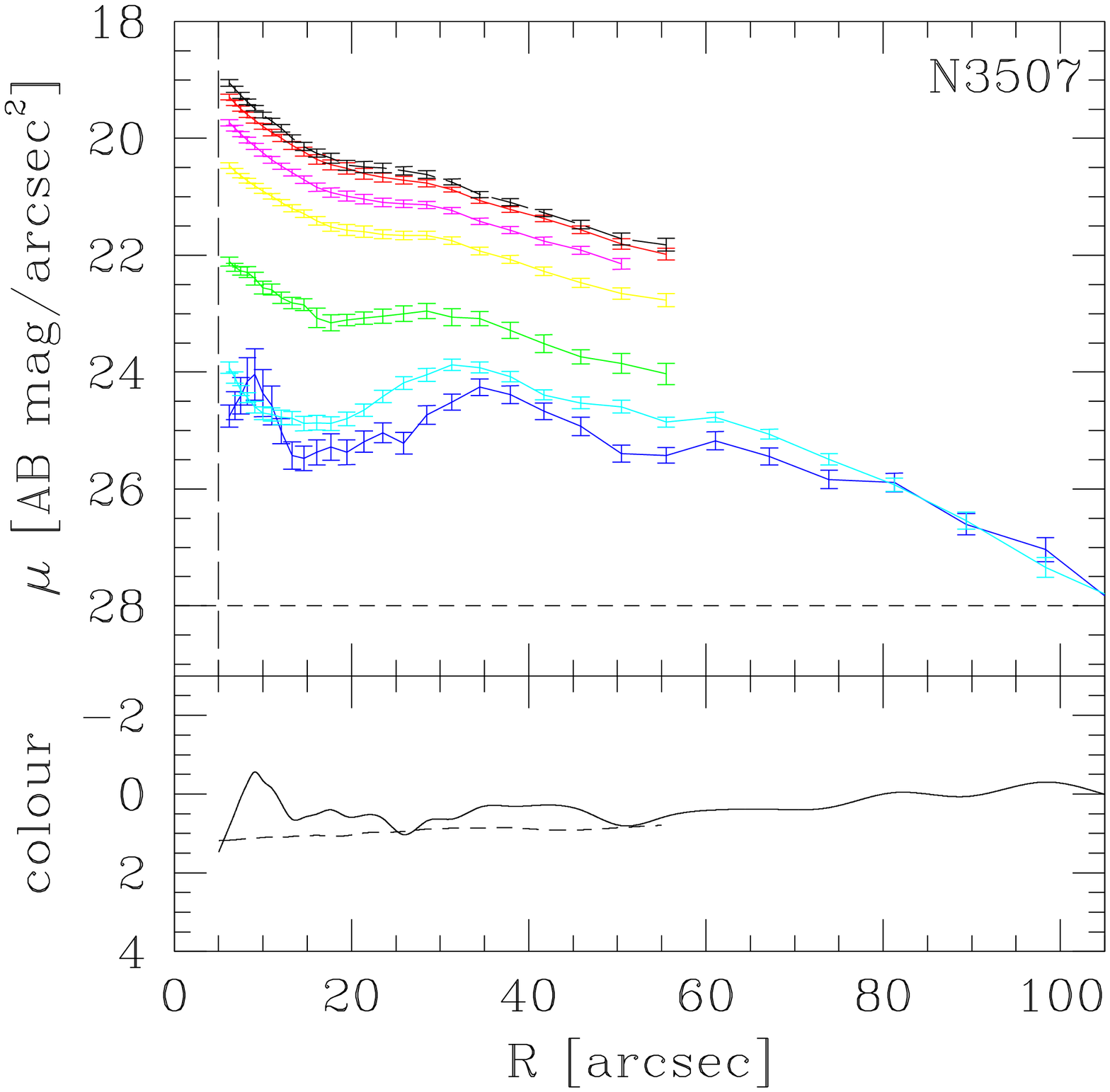}&\\
	 \hspace{-0.8cm}
	 \includegraphics[width=4.cm]{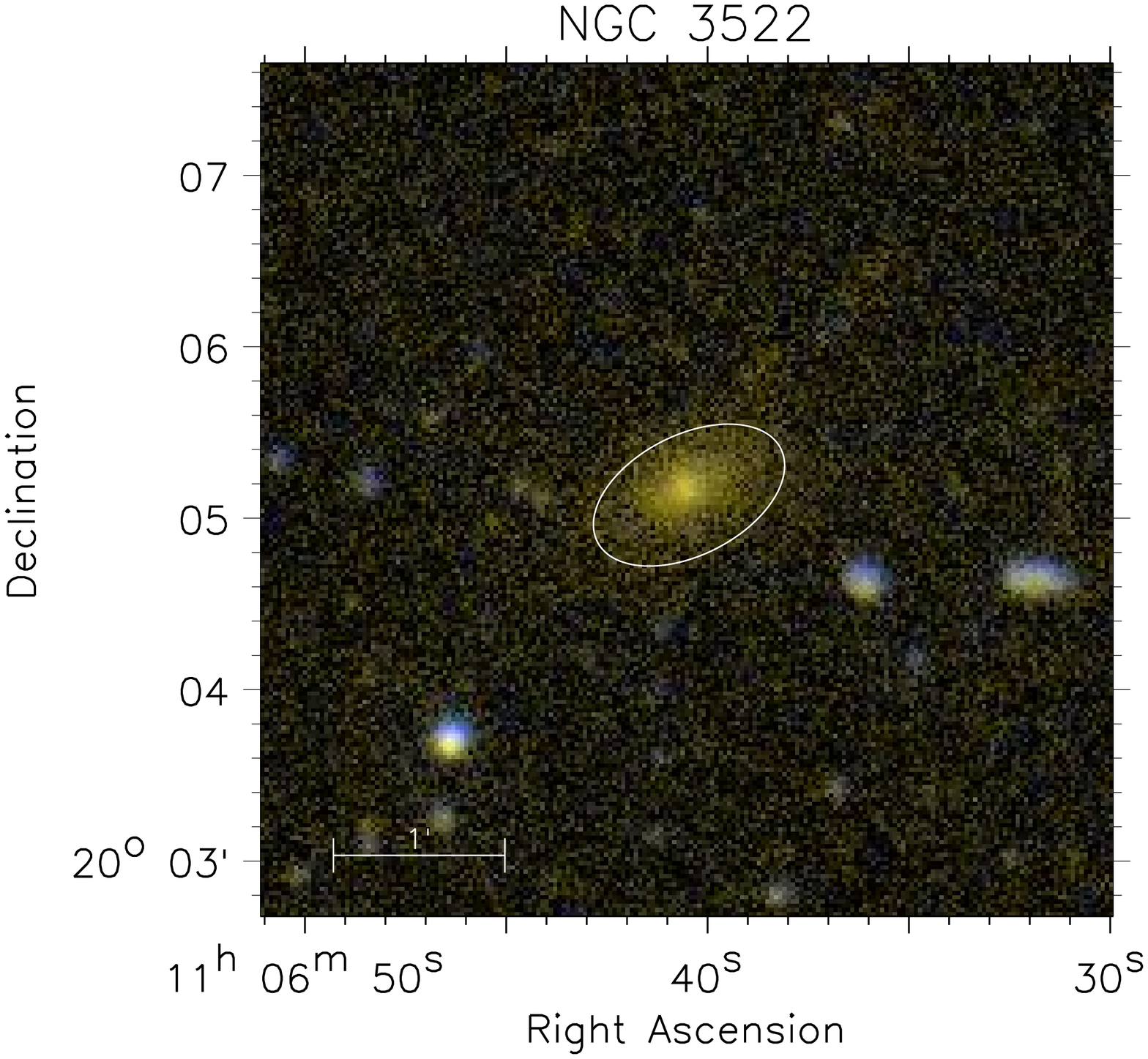}  &
	 \includegraphics[width=4.cm]{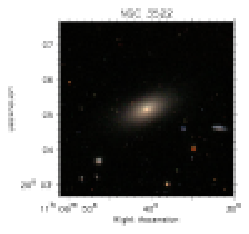}& 
	 \includegraphics[width=3.5cm]{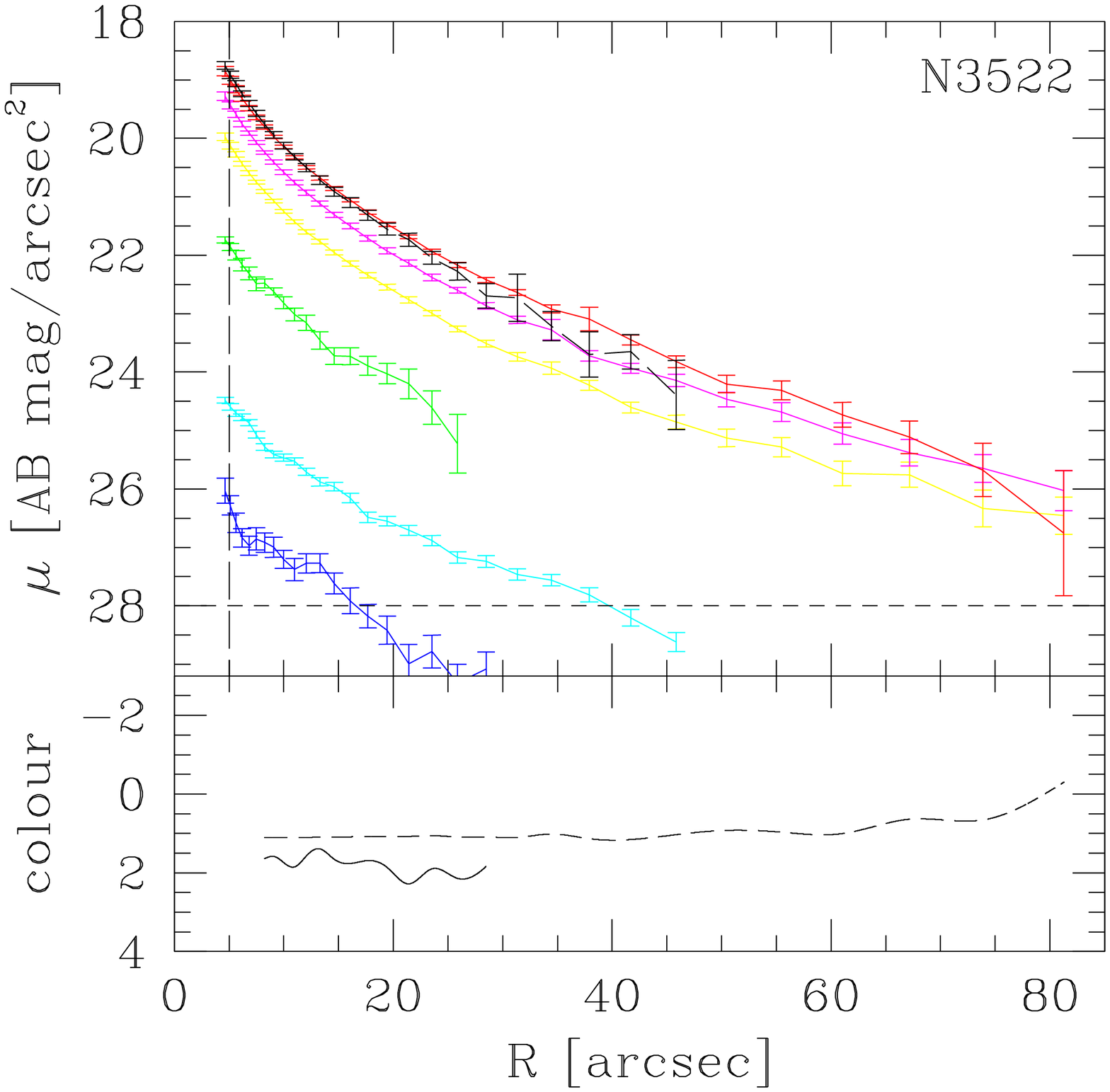}&
	 \includegraphics[width=4.cm]{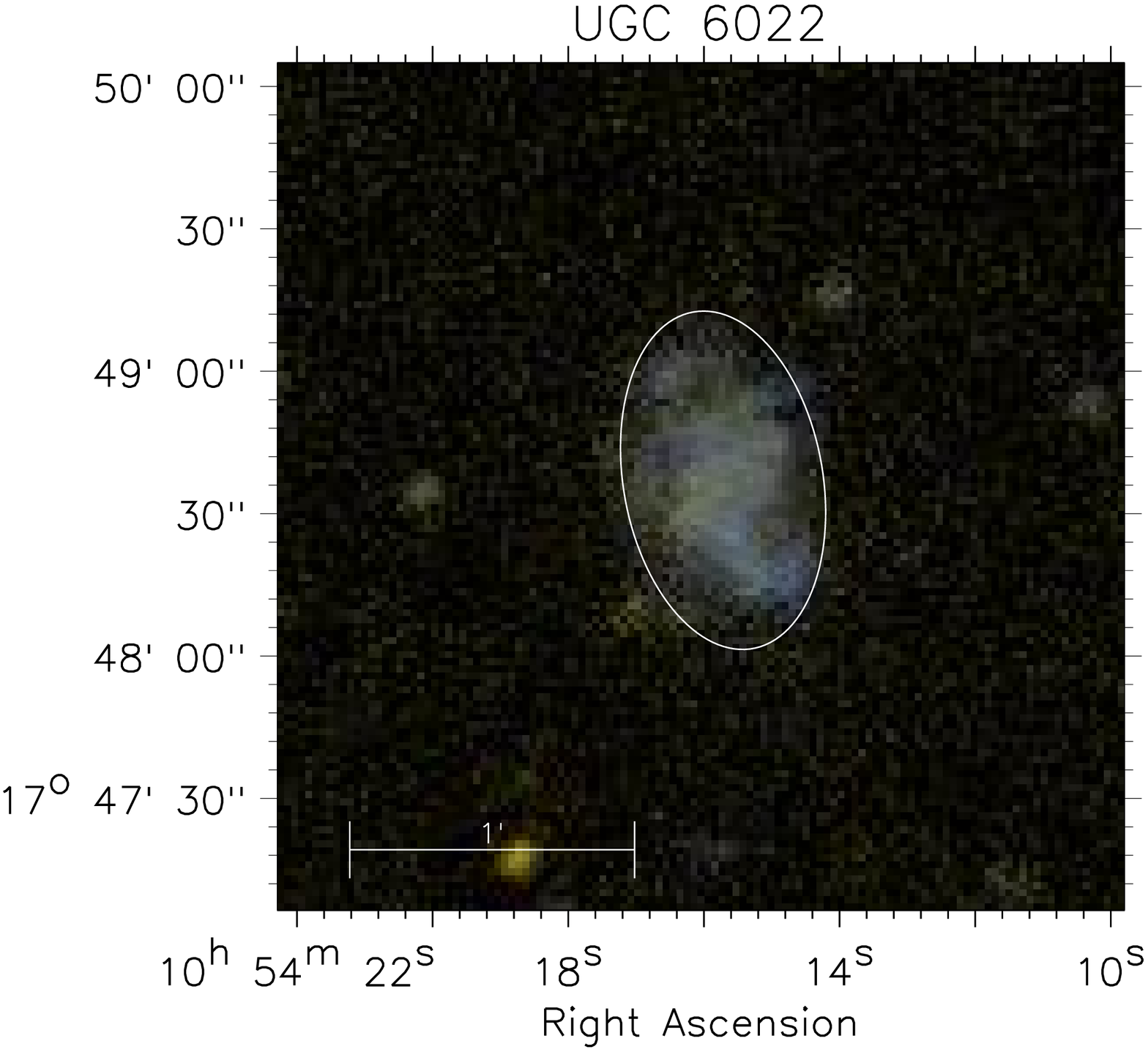}&
	 \includegraphics[width=4.cm]{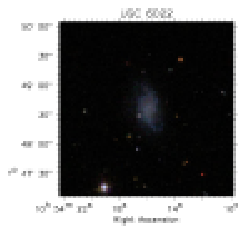}&
	 \includegraphics[width=3.5cm]{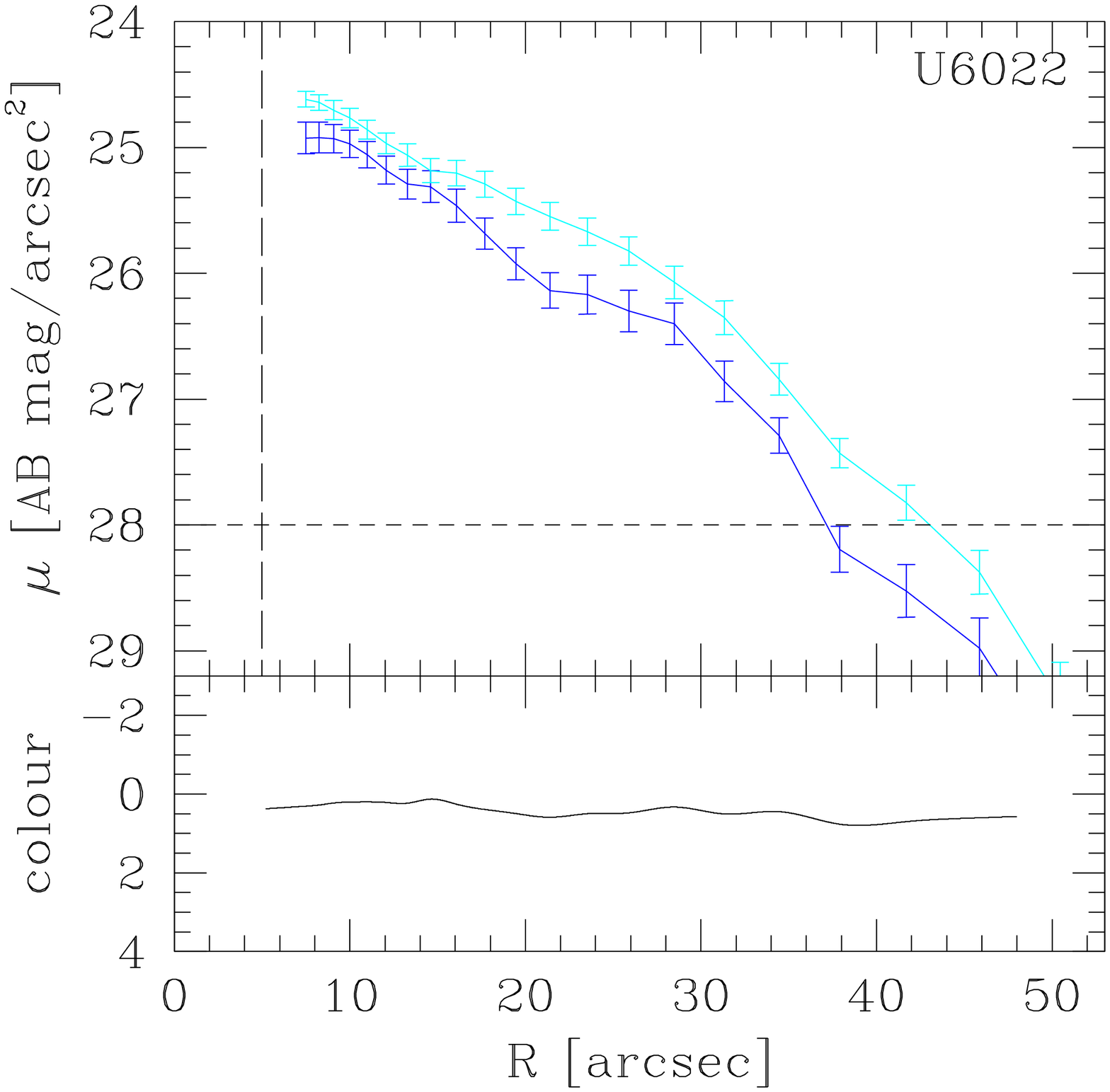}&\\
	 \end{tabular}
	  \caption{  Left: {\it GALEX } FUV and NUV  composite images of LGG~225 members.   
	  Middle: SDSS combined multiband images. Right: {\it GALEX} and SDSS surface brightness
	  profiles (top panel) with the vertical dashed line at 5\arcsec showing the  approximate FWHM of the
	{\it GALEX} point spread function, (FUV-NUV) (solid line) and ({\it g-i}) (dashed line) 
	 colors profiles (bottom panel) vs. $R$ the galactocentric distance along the semi-major 
	axis of the fitted ellipse.  The UV and optical images are on the same scale: the UV emission of NGC 3447/NGC 3447A,   
	 NGC 3455 and UGC 6035 extends much farther out than the respective optical images.} 
	 \label{225} 
	 \end{figure*}
	\end{landscape}
	
	\begin{landscape}
	\begin{figure*}[!h]
	\begin{tabular}{ccccccc}
	\hspace{-0.8cm}
	 \includegraphics[width=4.cm]{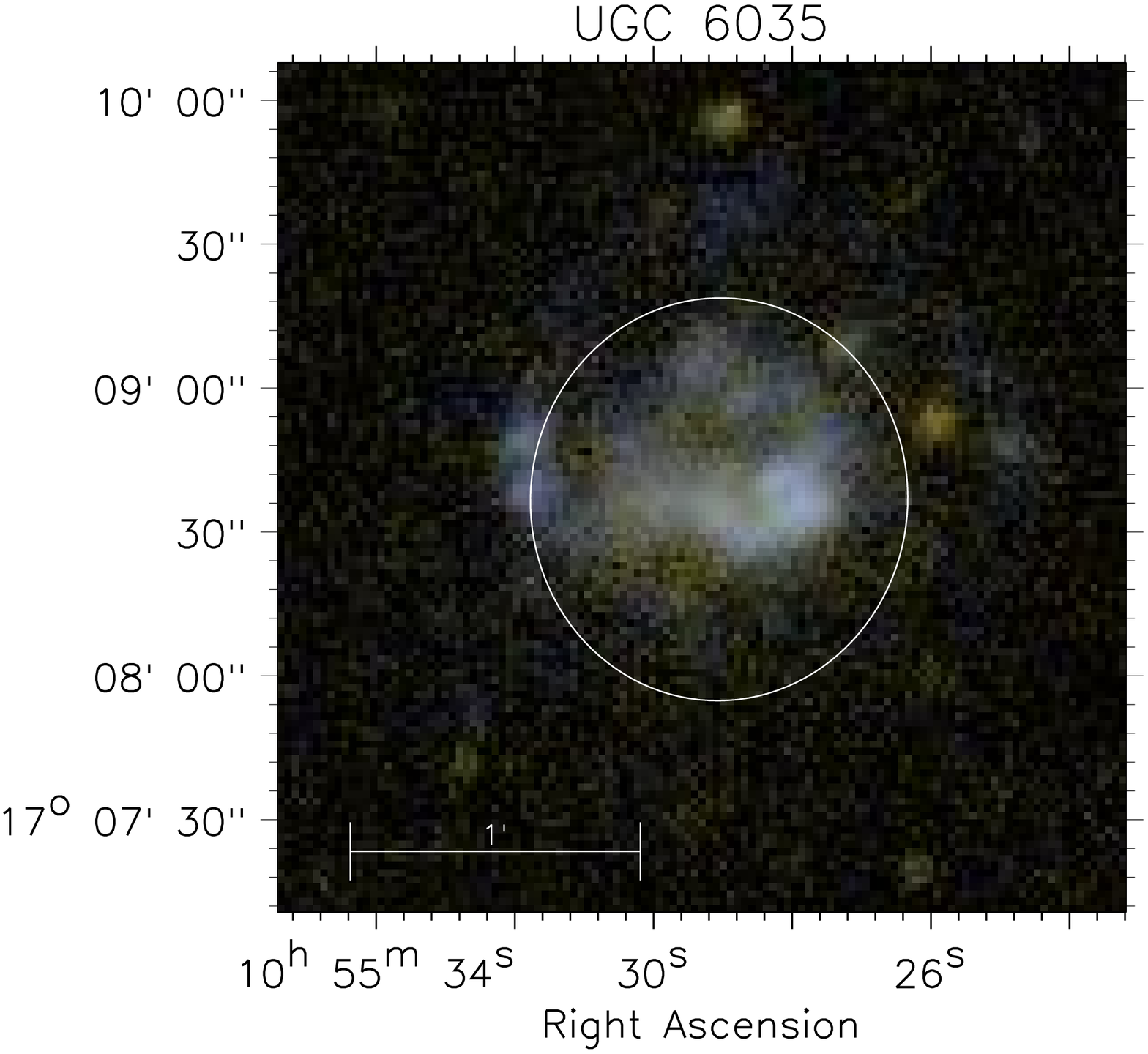}&
	 \includegraphics[width=4.cm]{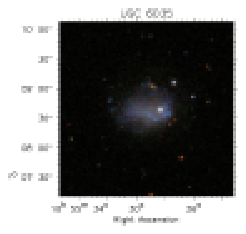}&
	\includegraphics[width=3.5cm]{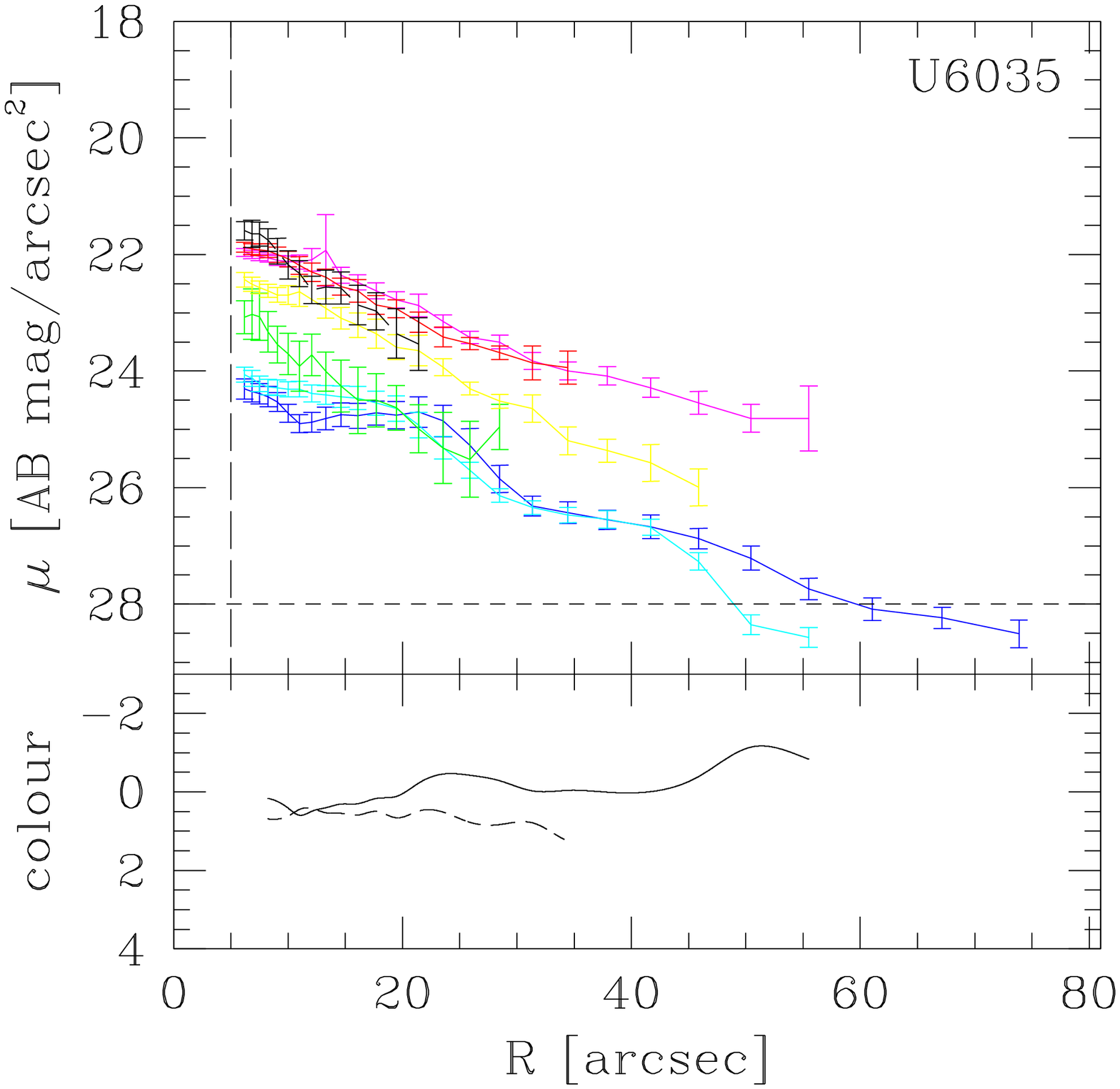}& 
	 \includegraphics[width=4.cm]{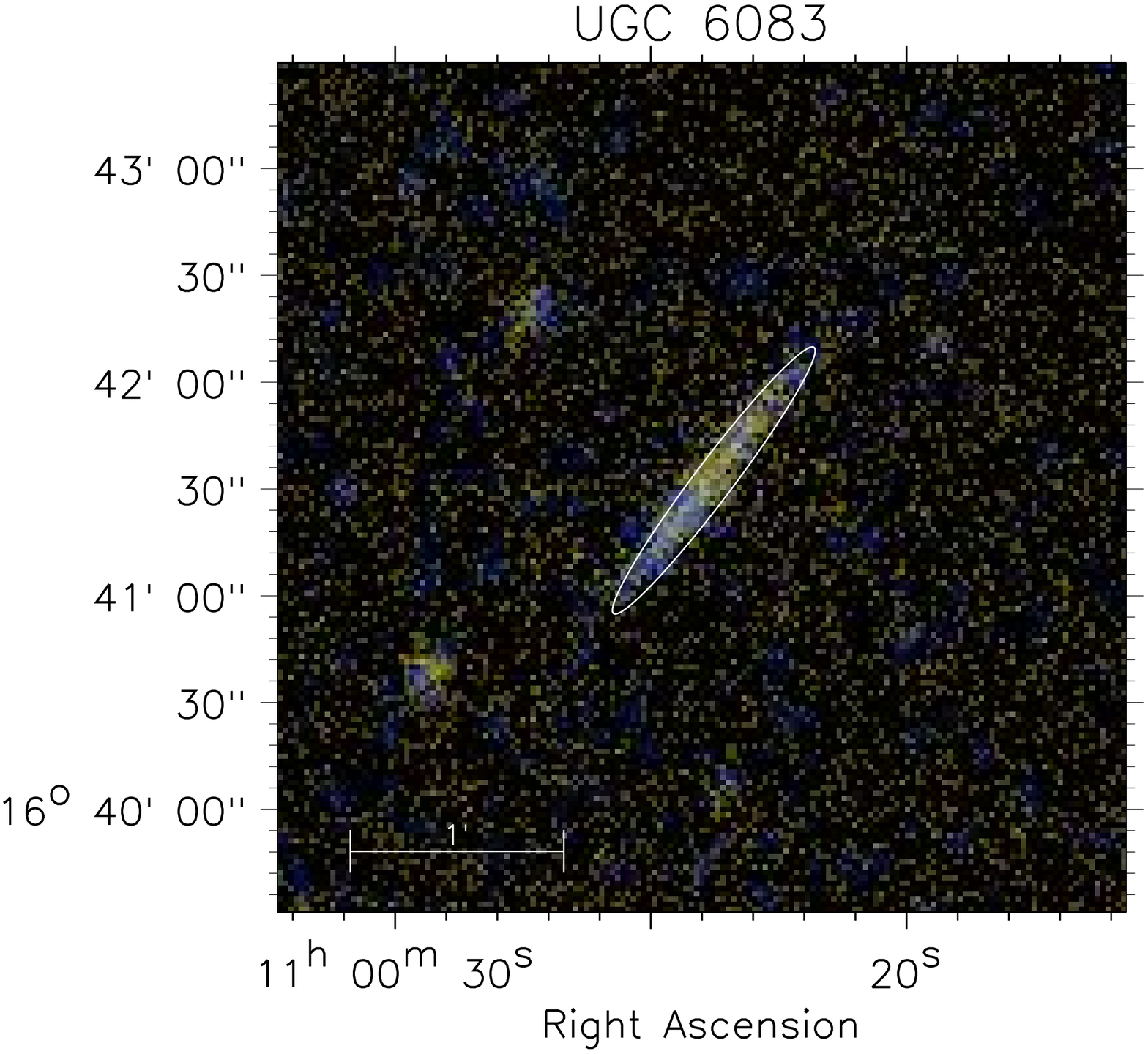} &
	 \includegraphics[width=4.cm]{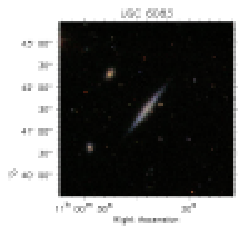} &
	\includegraphics[width=3.5cm]{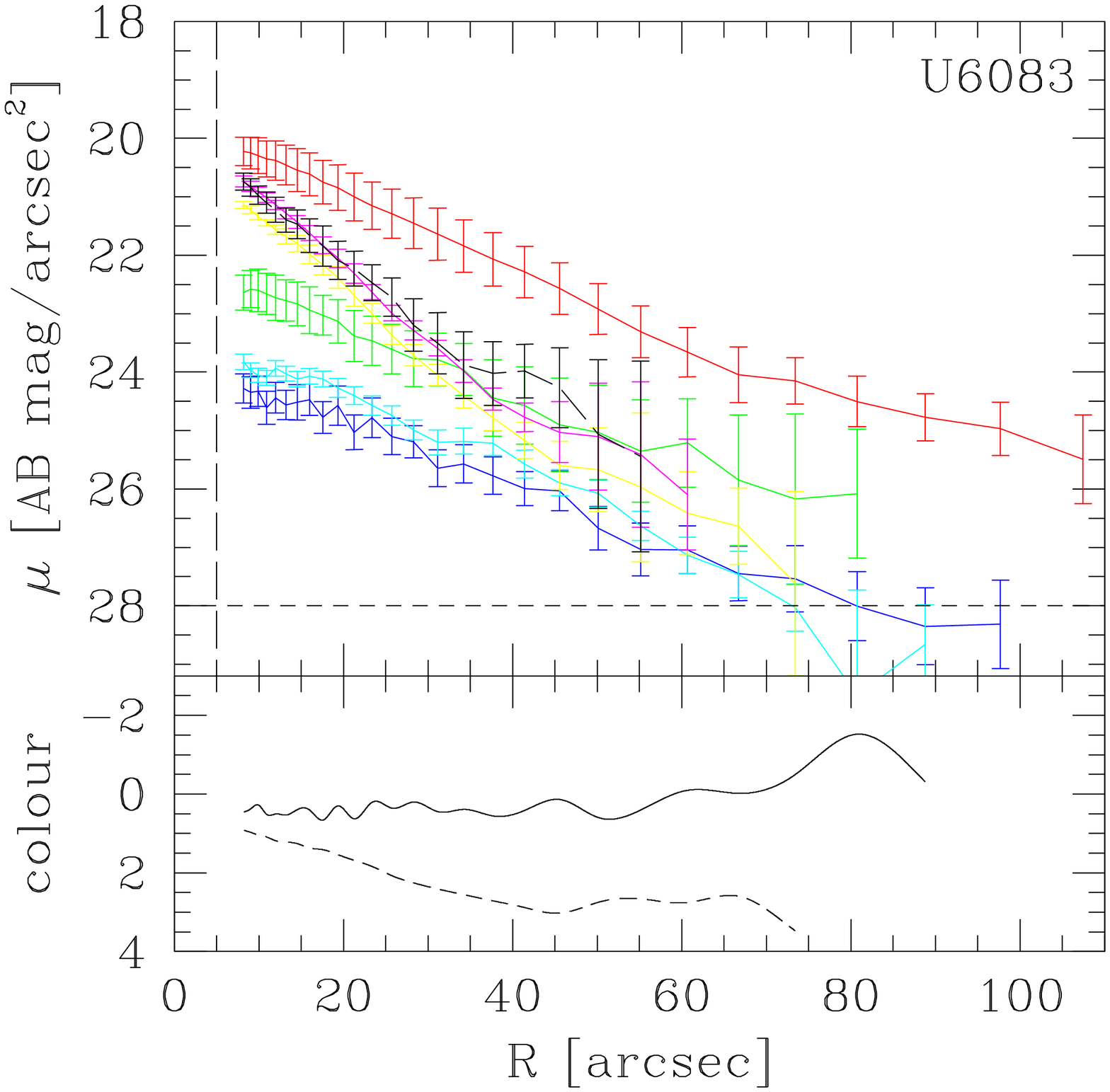}&\\
	 \hspace{-0.8cm}
	 \includegraphics[width=4.cm]{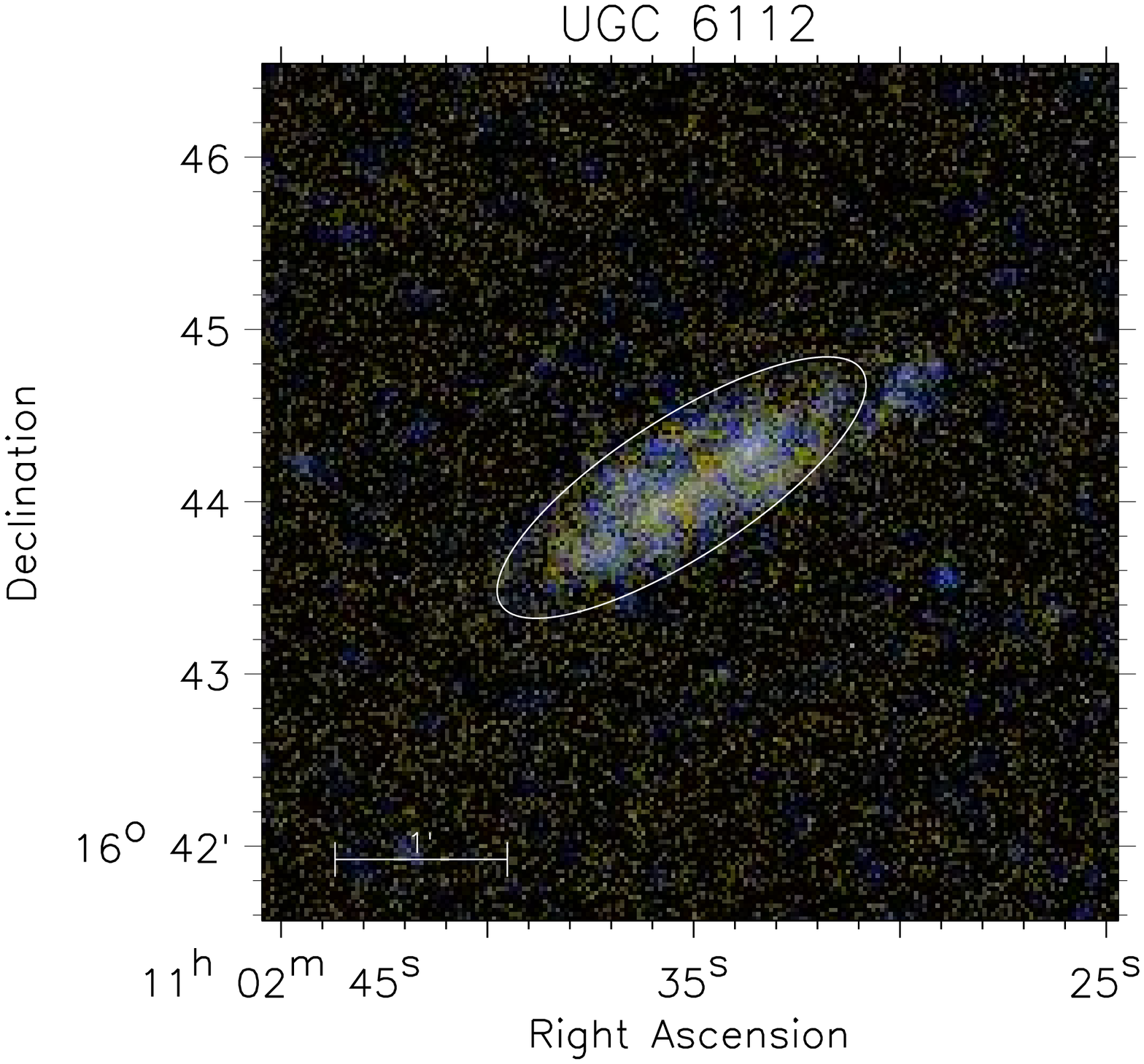} &
	 \includegraphics[width=4.cm]{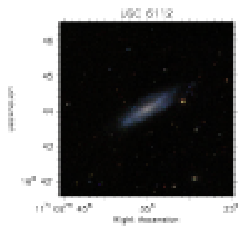}& 
	\includegraphics[width=3.5cm]{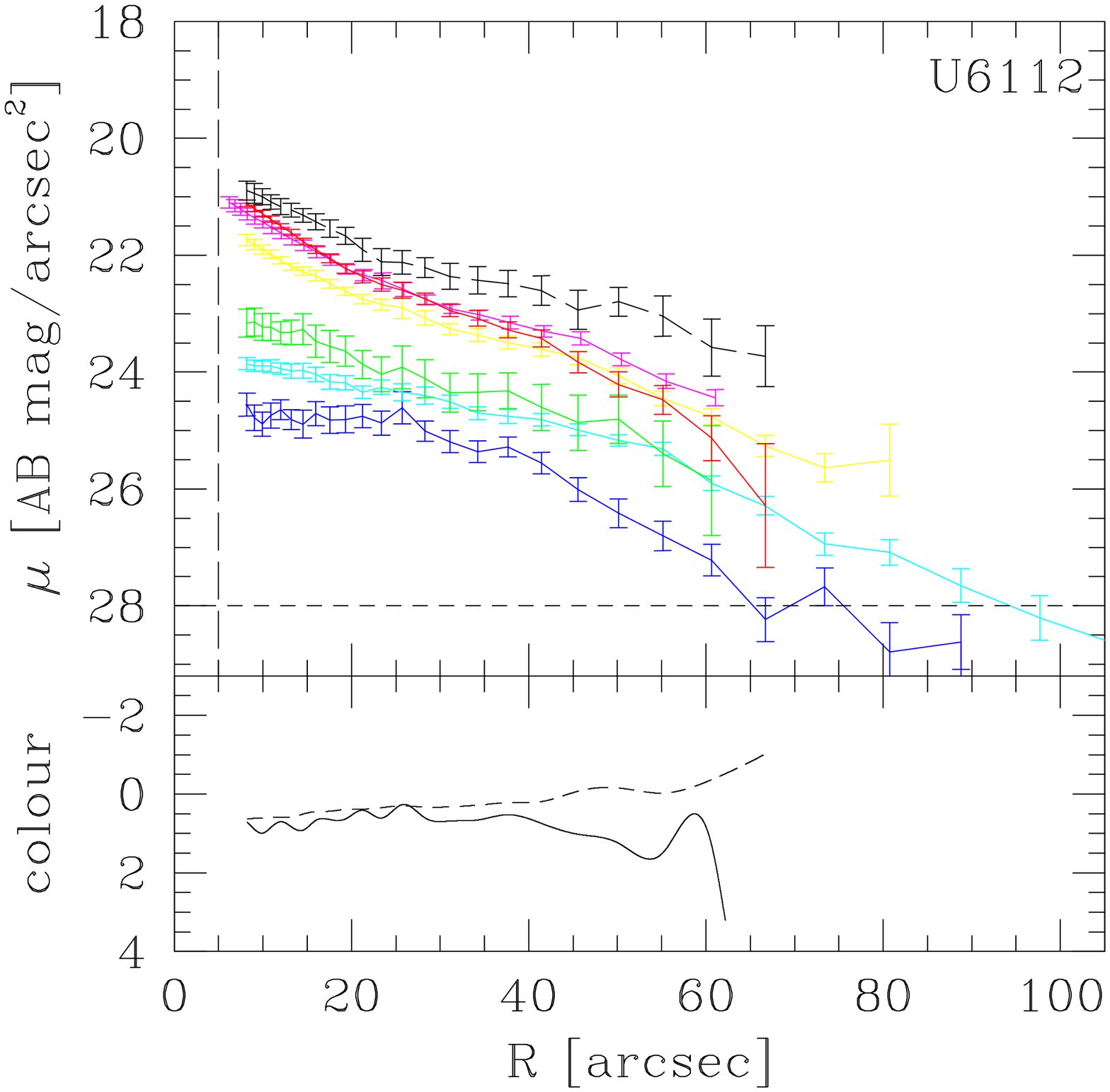}&
	 \includegraphics[width=4.cm]{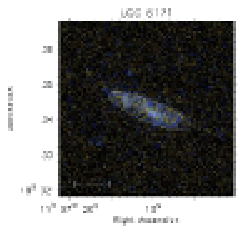}&
	 \includegraphics[width=4.cm]{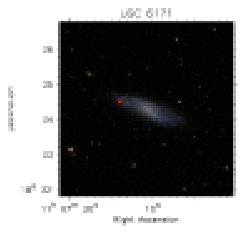}&
	\includegraphics[width=3.5cm]{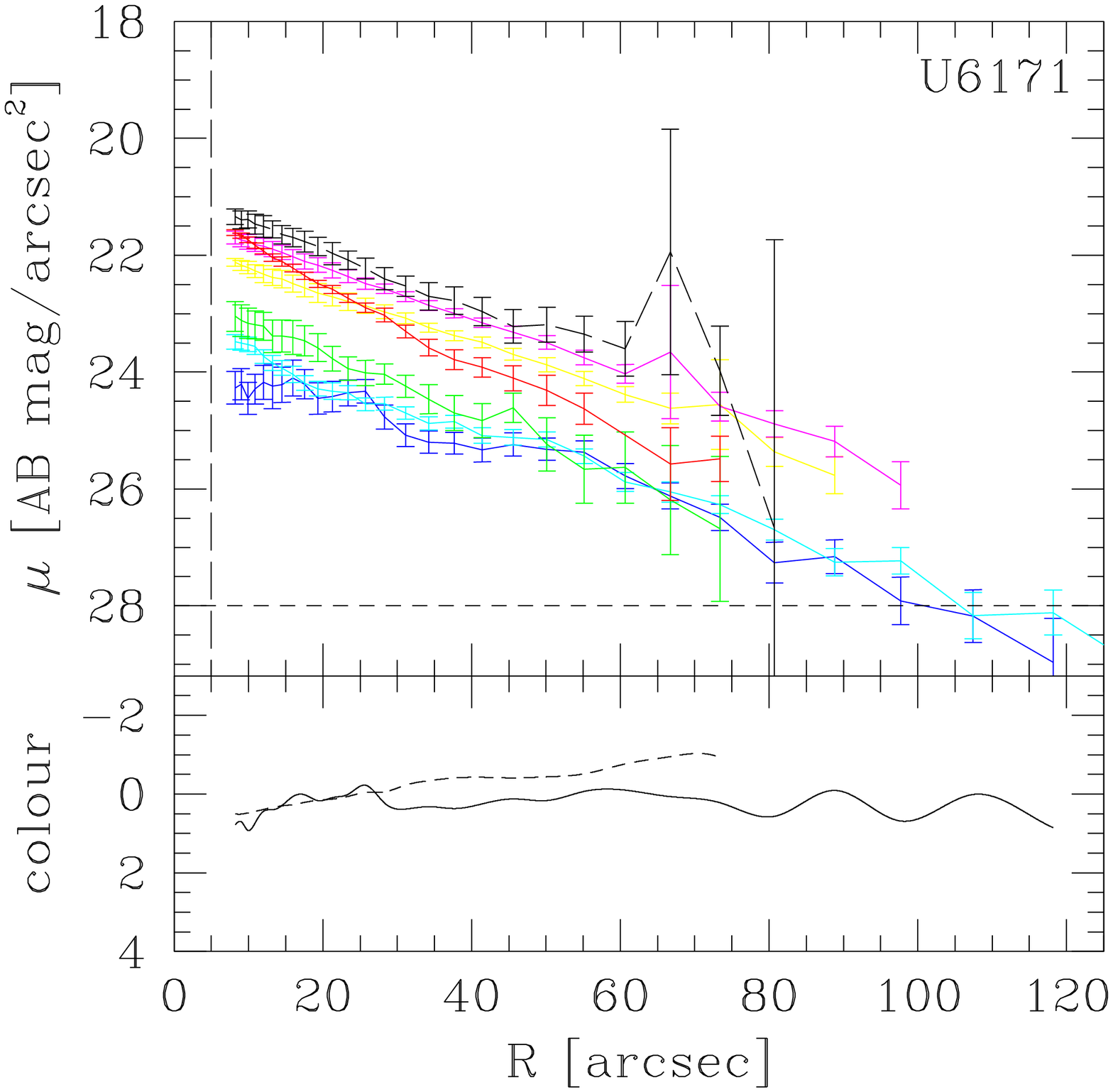}&\\
	 \hspace{-0.8cm}
	 \includegraphics[width=4.cm]{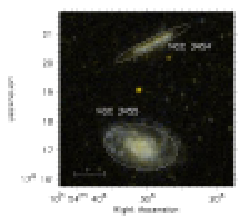}&
	 \includegraphics[width=4.cm]{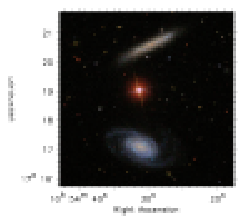}& 
	\includegraphics[width=3.5cm]{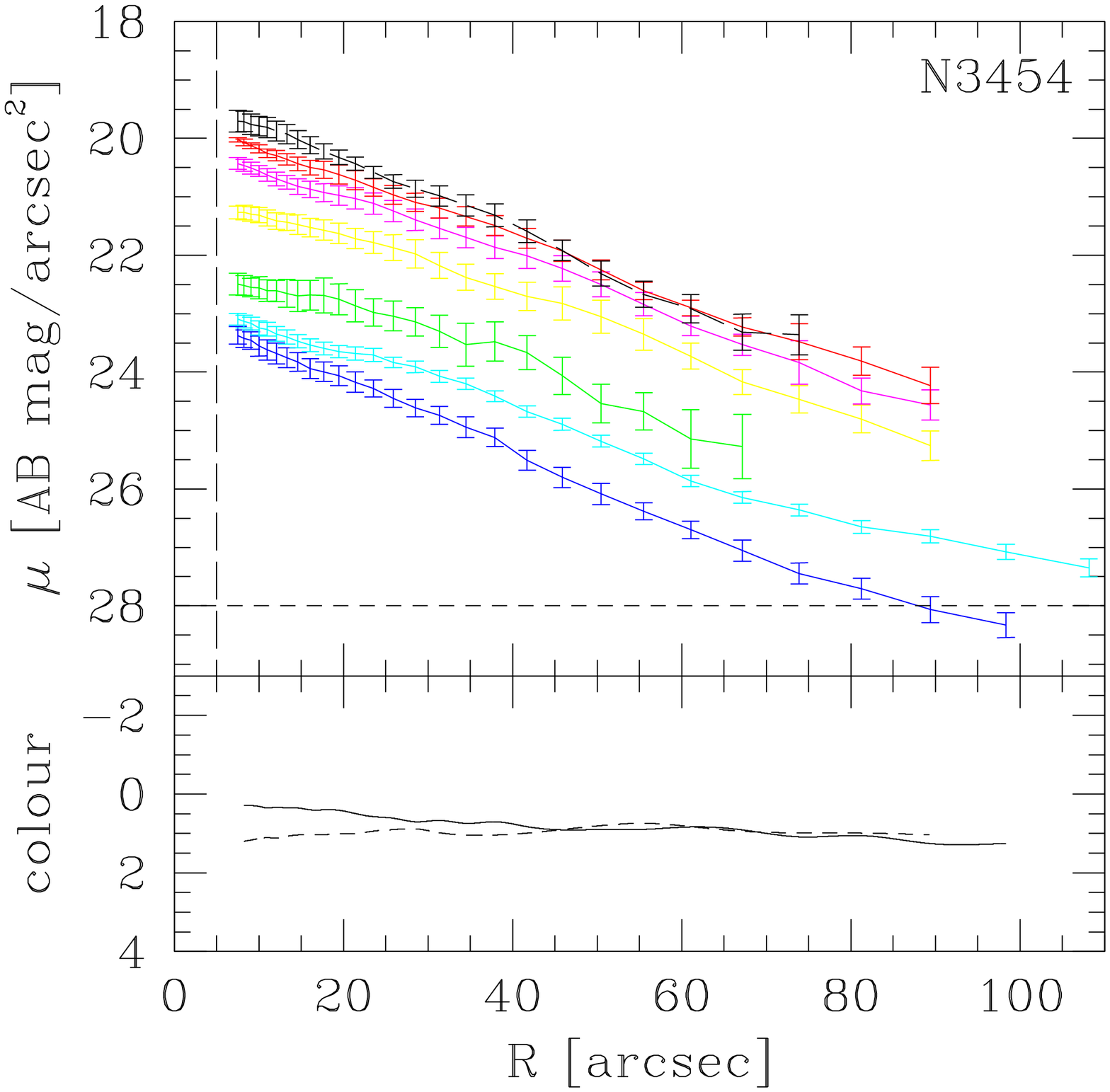}&
	\includegraphics[width=3.5cm]{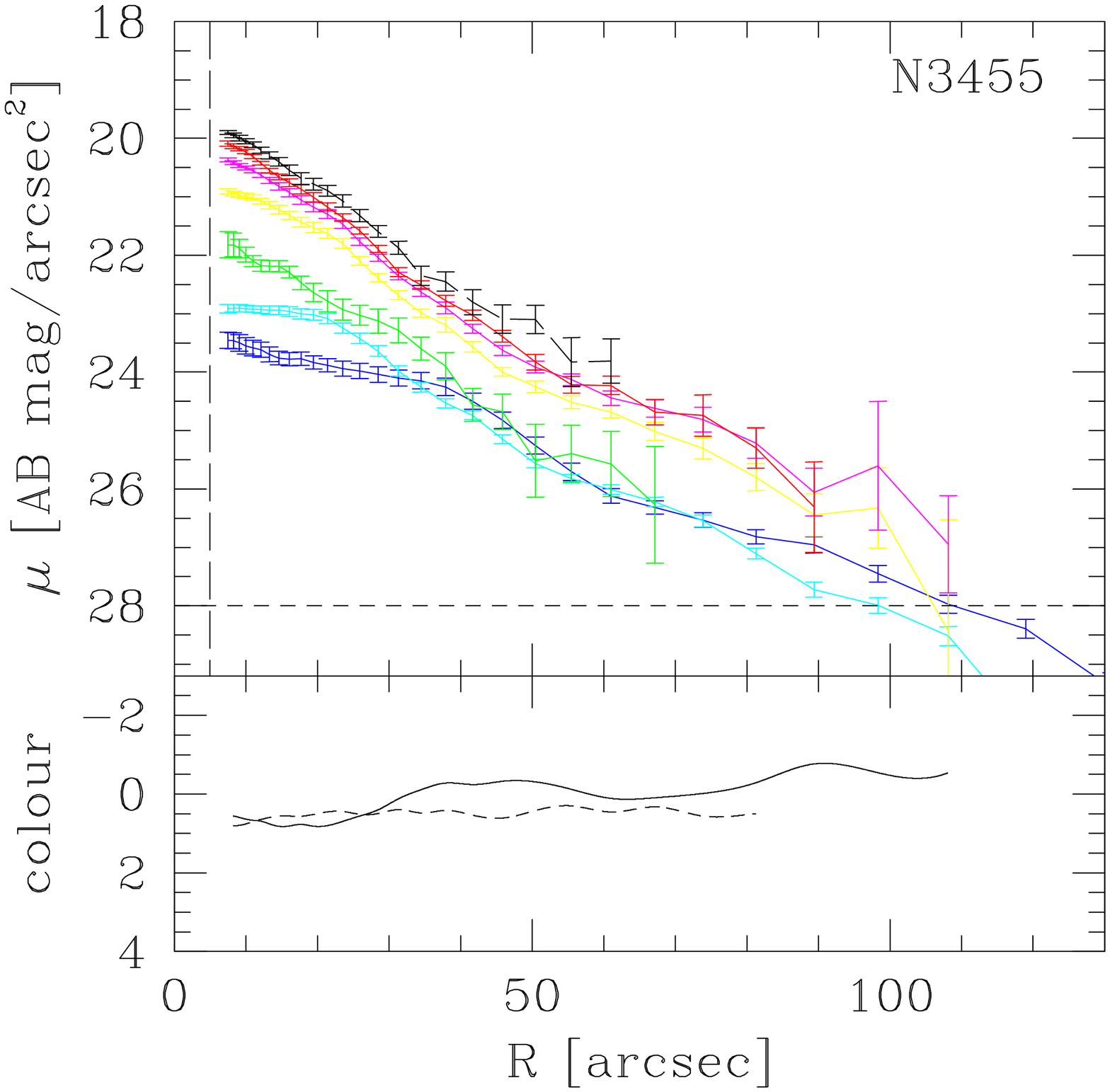} &&&\\
	 \end{tabular}
	  \addtocounter{figure}{-1}
	  \caption{continued.}
	 \label{225} 
	 \end{figure*}
	\end{landscape}

	\begin{table}[!t]
	\caption{ The {\it GALEX} photometry.}
	\label{tab:3}       
        \scriptsize
	\begin{tabular}{lllll}
	\hline\hline\noalign{\smallskip}
	Group & FUV & NUV & (FUV - NUV) \\galaxy & [AB mag] &[AB mag] 	& [AB mag] \\
	 \hline
	LGG 93 & & \\
	\hline
	 NGC 1249   & 14.10 $\pm$ 0.07 & 13.66 $\pm$ 0.04  & 0.44  $\pm$ 0.09	\\
	 NGC 1311   & 15.10 $\pm$ 0.08 & 14.68 $\pm$ 0.05  & 0.42  $\pm$ 0.10	\\
	 IC 1933    & 14.34 $\pm$ 0.07 & 13.96 $\pm$ 0.04  & 0.39  $\pm$ 0.08	\\
	 IC 1954    & 14.34 $\pm$ 0.07 & 13.87 $\pm$ 0.04  & 0.47  $\pm$ 0.08	\\
	 IC 1959    & 14.74 $\pm$ 0.08 & 14.42 $\pm$ 0.05  & 0.32  $\pm$ 0.09	\\
	\hline 
	LGG 127 & & \\
	\hline 
	 NGC 1744   &13.68 $\pm$ 0.07& 13.31 $\pm$ 0.04  & 0.37 $\pm$ 0.08   \\
	 NGC 1792   &14.10 $\pm$ 0.07& 13.31 $\pm$ 0.04  & 0.79 $\pm$ 0.08    \\
	 NGC 1800   &14.91 $\pm$ 0.08& 14.56 $\pm$ 0.05  & 0.35 $\pm$ 0.09    \\
	 NGC 1808   &14.68 $\pm$ 0.08& 13.94 $\pm$ 0.04  & 0.74 $\pm$ 0.09   \\
	 NGC 1827   &15.38 $\pm$ 0.09& 14.58 $\pm$ 0.04  & 0.80 $\pm$ 0.10   \\
	 ESO305-009 &14.97 $\pm$ 0.08& 14.66 $\pm$ 0.05  & 0.31 $\pm$ 0.09  \\
	 ESO305-017 &16.23 $\pm$ 0.09& 15.91 $\pm$ 0.05  & 0.31 $\pm$ 0.11  \\
	 ESO362-011 &15.14 $\pm$ 0.11& 15.52 $\pm$ 0.04  & 0.62 $\pm$ 0.12   \\
	 ESO362-019 &15.70 $\pm$ 0.21& 15.39 $\pm$ 0.12  & 0.32 $\pm$ 0.25   \\
	\hline
	LGG 225 & & \\
	\hline
	NGC 3370     &14.65  $\pm$ 0.08  &14.31  $\pm$ 0.04  & 0.34 $\pm$ 0.08  \\
	NGC 3443     &15.85  $\pm$ 0.07  &15.48  $\pm$ 0.04  & 0.37 $\pm$ 0.07  \\
	NGC 3447     &14.82  $\pm$ 0.05  &14.55  $\pm$ 0.03  & 0.27 $\pm$ 0.06  \\
	NGC 3447A    &15.83  $\pm$ 0.06  &15.72  $\pm$ 0.03  & 0.11 $\pm$ 0.06  \\
	NGC 3454     &17.22  $\pm$ 0.07  &16.60  $\pm$ 0.04  & 0.62 $\pm$ 0.08  \\
	NGC 3455     &15.34  $\pm$ 0.06  &14.77  $\pm$ 0.03  & 0.56 $\pm$ 0.06  \\
	NGC 3457     &20.23  $\pm$ 0.13  &17.76  $\pm$ 0.05  & 2.48 $\pm$ 0.14	\\
	NGC 3501     &16.67  $\pm$ 0.09  &16.12  $\pm$ 0.05  & 0.55 $\pm$ 0.11	\\
	NGC 3507$^a$ &14.55  $\pm$ 0.07  &14.13  $\pm$ 0.04  & 0.42 $\pm$ 0.08	\\ 
	NGC 3522     &19.92  $\pm$ 0.12  &18.03  $\pm$ 0.05  & 1.90 $\pm$ 0.13	\\  
	UGC 6022     &17.82  $\pm$ 0.07  &17.51  $\pm$ 0.04  & 0.31 $\pm$ 0.08\\  
	UGC 6035$^a$ &16.85  $\pm$ 0.06  &16.52  $\pm$ 0.04  & 0.33 $\pm$ 0.07	\\ 
	UGC 6083     &18.01  $\pm$ 0.13  &17.48  $\pm$ 0.07  & 0.53 $\pm$ 0.15  \\
	UGC 6112     &16.28  $\pm$ 0.09  &15.85  $\pm$ 0.05  & 0.42 $\pm$ 0.10  \\
	UGC 6171$^a$ &16.41  $\pm$ 0.09  &16.23  $\pm$ 0.05  & 0.18 $\pm$ 0.10  \\ 
	\hline
	\end{tabular}

	Note:  
	To avoid  overlap in the elliptical apertures, coordinates and major and minor axes of NGC 3447A and NGC 3447 have been
	positioned as shown in Figure \ref{225}.
	$^a$Magnitude of the foreground star have been subtracted. 
	Magnitudes are not corrected for galactic extinction. 
	\end{table}

	\begin{table*}[]
	\caption{SDSS photometry of LGG 225 members.}
	\label{3a}       
        \scriptsize
	\begin{tabular}{lllllllllcll}
	\hline\hline\noalign{\smallskip}
	 Galaxy &  u  & g & r & i & z  \\
	    & [AB mag]  &  [AB mag]  &[AB mag] &[AB mag] &[AB mag]     &    \\
	\hline
	NGC 3370 &      13.31 $\pm$ 0.06& 12.31 $\pm$ 0.02& 11.85 $\pm$ 0.03& 11.59 $\pm$ 0.03& 11.38 $\pm$ 0.06 \\   
	NGC 3443$^b$ &  15.91 $\pm$ 0.14& 14.79 $\pm$ 0.05& 14.33 $\pm$ 0.05& 14.11 $\pm$ 0.06& 13.90 $\pm$ 0.18 \\ 
	NGC 3447$^b$  & 14.47 $\pm$ 0.12& 13.22 $\pm$ 0.04& 12.90 $\pm$ 0.05& 12.78 $\pm$ 0.06& 12.06 $\pm$ 0.10   \\ 
	NGC 3447A&      15.34 $\pm$ 0.16& 14.89 $\pm$ 0.09& 14.71 $\pm$ 0.13& 14.65 $\pm$ 0.19& 14.32 $\pm$ 0.44  \\  
	NGC 3454 &      15.08 $\pm$ 0.12& 13.84 $\pm$ 0.03& 13.23 $\pm$ 0.03& 12.91 $\pm$ 0.04& 12.66 $\pm$ 0.09  \\  
	NGC 3455 &      13.98 $\pm$ 0.09& 13.02 $\pm$ 0.03& 12.66 $\pm$ 0.04& 12.50 $\pm$ 0.05& 12.31 $\pm$ 0.12  \\  
	NGC 3457 &      14.69 $\pm$ 0.06& 13.05 $\pm$ 0.02& 12.37 $\pm$ 0.02& 12.01 $\pm$ 0.02& 11.77 $\pm$ 0.04 \\ 
	NGC 3501 &      14.41 $\pm$ 0.07& 13.06 $\pm$ 0.02& 12.40 $\pm$ 0.02& 11.99 $\pm$ 0.03& 11.68 $\pm$ 0.05 \\                
	NGC 3507$^{a,b}$ & 12.16  $\pm$ 0.02& 10.75  $\pm$ 0.02& 10.18  $\pm$ 0.02& 9.78  $\pm$ 0.02& 9.62  $\pm$ 0.03 \\   
	NGC 3522 &      15.16 $\pm$ 0.11& 13.59 $\pm$ 0.03& 12.88 $\pm$ 0.02& 12.52 $\pm$ 0.03& 12.37 $\pm$ 0.06  \\
	UGC 6022 &      16.94 $\pm$ 0.35 & 15.84 $\pm$ 0.11& 15.52 $\pm$ 0.14& 15.38 $\pm$ 0.19& 15.07 $\pm$ 0.51  \\
	UGC 6035$^a$ &  13.77 $\pm$ 0.11& 12.90 $\pm$ 0.04& 12.21 $\pm$ 0.06& 11.75 $\pm$ 0.05&10.92 $\pm$ 0.13 \\
	UGC 6083 &      16.31 $\pm$ 0.17& 15.17 $\pm$ 0.05& 14.71 $\pm$ 0.06& 14.43 $\pm$ 0.07&14.35 $\pm$ 0.23 \\   
	UGC 6112$^b$  & 15.22 $\pm$ 0.14& 14.29 $\pm$ 0.05& 13.94 $\pm$ 0.06& 13.76 $\pm$ 0.08&13.75 $\pm$ 0.31 \\
	UGC 6171$^a$ &  15.50 $\pm$ 0.15& 13.41 $\pm$ 0.05& 13.03 $\pm$ 0.06& 12.79 $\pm$ 0.06&12.69 $\pm$ 0.14 \\
	\hline
	\end{tabular}

	Note:$^a$ Magnitude of the foreground star have been subtracted. $^b$ Galaxy is near SDSS edge, magnitudes have been computed 
	in a reduced area (elliptical aperture with a=0.5\arcmin, b=0.3\arcmin~ for NGC 3443, with a=1.6\arcmin, b=0.8\arcmin 
	~for NGC 3447, with a=1\arcmin, b=0.4\arcmin ~ for UGC 6112 and aperture of radius 0.9\arcmin ~ for NGC 3507 respectively,
	and scaled to the total area using the UV ratio to derive their mass.
	      
	\end{table*}

	\section{Morphology and photometry}

	\subsection{UV morphology}

	Figures \ref{93} and \ref{127}  show the  composite {\it GALEX} FUV and NUV images and the light
	profiles of individual galaxies members of LGG 93 and LGG 127. 
	Figure \ref{225} shows the  composite {\it GALEX} and SDSS 
	images, surface brightness profiles, (FUV - NUV) and {\it g - i} colors of LGG 225 galaxies. \\ 

	\noindent \underbar{LGG~93}~~~~~ This is the poorest group of our sample and the analysis
	is exclusively based on data from the public {\it GALEX} archive.
	 
	All galaxies are late-type spirals seen at different, although high, inclination angles 
	(58$^{\circ} \leq i \leq$ 90$^{\circ}$). The FUV and NUV images 
	are basically indistinguishable.

	IC 1954 shows a relatively small bulge and extended, S-shaped, inner arms. The morphology
	appears undistorted although there is evidence of multiple arms. IC 1933 is classified as
	"Sd" by \citet{Sengupta06}. The galaxy in both UV bands appears flocculent without a clear
	spiral arm structure.

	The spiral NGC 1249 shows a bar in both NUV and FUV bands, multiple arms and
	signatures of asymmetry. IC 1959 and NGC 1311 are viewed edge-on. In both galaxies the disk
	structure appears undistorted. \citet{Meurer06} describe IC 1959 and NGC 1311 as galaxies
	with an elongated shape mixed with gas and a large number of regions of active star formation,
	also prominent in  H$_\alpha$.  The presence of dust in NGC 1311 gives a
	`broken' appearance to its body, in both  the NUV and FUV images.
	The UV images of  IC 1959 and NGC 1311 do not show clear evidence of bars, 
	while they are classified as barred in the RC3 catalog.
	 
	\bigskip
	\noindent \underbar{LGG~127}
	~~~~~Nine identified members are shown in 
	 Figure \ref{127}. NUV and FUV morphology 
	of all the galaxies are very similar, suggesting that the stellar populations
	are young.

	Both the FUV and NUV images of NGC 1744   are very similar to the R-band  and 
	H$_{\alpha}$ images in \citet{Meurer06}. The galaxy has
	a bar from which  grand design, open spiral arms depart. 
	Star forming regions  are visible in all parts of the galaxy. 
		A bar structure is present also in the spiral ESO305-009. 

	NGC~1800 is classified Sd in the RC3 catalog but does not show signatures of arms in 
	either UV band. Rather it has a bar-like structure in both NUV and FUV  images.

	NGC 1792 and ESO362-019   have  a bulge-less flocculent appearance,
	while NGC~1808 has an outer  ring-like structure and an 
	extended multiple-arms structure.

	None of galaxies show unambiguous  morphological distorsions due to  
	interaction, while most of them show outer asymmetric structures and bars which are often associated 
	to past interactions events \citep[see simulations of][]{Noguchi87}. 
	 
	\bigskip
	\noindent \underbar{LGG 225} ~~~~~
	Our two {\it GALEX} fields  cover  eight members of the group LGG 225, including its brightest
	galaxy population. Other seven members are included in   
	archival images (Table~2)  with significantly shorter exptime, and 
	$\sim$ 3 mag brighter detection limit. 

	NGC~3447A (FUV-NUV=0.11$\pm$0.08, see next Section) 
	is the bluest galaxy and is strongly interacting with 
	NGC~3447. A bar is still visible in NGC 3447, while both the underlying disk and the
	(multiple?) arms are tidally distorted. Interaction is so strong that for NGC~3447A 
	it is even difficult to figure out  whether   the original morphological type was  irregular or late-type
	spiral. 

	NGC~3443  has no  nearby companions, nevertheless
	the northern half of the galaxy is completely different from the southern one. 

	NGC~3454 and NGC~3455 are close both in projection and in  redshift space (Table~1). 
	The first galaxy displays a clean edge-on spiral morphology while the outer eastern 
	arms of NGC~3455 appear tidally distorted.

	UGC 6022 appears irregular and distorted.
	UGC~6035 has a foreground star  superimposed on its main body, it appears also very
	irregular and probably distorted by an interaction event.

	NGC~3501, UGC~6083, UGC~6112 and UGC~6171 are seen nearly
	 edge-on and show signature of distortion as in the case of NGC~3370 and
	  NGC~3507, this latter seen, in contrast, nearly face-on.

	The two early-type galaxies members of the group, NGC 3457 and NGC~3522
	do not show signatures of distortion. 
	 
	Summarizing, although cases of morphological distortion/asymmetries and multiple arms are
	present in all groups, only in LGG~225 there is  a direct evidence of tidal
	distortion, as in the cases of pairs NGC~3447/NGC~3447A  and NGC~3454/NGC~3455.
	 
        Furthermore the UV emission in NGC 3447/NGC 3447A, in NGC 3455 and UGC 6035 extends 
        much farther out than the respective optical images (Figure \ref{225}). 

	\subsection{UV and optical photometry}

	Table~3 gives the {\it GALEX} FUV and NUV magnitudes of the observed
	members of the three groups. AB magnitudes  were measured
	from GALEX background-subtracted intensity images, within the 
	ellipses shown in Figures \ref{93}, \ref{127} and \ref{225}.  The major and minor axes as well as the P.A.
	of the ellipses   
	are reported in Table 1, columns 8, 9 and 10. 

	FUV and NUV magnitudes were computed as m(AB)$_{UV}$ = -2.5 $\times$ log CR$_{UV}$ + ZP
	where CR is the dead-time corrected, flat-fielded count 
	rate, and  the zero points  ZP=18.82 and ZP=20.08 in FUV and NUV 
	respectively \citep{Morrissey07}.
	In order to estimate the errors on UV magnitudes, we propagated
	 the Poisson statistical errors on source and background counts. 
	In addition to the statistical error, we added an uncertainty to account 
	for systematic inaccuracies in the zero-point of the absolute calibration of 0.05 and 
	0.03 magnitudes for FUV and NUV respectively  \citep{Morrissey07}.
	For the three galaxies in our sample (NGC 1800, NGC 1808 and NGC 3522) in common with the sample
	of \citet{dePaz07}, the measured magnitudes are consistent within errors with those measured  by 
	\citet{dePaz07} in d$_{25}$. 

	For LGG~225, we registered the SDSS  images (corrected frames 
	subtracted by the `soft bias' of 1000)  to the corresponding {\it GALEX} 
	ones using the IRAF tool {\tt sregister}. 
	We then computed the SDSS magnitude in the u, g, r, i and z bands in 
	the same elliptical apertures used for the UV.
	We converted SDSS counts to magnitudes following the recipe provided
	in {\tt http://www.sdss.org/dr7/algorithms/fluxcal.html\# \\counts2mag}.
	Table \ref{3a} lists the SDSS AB magnitudes of the  LGG~225 members. 
	Due to the proximity of some  galaxies to the SDSS image boundaries, for six
	galaxies the UV elliptical apertures were reduced to be the same of the SDSS ones 
	(see note in Table \ref{3a}).
	 
	Surface photometry was also carried out on the background subtracted
	{\it GALEX}  and  SDSS images for LGG 225 with the {\tt ELLIPSE} 
	fitting routine in the {\tt STSDAS}
	package of {\tt IRAF}. {\tt ELLIPSE} computes a Fourier expansion for each successive
	isophote \citep{Jedrzejewski87}, resulting in the surface photometric
	profiles.  The UV and SDSS luminosity and (FUV-NUV) and ($g - i$) color profiles    
	are plotted in the right panels of Fig. \ref{225}. The profiles are plotted 
	versus  galactocentric distance along the semi-major axis.

	\begin{figure}[!ht]
	\includegraphics[width=6.5cm,angle=-90]{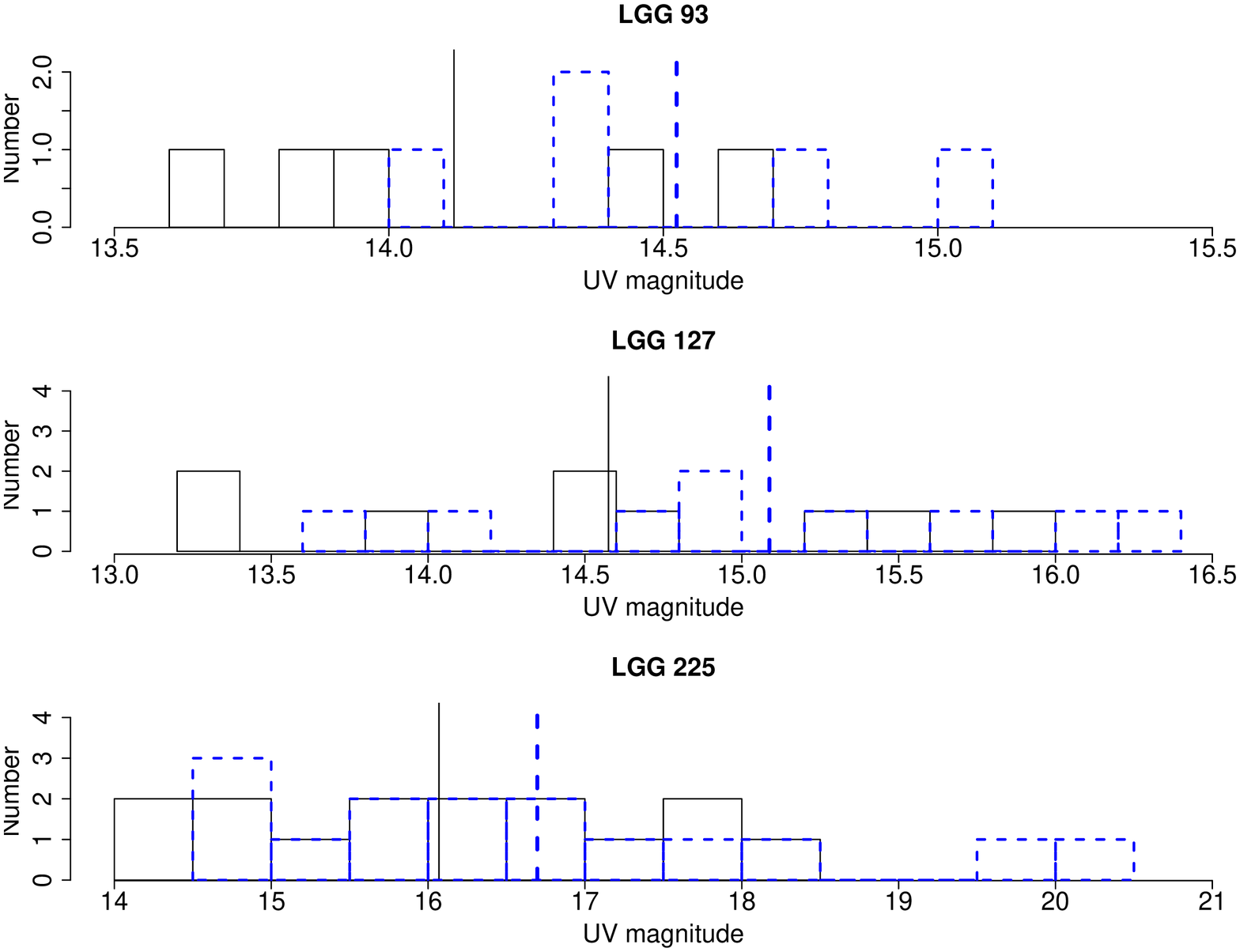} 
	 \caption{NUV (black solid lines) and FUV (blue dashed lines) AB magnitude
	 distributions of the group members and their luminosity weighted mean value.    
	}
	\label{mag}
	\end{figure}

	\begin{figure*}[!ht]
	 \includegraphics[height=15.5cm,angle=-90]{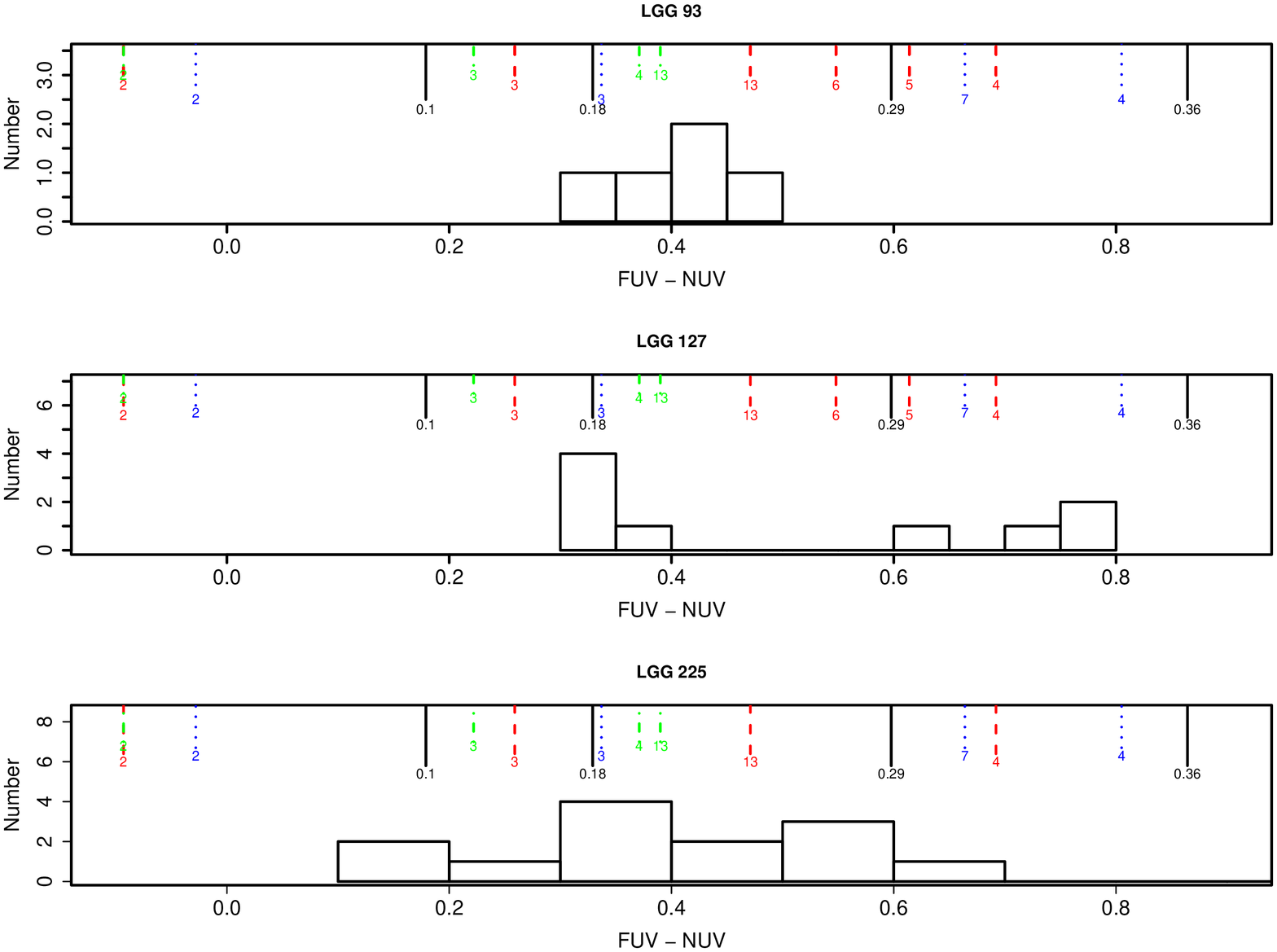} 
	 \caption{(FUV-NUV) distribution of the spirals and irregulars of the three groups. We also show   
	 (FUV-NUV) colors from SSP (with solar metallicity (black solid lines), and from  GRASIL spirals
	 with inclination 45$^\circ$ (green dotted dashed lines)  and inclination 90$^\circ$ 
	 (red dashed lines)  to appreciate the difference due to internal dust extinction.  
	  Models with inclination 90$^\circ$ are also show  with an additional reddening of E(B-V)=0.3
	  (R$_v$=3.1 blue dotted lines). Ages of the models are in Gyrs.  
	}
	\label{col}
	\end{figure*}

	\begin{figure}
	\includegraphics[width=6.5cm,angle=-90]{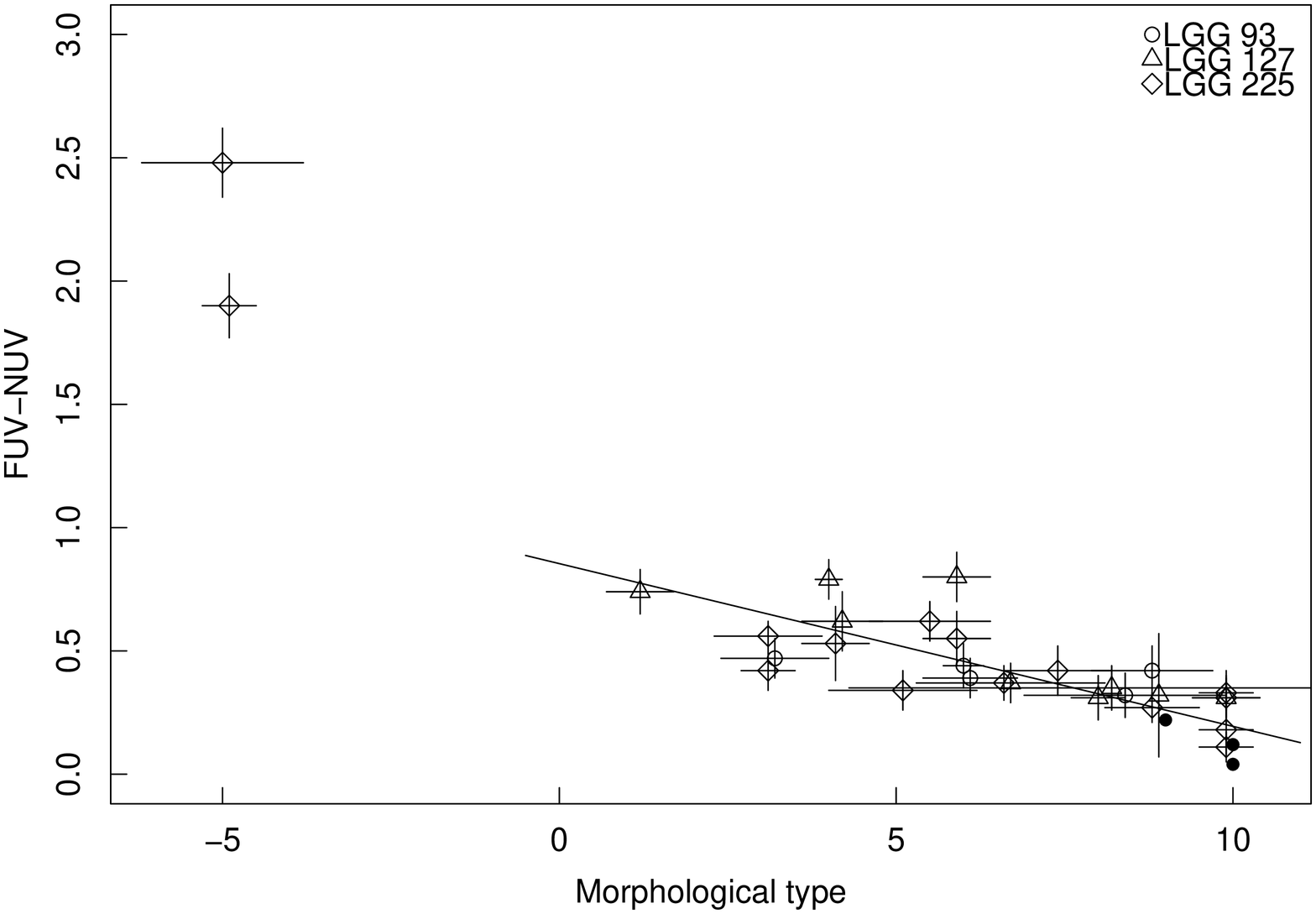}
	 \caption{The (FUV-NUV) color as a function of morphological
	 type for all group members. The solid line represents the best linear fit for types
	 T=-0.5 or later (i.e. spiral galaxies and irregular) in \citet{dePaz07}.   Points on the upper
	 left side of the plot are the two elliptical galaxies NGC~3457 and NGC~3522.  Filled dots 
	represent three irregular LG galaxies  in the sample of \citet{dePaz07}.
	}
	\label{type}
	\end{figure}

	Figures~\ref{mag}  and   \ref{col} show the distribution of NUV and FUV magnitudes and  
	(FUV-NUV) colors   for the three groups. 
	The five members of  LGG 93 have very
	similar total apparent magnitudes in both UV bands 
	($\langle NUV_T \rangle$=14.12$\pm$0.42,
	$\langle FUV_T \rangle$=14.52$\pm$0.40) and their 
	(FUV-NUV) colors (see Table~3) show a small
	dispersion (0.41$\pm$0.06), IC~1954 being the reddest galaxy 
	(FUV-NUV=0.47$\pm$0.08) and
	IC~1959 the bluest (FUV-NUV=0.32$\pm$0.09).
	The mean NUV and FUV  magnitudes  of 
	LGG~127 and LGG~225  are $\langle NUV_T \rangle$=14.58$\pm$0.93, 
	$\langle FUV_T \rangle$=15.09$\pm$0.87 and 
	$\langle NUV_T \rangle$=16.07$\pm$1.27, 
	$\langle FUV_T \rangle$=16.70$\pm$1.73 respectively. 
	The (FUV-NUV) colors of LGG~127  (Table~3)
	show a large dispersion, NGC~1827 being  the reddest galaxy
	with (FUV-NUV)=0.80$\pm$0.10 and ESO305-009 and ESO305-017  the bluest 
	with (FUV-NUV)=0.31$\pm$0.11.  
	Also for LGG 225, the  (FUV-NUV) colors show a large dispersion, 
	the two ellipticals (NGC~3457 and NGC~3522) being 
	the UV-faintest and reddest galaxies in the sample, however their magnitudes
	have  a large uncertainty due to the short exposure times.
	For comparison, we also show in Figure {\ref{col} the synthetic colors for
	 Single Stellar Population (SSP) at representative ages   and for GRASIL models of spirals
	at 2 inclinations and extinctions, computed  as described in the next Section. 

	In Figure~ \ref{type} we plot the (FUV-NUV) colors as a 
	function of  morphological type for all group members.
	  We also include  three irregulars belonging to our LG \citep{dePaz07}. 
	 The solid line represents the best linear fit for spirals obtained by
	\citet{dePaz07}. Spiral members (FUV-NUV) colors   
	agree well with the best fit,  early-spiral are redder 
	than late-type spirals and irregular galaxies. 
	The two early-type members  are located
	in the upper left region of the plot.

	\begin{figure*}[!ht]
	\begin{tabular}{cc}
	\vspace{-0.35cm}
	\includegraphics[width=6.cm,angle=-90]{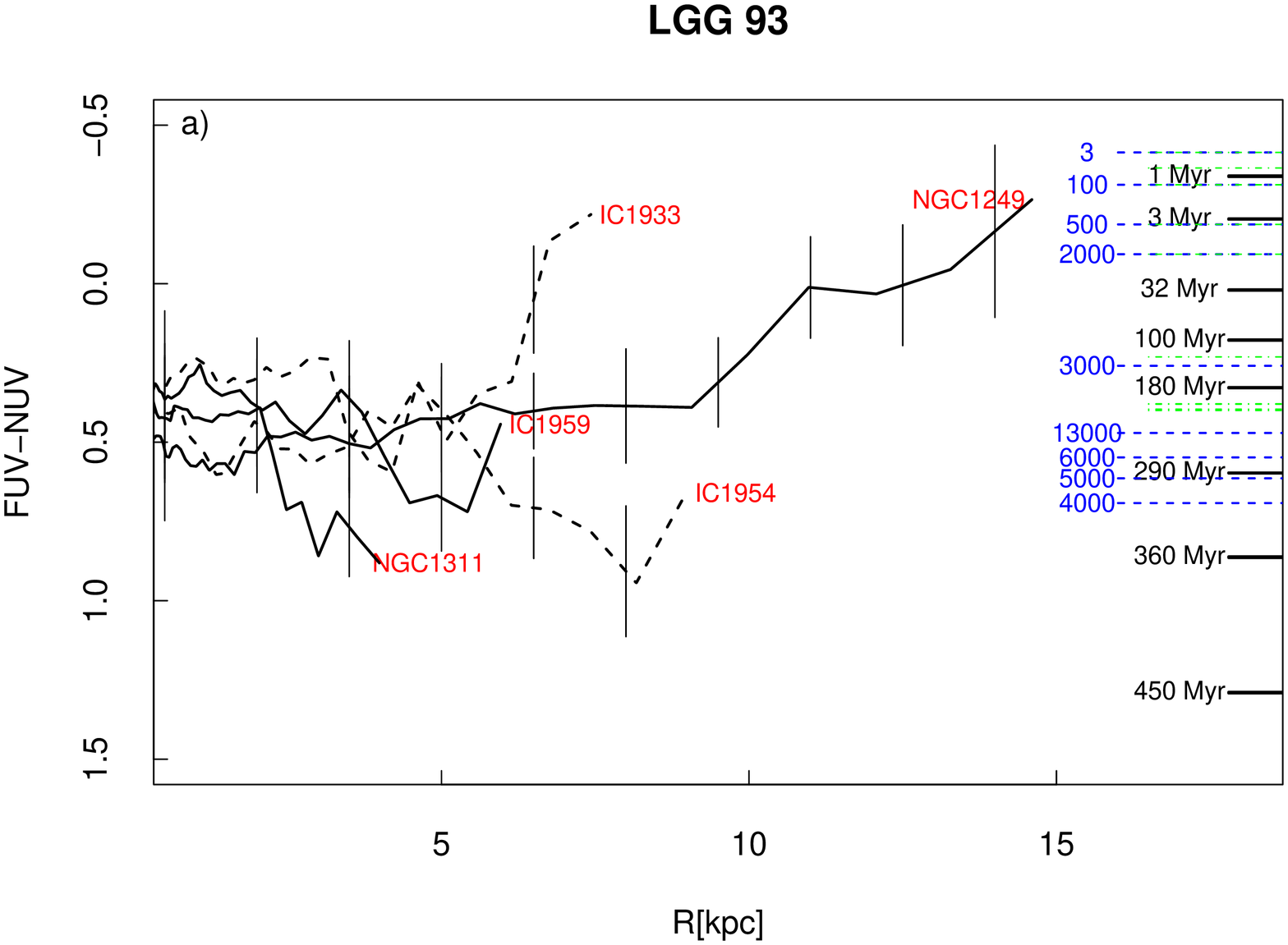} & 
	\vspace{-0.35cm}
	\includegraphics[width=6.cm,angle=-90]{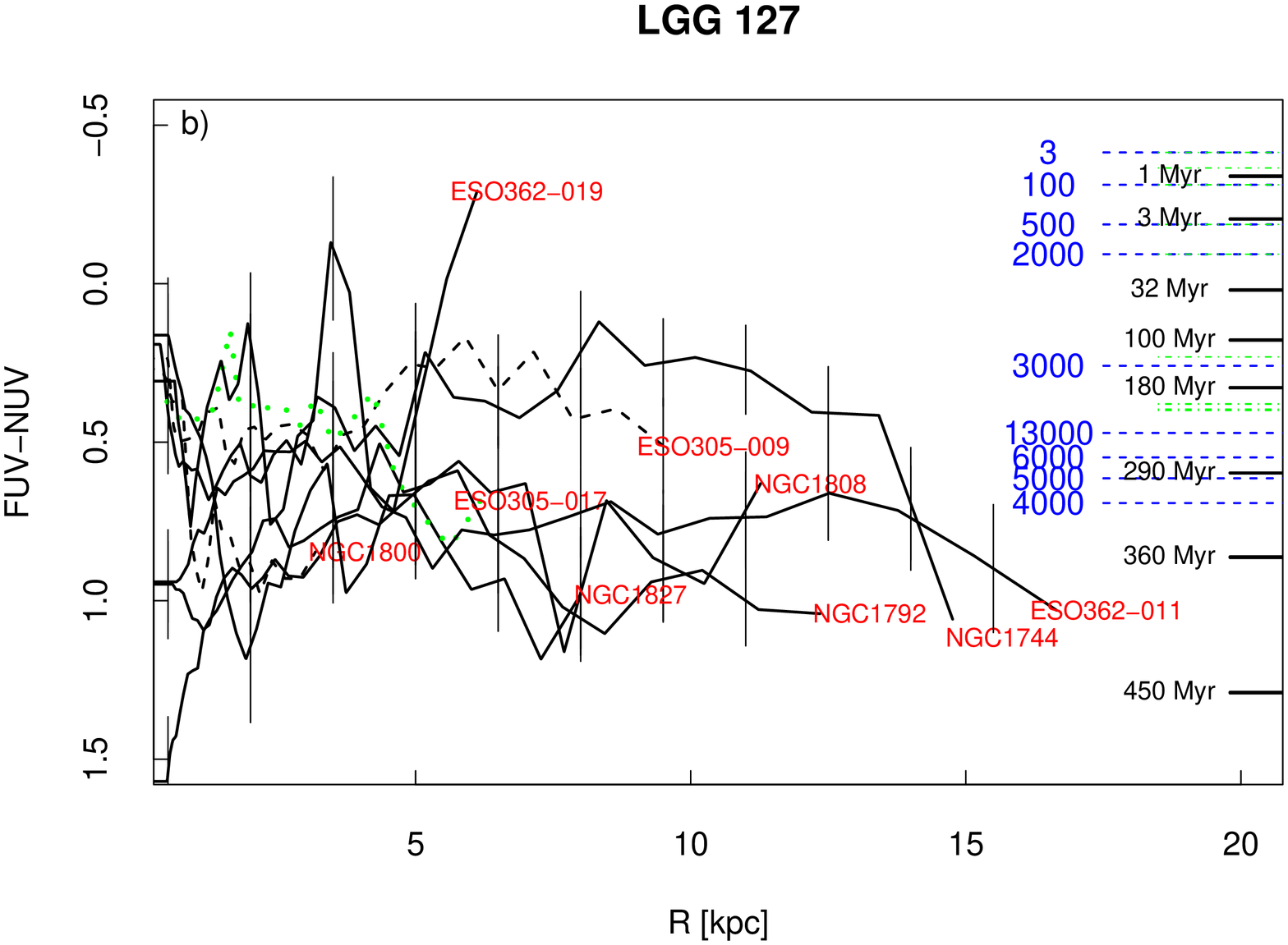} \\
	\vspace{-0.35cm}
	\includegraphics[width=6.cm,angle=-90]{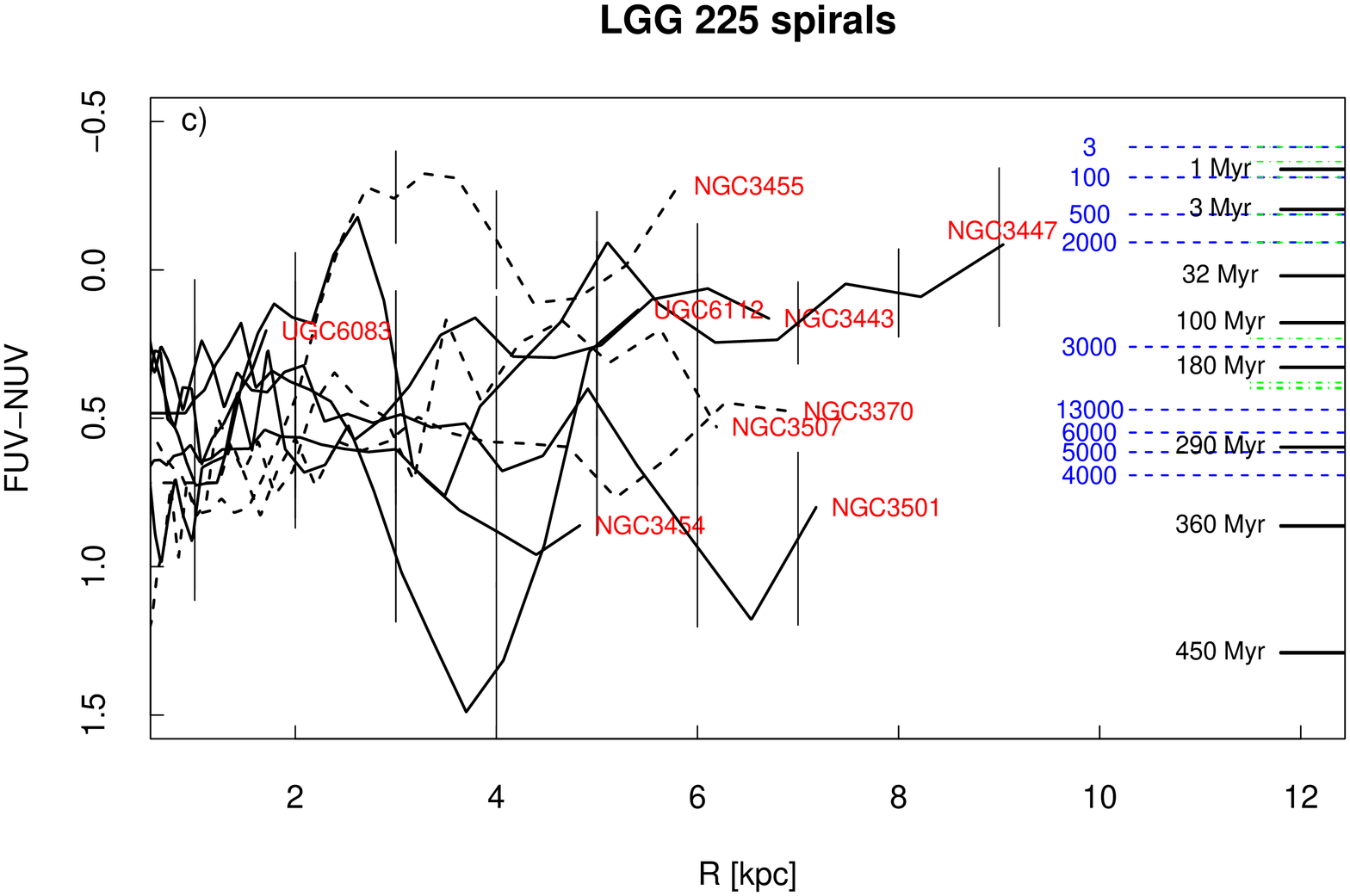} &
	\vspace{-0.35cm}
	\includegraphics[width=6.cm,angle=-90]{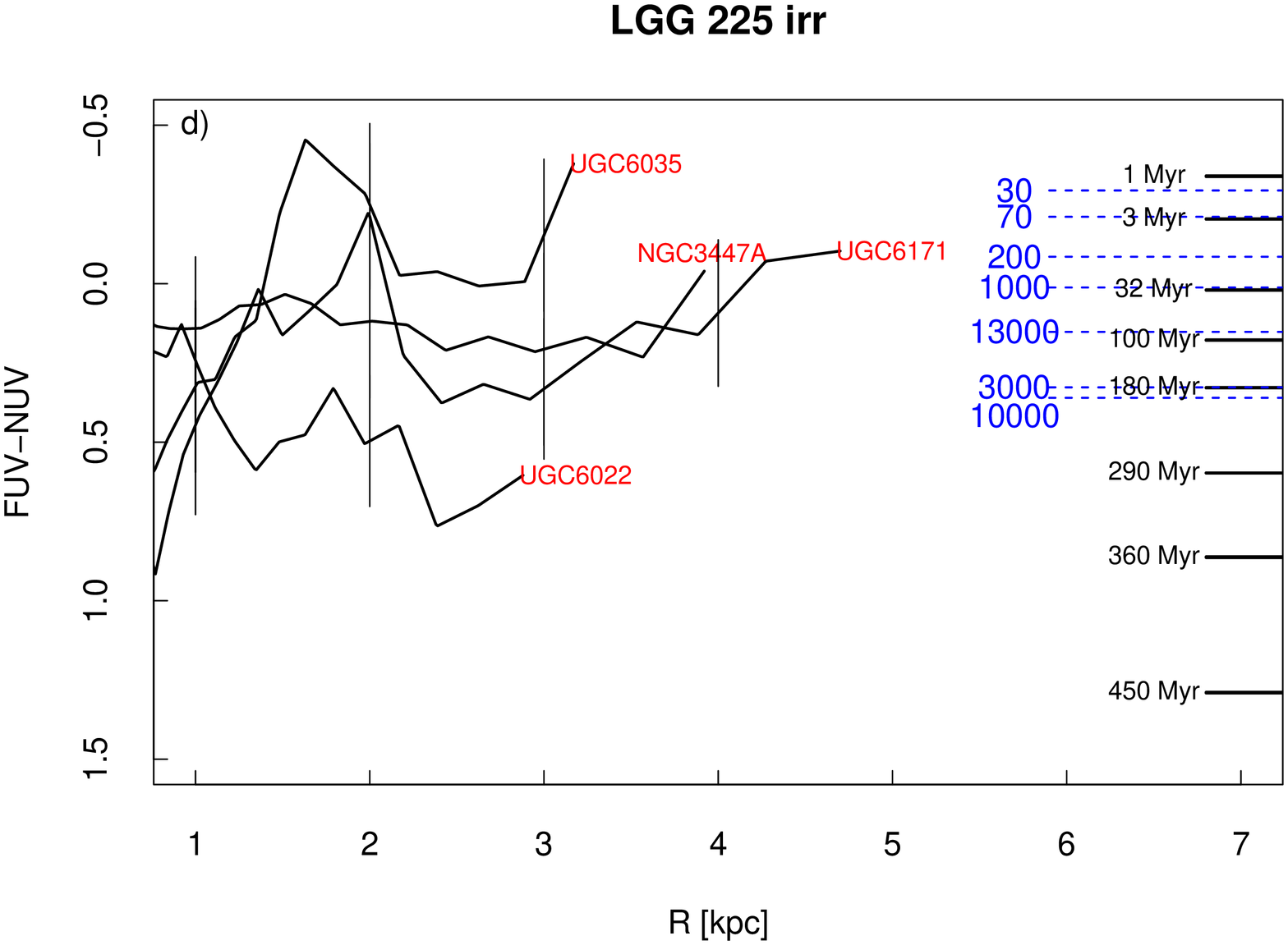}\\
	\vspace{-0.35cm}
	\includegraphics[width=6.cm,angle=-90]{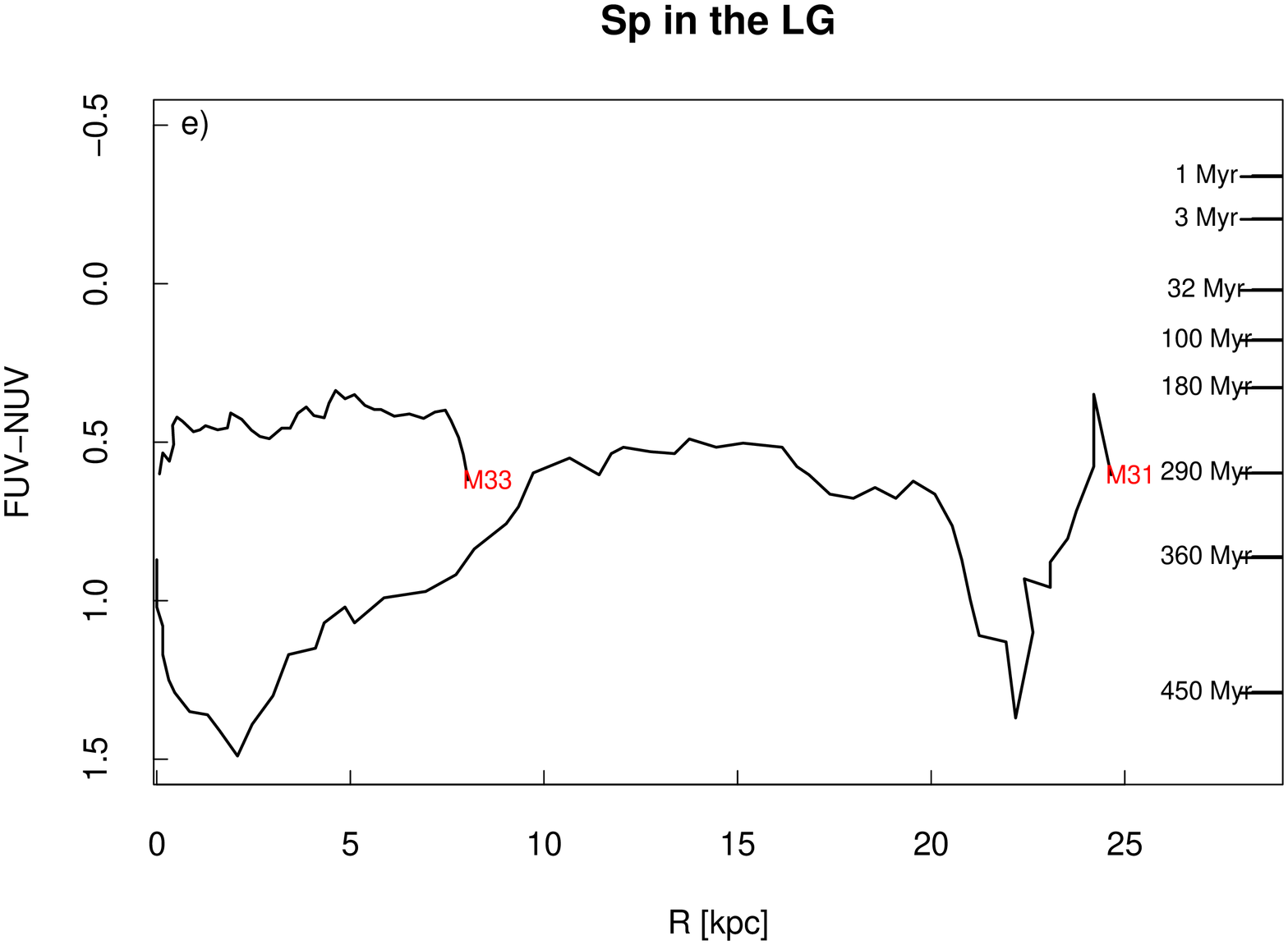} & 
	\hfill
	\includegraphics[width=6.cm,angle=-90]{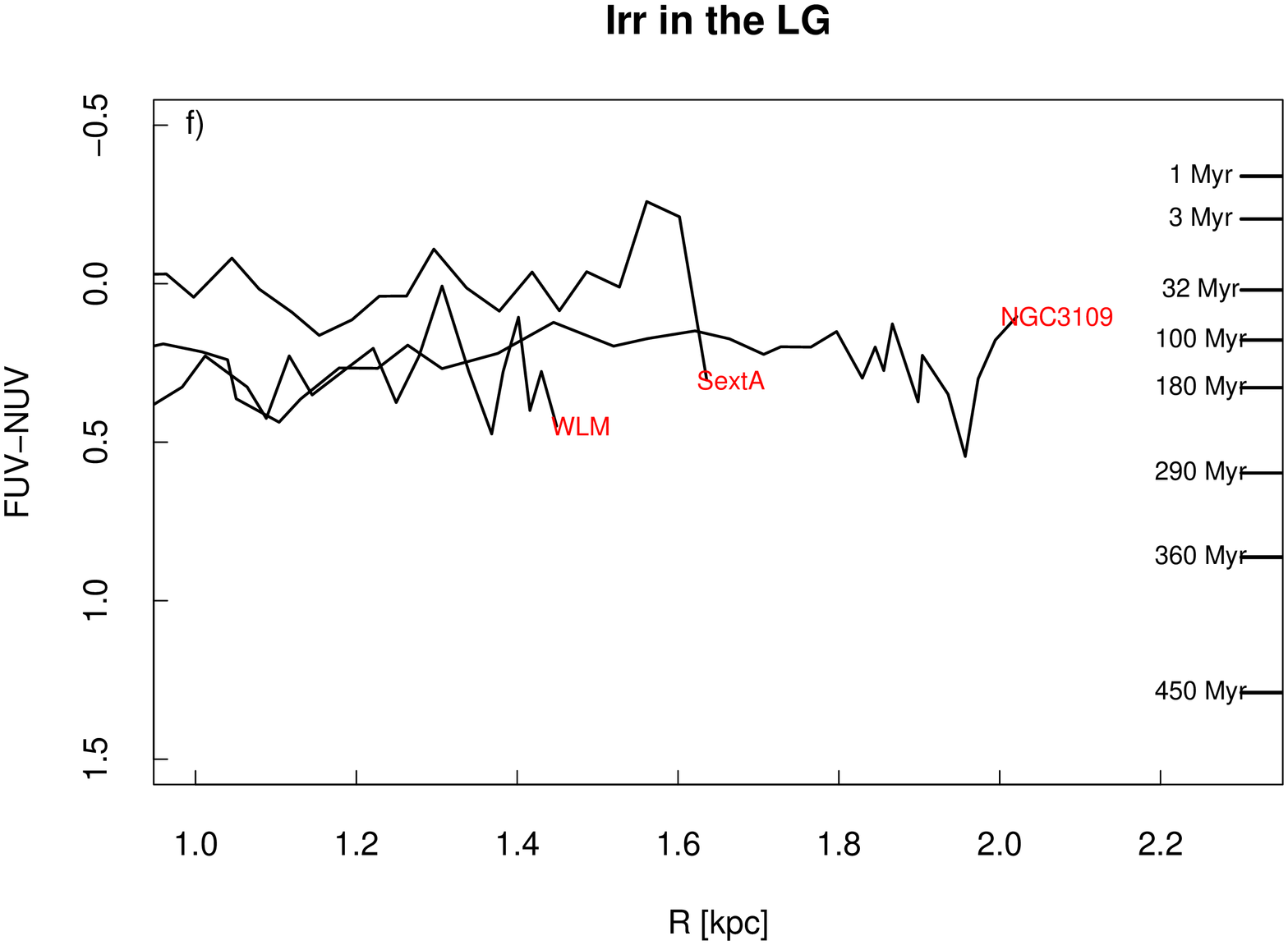}\\
	\end{tabular}
	\caption{Radial (FUV-NUV)  color profiles ($\pm$1 $\sigma$ error bar) 
	of spiral and irregular galaxies of the three groups compared to (FUV-NUV) 
	of spirals and irregulars in the LG. Solid lines show spirals with 
	inclination greater than 60$^\circ$  and dashed lines spirals with inclination
	less than  60$^\circ$. (Panel $a$) Spiral galaxies in LGG~93. (Panel $b$) spiral galaxies
	and the only one irregular (green dotted line) in LGG~127. 
	(Panel $c$) and $d$) spiral and irregular galaxies in LGG~225. (Panel $e$)
	(FUV-NUV) color profile of M~31 and M~33 \citep{Thilker05} and (panel $d$) of 
	three irregular galaxies in the Local  Groups (Gil de Paz et al. 2007).
	Synthetic color for instantaneous 
	burst population of varied age 
	are indicated on the right side 
	of each plot. We also show colors from GRASIL models for spirals with inclination=90$^\circ$
	(blue dashed lines) and inclination=45$^\circ$ (green dotted lines) and for irregulars.
	Reddening for foreground extinction has no effect on the GALEX (FUV-NUV) 	
	color if the dust is "typical Milky Way dust" with Rv=3.1
	 \citep[e.g.][]{Bianchi09}, but may significantly redden the color if UV-steeper extinction curves apply.
 } 
 
 \label{pr}
\end{figure*}

\begin{figure*}[!tt]
 \includegraphics[width=6.5cm,angle=-90]{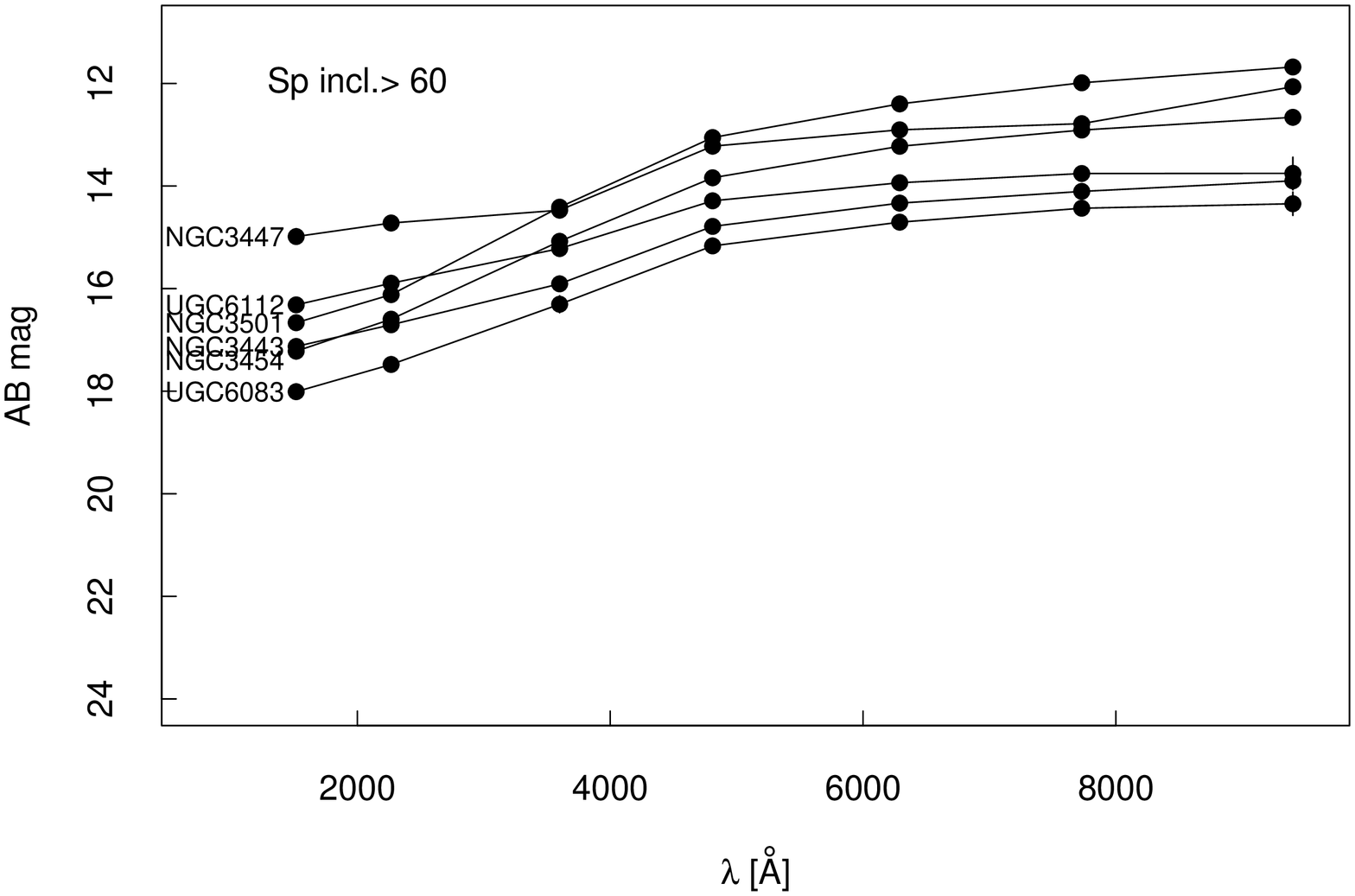}
\vspace{-1.2cm}
\includegraphics[width=6.5cm,angle=-90]{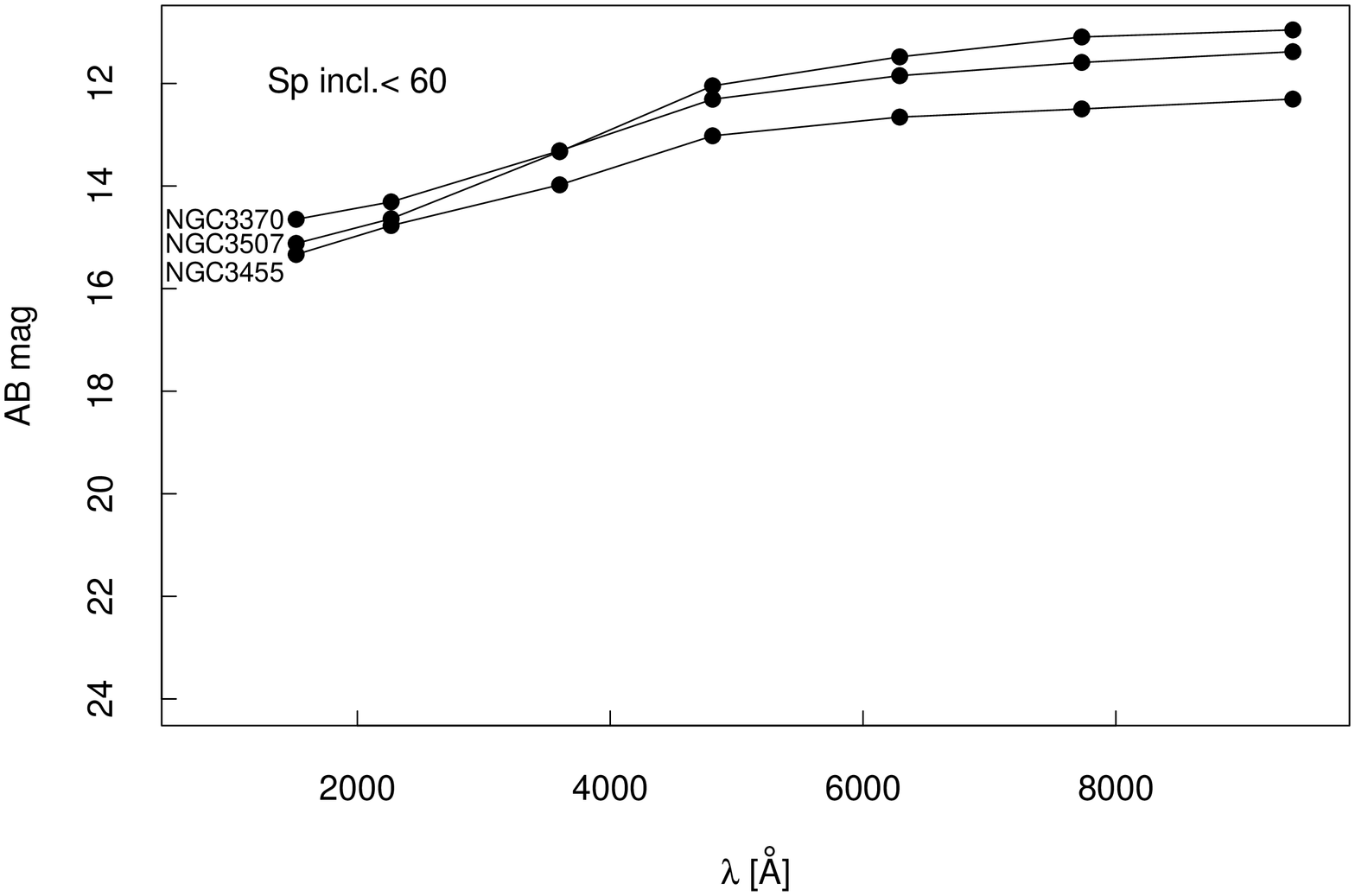} \\
\vspace{-0.4cm}
\includegraphics[width=6.5cm,angle=-90]{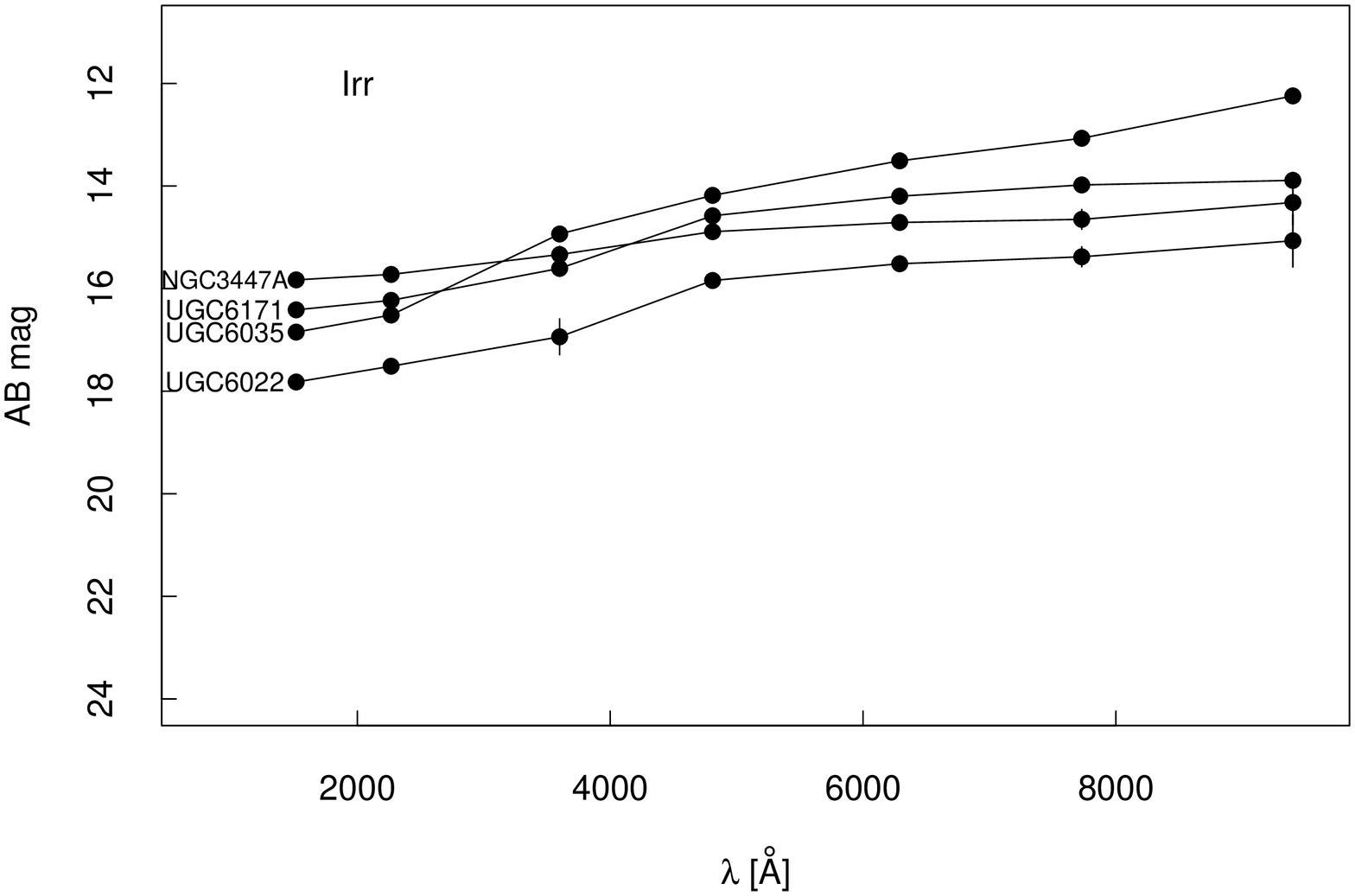}
\includegraphics[width=6.5cm,angle=-90]{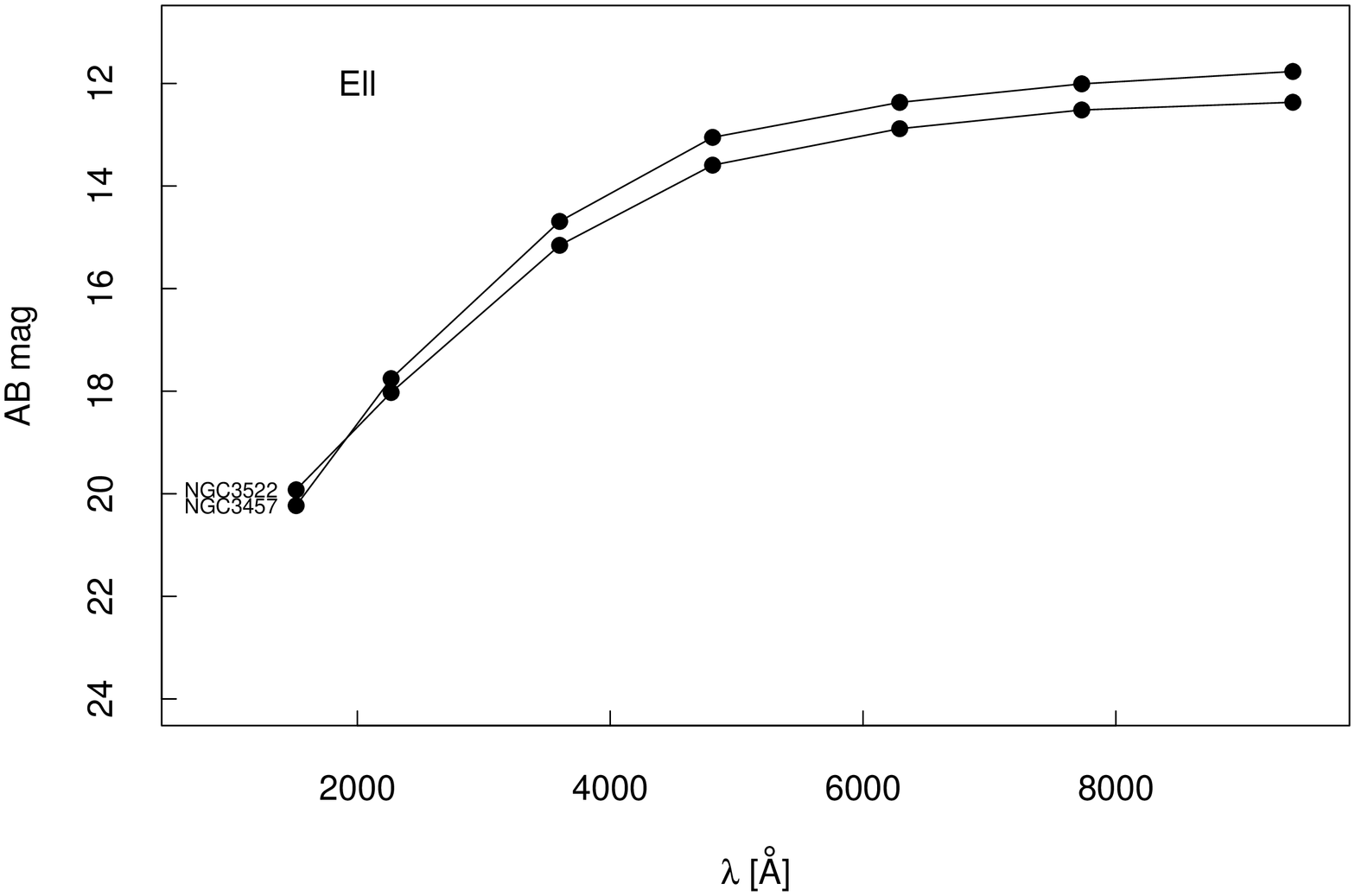} 
\caption{SEDs (GALEX FUV, NUV, and  SDSS {\it u g r i z} photometry)
 for spirals with inclination $>60^\circ$ and $<60^\circ$ (left and right top panels), 
 for irregulars (left bottom panel) and ellipticals (right bottom panel) of LGG 225. 
 Points represent aperture photometry obtained in the {\it GALEX} FUV and NUV 
 and SDSS $ u, g, r, i, z$ bands.  Errors are typically less than bullets.   
 }
\label{SED}
\end{figure*}

\begin{figure*}[!h]
\vspace{-1.cm}
\includegraphics[width=7.cm,angle=-90]{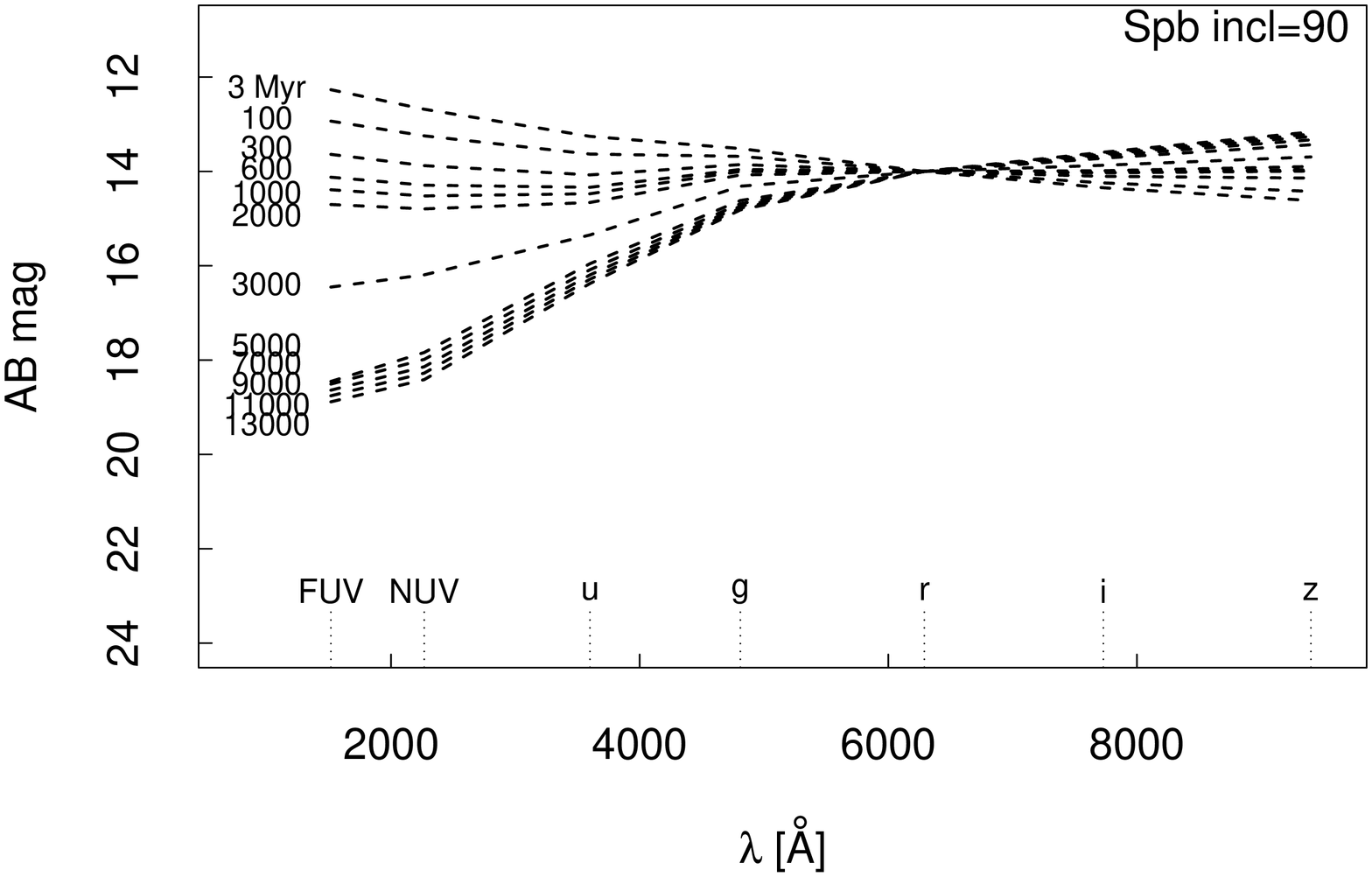} \vspace{-1.5cm} \includegraphics[width=7cm,angle=-90]{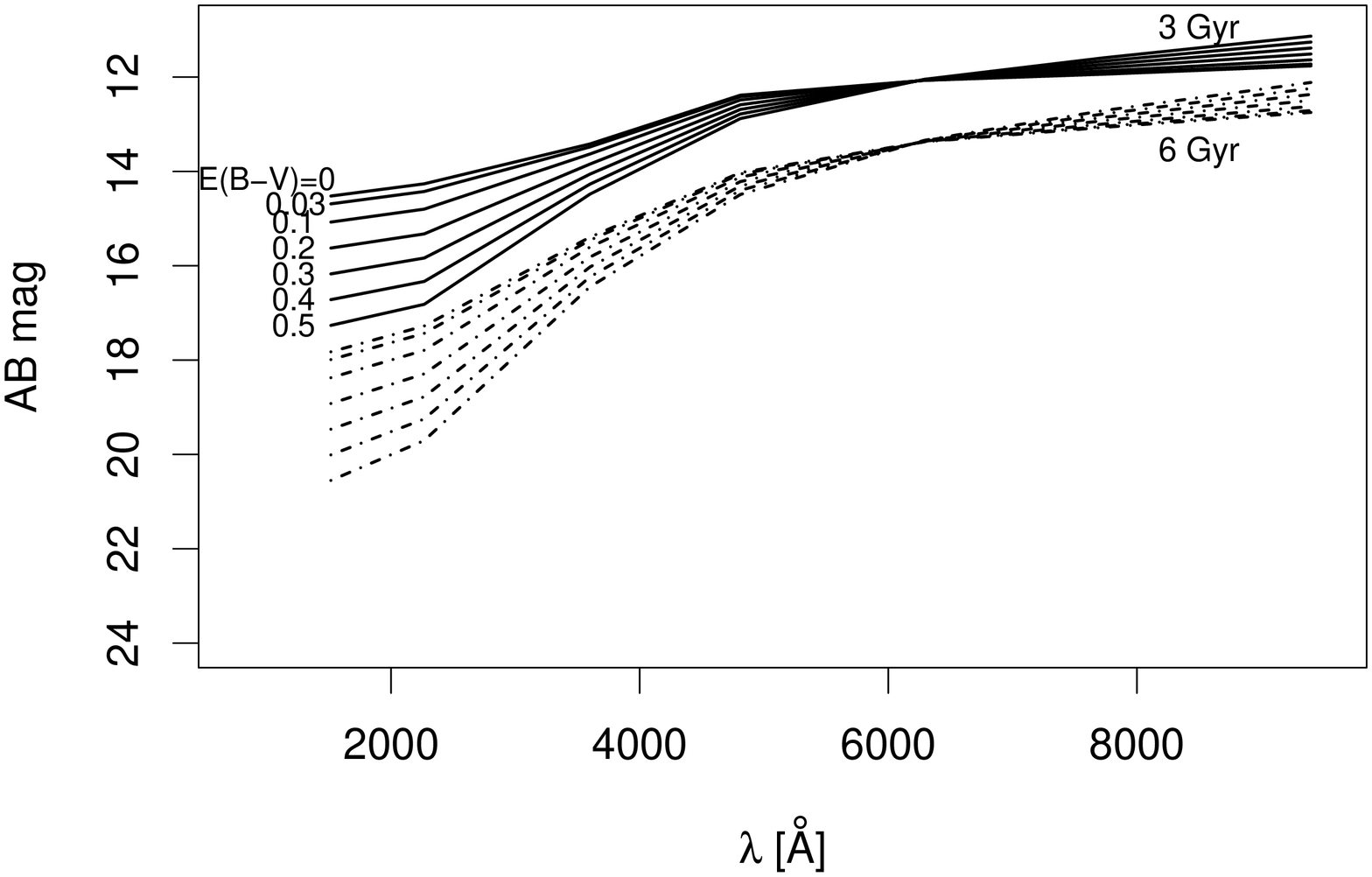}\\ 
\includegraphics[width=7cm,angle=-90]{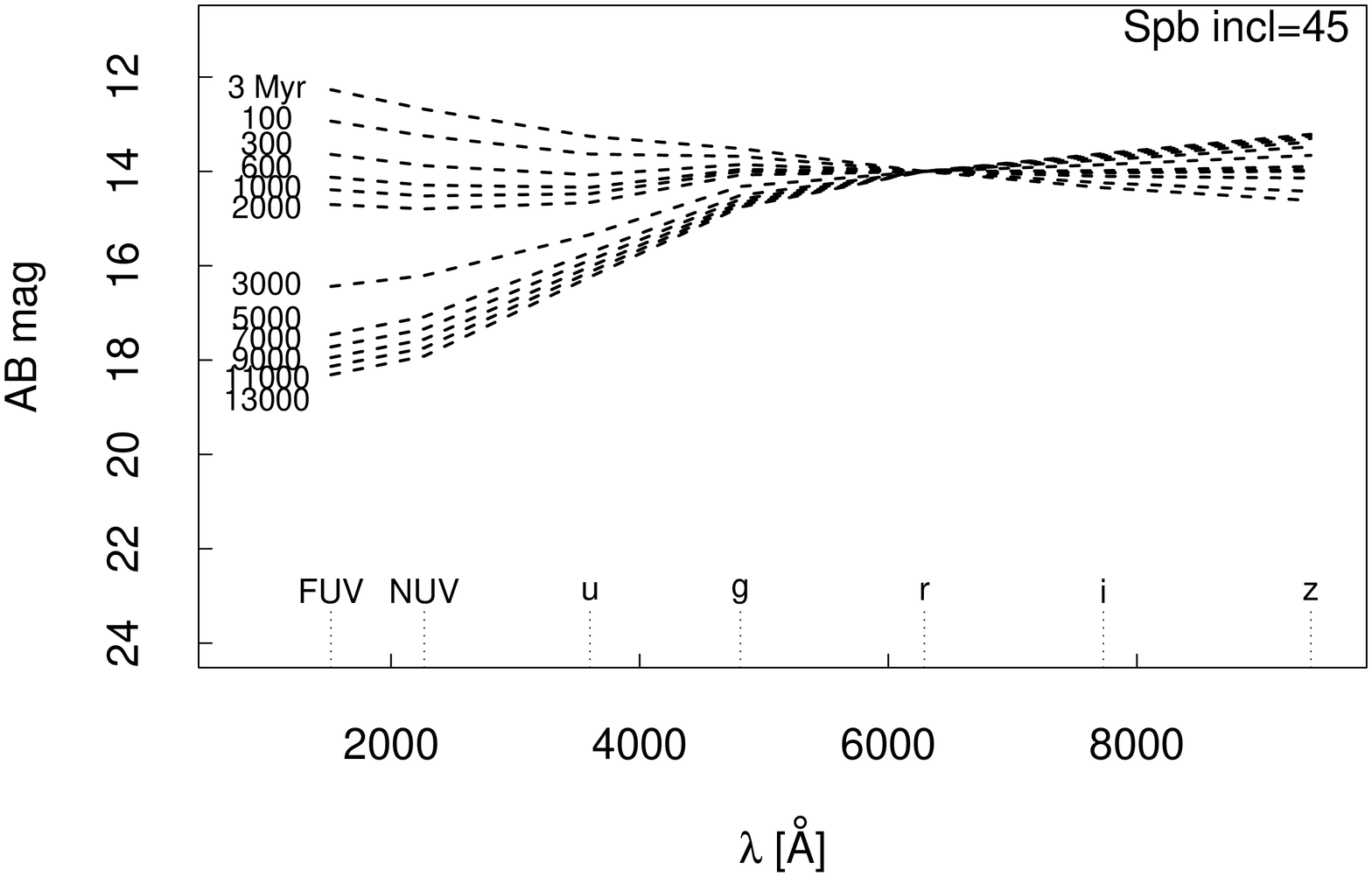} \vspace{-1.5cm}  \includegraphics[width=7cm,angle=-90]{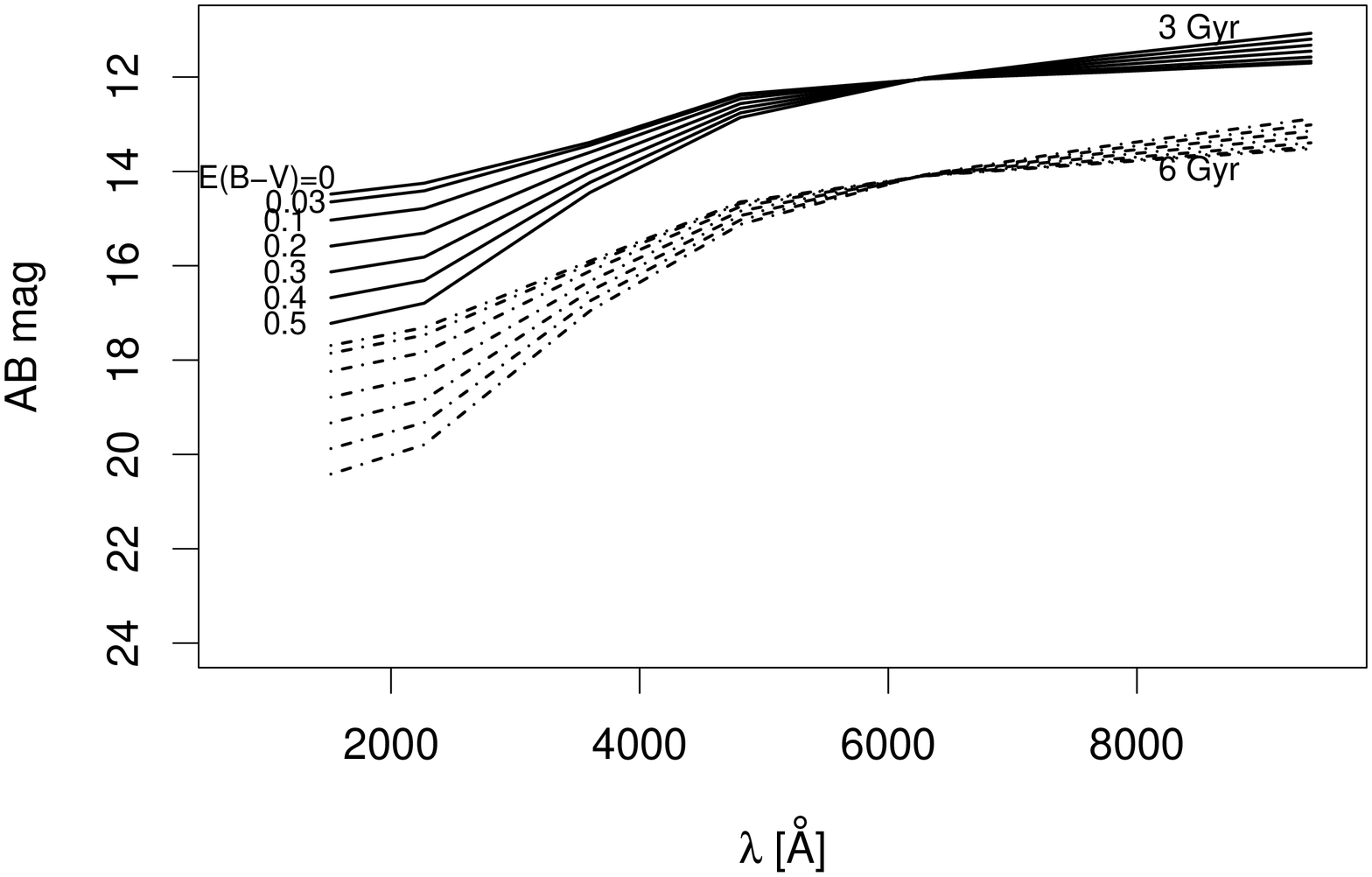}\\ 
\includegraphics[width=7cm,angle=-90]{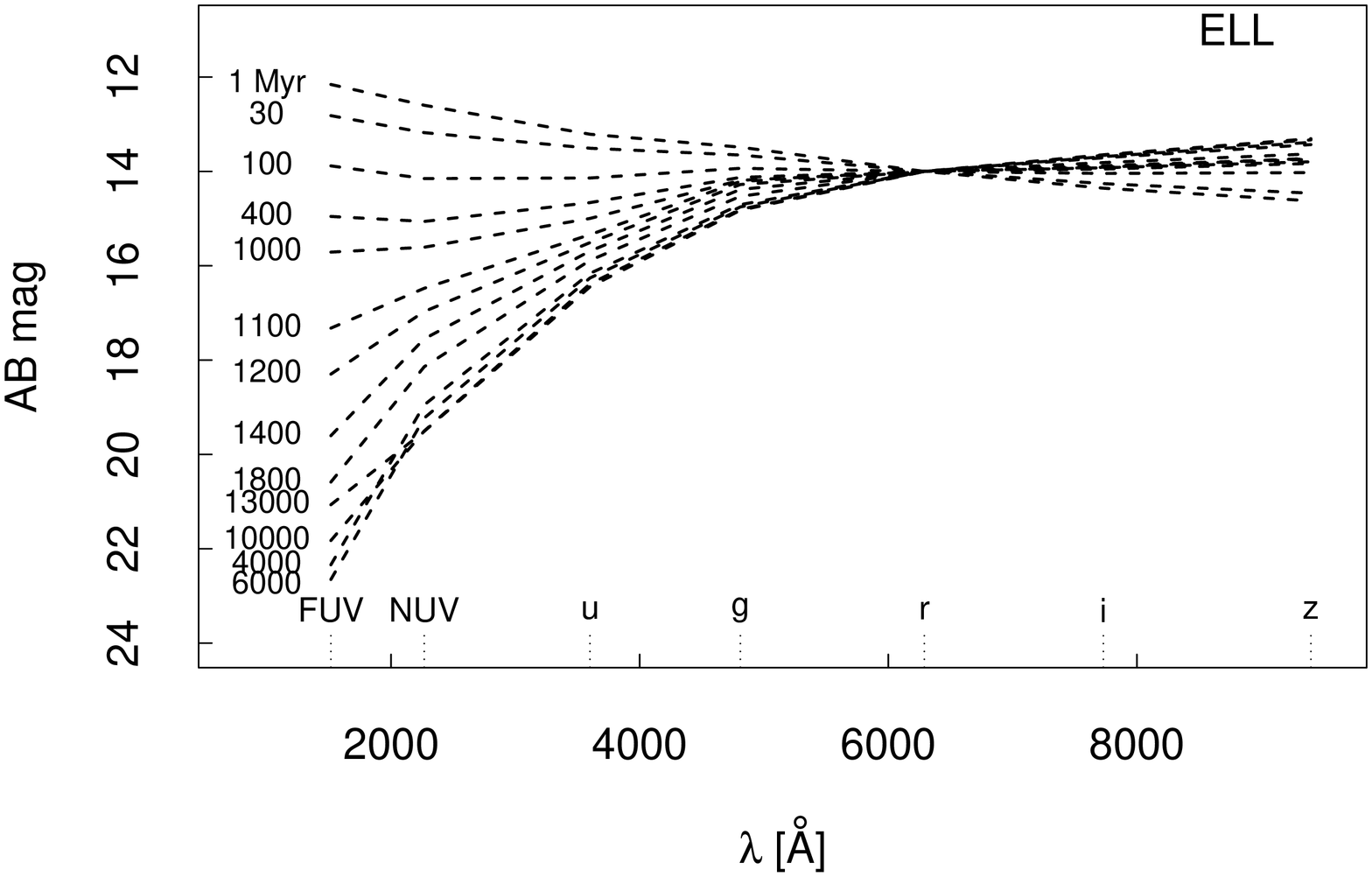} \vspace{-1.5cm}\includegraphics[width=7cm,angle=-90]{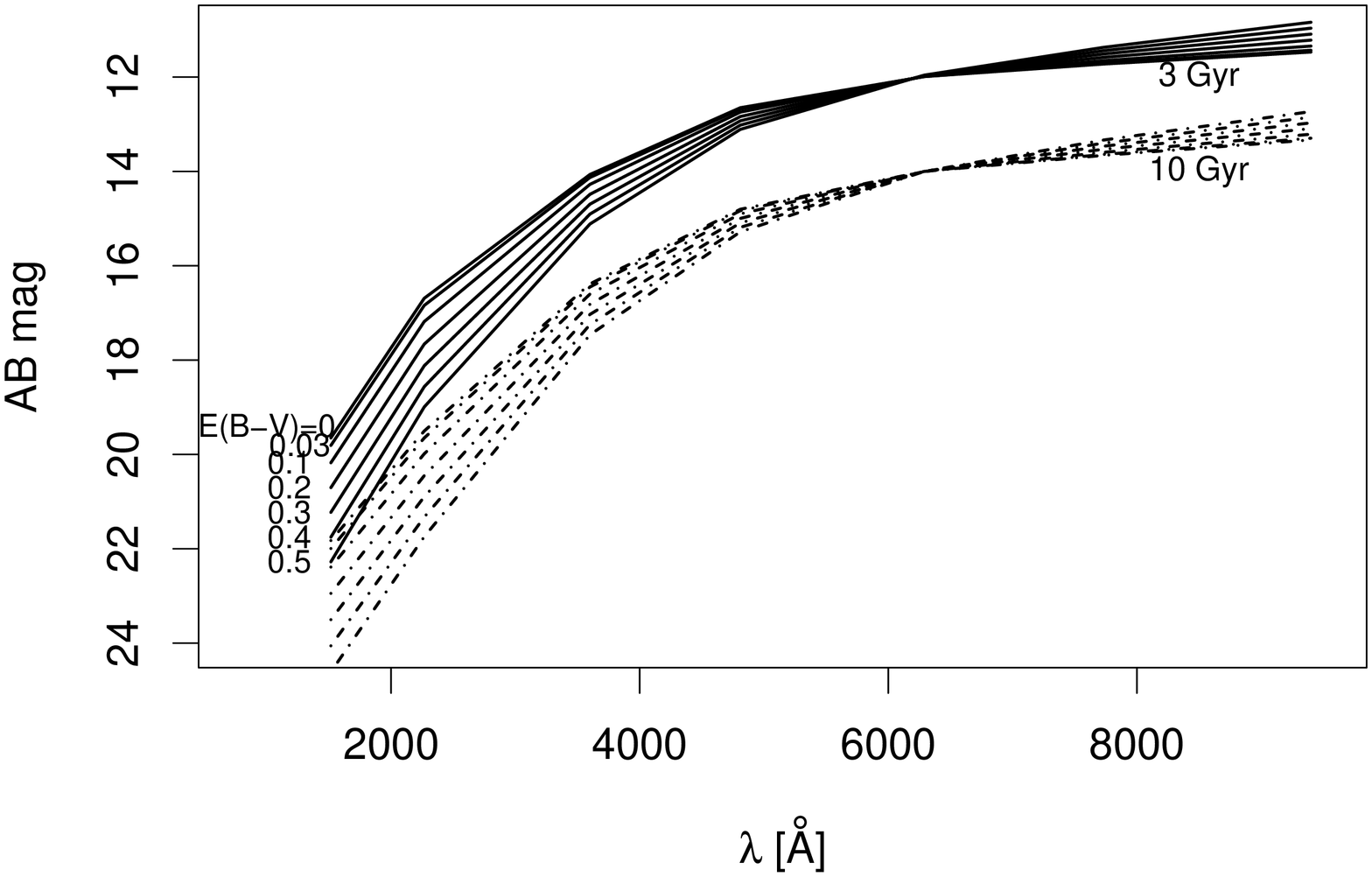}\\
\includegraphics[width=7cm,angle=-90]{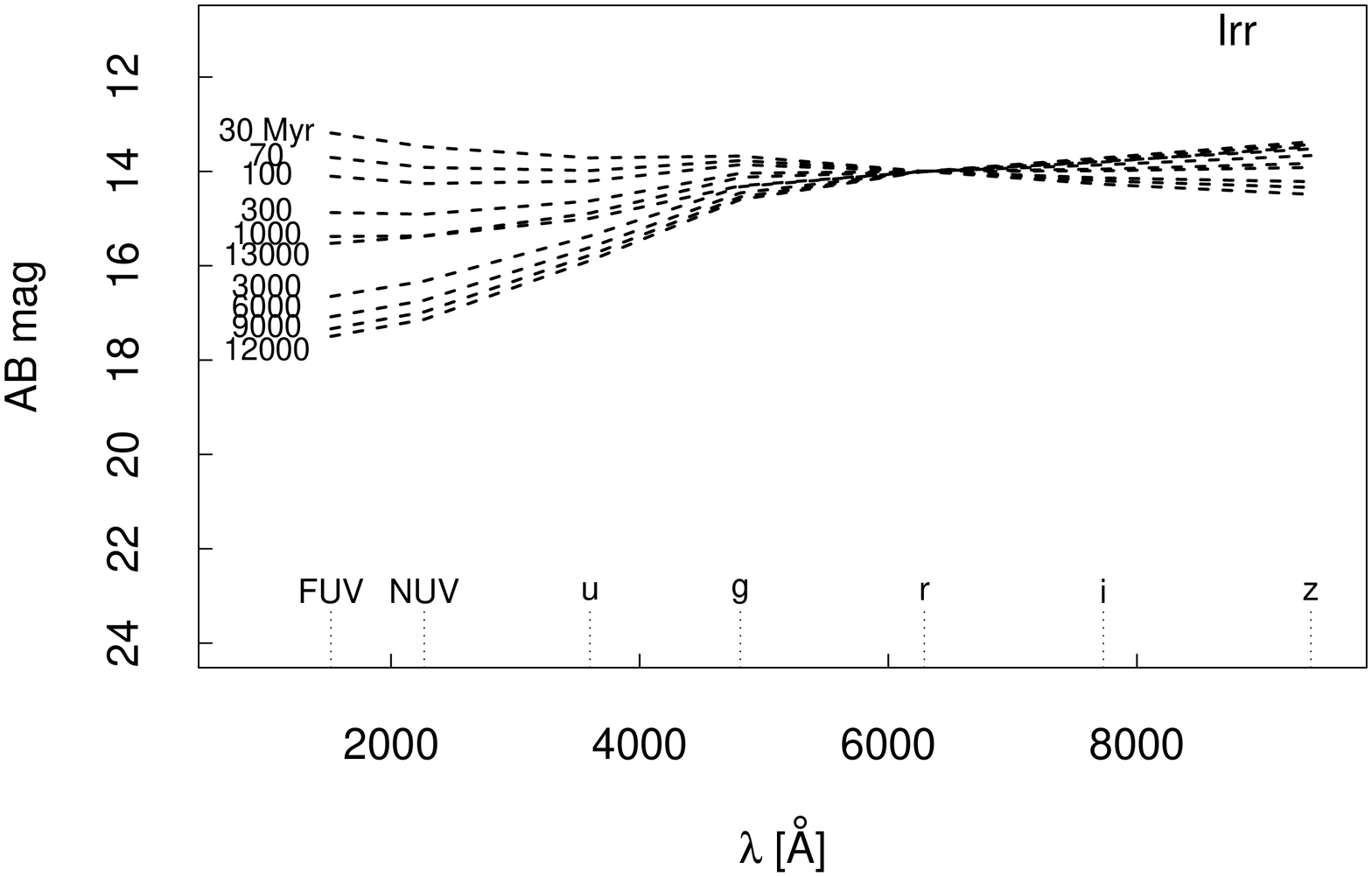} \includegraphics[width=7cm,angle=-90]{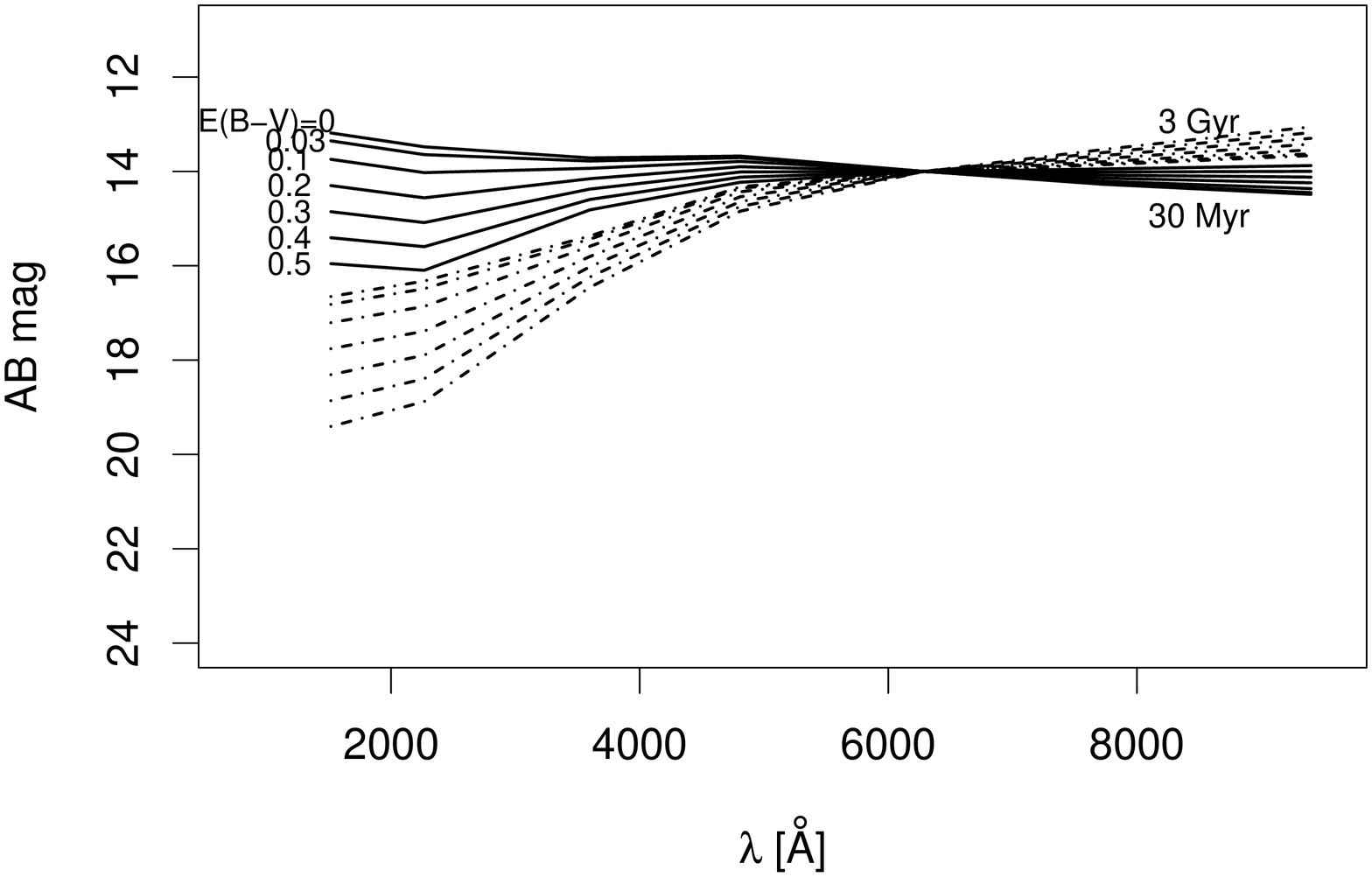}\\
   \caption{Left: GRASIL SED models for different ages in Myrs adopting SFHs (Table 5) for
 spiral with inclination of 90 and  45 degrees, elliptical and irregular. 
 Right:the effect of the extinction with R$_v$=3.1 on   model SEDs for some representative ages are shown for each SFH examined.}  
  \label{Mod}
\end{figure*}

\section{Analysis}
 
\subsection{FUV - NUV color profiles in member galaxies }   
In Figure \ref{pr}  (top and middle panels) we show (FUV - NUV) 
radial color profiles of the 
three groups. 
We evaluated color profiles within   isophotes out to radii 
where the uncertainty in the surface
brightness was less than 0.3 mag/arcsec$^2$.
The color profile of galaxies 
of each  group is plotted as a function of the linear projected distance from 
the galaxy center.   
For comparison, synthetic colors for SSP and for spirals computed with the
 GRASIL model code
 \citep{Silva98} at representative ages  
are also shown (see next Section for spirals that account for internal reddening). 
 Spirals  with  
inclination larger or smaller than 60$^{\circ}$  are plotted 
in Figure \ref{pr}
with solid and dashed lines respectively. In  panel b)  of the same figure, the green dotted
line marks the only irregular in  LGG 127.  
For comparison, in the  bottom panels of Figure \ref{pr}, 
we show the (FUV-NUV) color profiles  of M31 and M33 presented 
by Thilker et al. (2005) and of three irregulars in the LG  \citep{dePaz07}. 
Extinction certainly affects the color: a signature is clearly
visible in the  `broken' appearance of  NGC~1311's disk (see Figure \ref{93}).
The (FUV-NUV) color is essentially reddening free for average Milky Way extinction 
with Rv=3.1  \citep{Bianchi05, Bianchi07} and thus gives a direct indication of age. 
Panel $a$) in Figure \ref{pr} shows color profiles of spirals in the LGG~93 group. All
galaxies have similar (high) inclination. 
Most of the 
galaxy centers in LGG 93 show indication of young stellar populations.
NGC~1249 and IC~1933 color profiles became bluer in the outskirts
suggesting the presence of young stellar populations. 
Color profiles of the group LGG~127 are shown in   panel  $b$) of
Figure~\ref{pr}.   
The general trend of the (FUV-NUV) radial profiles in  LGG 127 spirals, 
in particular that of NGC~1744 
and NGC~1792, is not very dissimilar from that of M33.
(FUV-NUV) colors profiles in LGG225  suggest
that irregular galaxies (panel d)  host  younger stellar populations than
spiral members (panel c).   
Blue color profiles  of the  interacting pair NGC~3454/NGC~3455, and NGC 3447/NGC 3447A, 
suggest that  recent star formation events could be triggered by tidal interaction. 
This could be also the case of the strongly distorted galaxy UGC~ 6035, whose  
(FUV-NUV) color  suggests the presence of very young stellar population.
The (FUV-NUV) color profiles  of 
three irregulars of the LG, studied by \citet{dePaz07}, namely WLM (IB, RC3 Type =10), 
Sextans~A (IB RC3 Type=10) and NGC 3109 (SBS9  RC3 Type=9) show a 
color profile quite flat (Figure \ref{pr}  panel f)) similar to UGC~6171 and NGC~6022 
in LGG~225.
 
\subsection{The spectral energy distributions of LGG225 galaxies}

LGG225 is the only group for which we have measurements of spectral energy 
distributions (SEDs)  from UV to near-infrared (NIR).  
Figure \ref{SED}  shows the SED of LGG225 members arranged by
morphological types, top panels  show the spirals with
inclination larger and smaller than 60$^{\circ}$ respectively, and bottom panels 
irregulars and ellipticals.
The SEDs of the spiral galaxies are qualitatively similar, especially at optical 
wavelengths,  but show a range of slopes
at near-UV and far-UV  wavelengths, indicative of a range of `ages' of the
composite population, i.e. of a varying relative contribution of the UV-emitting
younger populations.  
Note the `broken' appearance of the SED 
of NGC 3447.   
Most of the irregulars also show a similar SED in optical bands while  UV bands have  different slopes. 
In addition to the amount of recent star formation,
a major factor affecting the SED  in the UV is the extinction by interstellar dust. 
As expected, the SED of the spirals seen edge-on (Figure \ref{SED} top left panel) 
appears in general 
`redder' than the spirals with low ($<$60$^{\circ}$) inclination
(Figure \ref{SED} top right panel).  
In order to interpret 
the effects of evolution (`age' of the integrated population) 
and extinction  we computed grids of models  
using the GRASIL  code \citep{Silva98}.  
  
 \subsection{Model grids}
 Full details of GRASIL code are given in \citet{Silva98}. Here we briefly 
 describe the main characteristics. 
The code works in two steps: first, the history of the 
star formation rate, the Initial Mass Function (IMF), the metallicity and 
the residual gas fraction are assumed.
The main parameters of this step are the baryonic mass of the galaxy, 
the gas infall time-scale (t$_{inf}$) and the 
star formation efficiency ($\nu$) of the assumed Schmidt law. We adopted the Salpeter IMF
and the typical parameters regulating the star formation history (SFH) for spiral, 
irregular and elliptical types 
from \citet{Silva98}.\footnote{available on the web 
at \\ http://adlibitum.oat.ts.astro.it/silva/grasil/modlib/modlib.html.}
The main SFH  parameters for the different morphology types are 
given in Table \ref{che}, all other parameters of the code have been kept
to their default values.   
In the second step, the integrated SED  is computed taking into account all 
the stars and the gas at  any given age (t$_G$).  An important feature of GRASIL
is that it includes the effect of age-dependent extinction with young stars being 
 more affected by dust. In particular, it takes into account several environments with different dust 
 properties and distributions, such as the AGB envelopes, the diffuse interstellar medium and the
 molecular clouds.
New stars are born in Molecular Clouds (MC) and progressively dissipate them, 
the fundamental parameter that describes this process 
is the time-scale t$_{esc}$. 
The geometry of the spiral galaxies in GRASIL is described as a 
superposition of an exponential disc component
and a bulge component with a King profile for spirals. For ellipticals and irregulars 
a spherically symmetric distribution for both stars and dust with a King profile is adopted.  
The total gas mass (diffuse and MCs) of the galaxy at a 
given age come from the chemical 
evolution model. 
The relative fraction of molecular gas is a free parameter of the code (f$_{mc}$)
and is set to zero for early-type galaxies.

\subsection{Comparison of the observed SEDs with GRASIL models}
From the model spectra we computed synthetic broad-band 
magnitudes in the {\it GALEX}  FUV, NUV  and SDSS {\it u, g, r, i, z}  filters over a large
range of ages.
The resulting SEDs at some representative ages  are shown 
in Figure \ref{Mod}  to illustrate the effects 
of evolution (left panels) and extinction (right panels).
As said in Sect. 5.3,  GRASIL models take into account internal extinction by interstellar dust,
in computing the emerging flux. 
In  figure \ref{Mod} (left panels)
the model magnitudes are shown without additional foreground
extinction.  
In the right panel of the same figure, we  show  the effects of additional foreground extinction 
applied to the models of some representative  ages.
The comparison of the left panels with the right panels 
shows the advantage of using UV plus optical bands. 
 Evolution  and extinction affect   in
different way the SED  in the UV part of the spectrum (FUV, NUV and  u).

The comparison of  the observed SEDs in  Figure \ref{SED} (top panels)
to the  model grids
in Figure \ref{Mod}  indicates that ages younger than 3 Gyrs and older 
than  6 Gyr 
can be excluded for spirals 
if no additional extinction (to the GRASIL estimated internal extinction) is assumed. 
In both cases, thanks to the UV part of the SED, it is possible exclude extinction 
(Rv=3.1) greater than E(B-V)=0.1. 
 
While  Figure \ref{Mod}  illustrates the effects of population evolution and 
dust extinction, for a typical SFH of each morphological type, in practice 
we used  the model grid from a few Myr to 13 Gyr  to estimate by $\chi^2$ 
fitting the best `age' of the composite population
for each observed SED, and we used the best-fit model result to estimate the 
galaxy current stellar mass. 
We performed the SED model fitting in two ways:
first,  assuming
  foreground extinction as given in Table 1, which is minimal, in addition to the
internal extinction  estimated by GRASIL  and, second, 
by treating the `age' of the population and a foreground 
extinction component, both as free parameters. As we do not attempt in this work
to derive the exact SFH history of the galaxies, we consider the results from
these two options  to bracket the range of possible solutions with a good
approximation. This is supported by the fact that a very different SFH we
explored (the ``Irr" type as described by \citet{Silva98} and given in table 5) 
cannot produce satisfactory fits to the observed  SEDs of spirals (Figure \ref{SED} left panel).
The overall results characterize the galaxy populations of the LGG 225 group:
the composite populations of spirals have evolutionary times spanning
between a few Gyrs to 6 Gyrs, and the total mass (sum of the stellar mass
of all galaxies measured) is estimated between  5  and 35 $\times 10^{10}$ M$_{\odot}$,
with the assumed SFH in our model analysis.
 
Lacking SDSS observations for LGG 93 and LGG 127, we used FUV, NUV from Table 3 and B  
magnitudes from  Table 1 to estimate the stellar mass using  the same GRASIL grids.
The estimated  total stellar masses are  $\approx$ 4 $\times 10^{9}$ M$_{\odot}$  
for LGG 93 and $\approx$ 4 
$\times 10^{10}$ M$_{\odot}$  for LGG 127.  
 Uncertainties  affecting derived masses 
of LGG 93 and LGG 127 are larger than those derived for LGG 225, not only because we use only three  bands
in the SED fits but also because the B magnitude is  the total magnitude while FUV and NUV are d$_{25}$ magnitudes.

\subsection{Are there rejuvenation signatures in the Elliptical members of LGG225?}

Local early-type galaxies (hereafter ETGs) are
considered the  fossil record of the processes of galaxy formation
\citep[see for a review:][]{Renzini06}.
At the same time, there is growing evidence that the assumption of passive
evolution of ETGs may be too simplified, especially for the ETGs population
in low density environments (hereafter LDEs). 

Several studies point to the galaxy environment as a 
possible ingredient in their evolution \citep[see e.g.][]{Clemens09}. 
At odds with cluster counterparts, ETGs in LDEs frequently show significant
signatures of relatively recent activity as indicated by the
presence of distorted morphologies \citep{Reduzzi96, Colbert01} 
and of kinematical sub-components  \citep[][and reference therein]{Em04}. 
The analysis of line-strength indices \citep{Trager00, Longhetti00, Ku00, 
Thomas03, Denicolo05, Annibali07, Rampazzo07}  shows that ETGs in LDE have a
large dispersion in their luminosity weighted age, from 
a Hubble time to $\approx$ 1 Gyr.  About 40\% of ETGs 
in the \citet{Annibali07} sample have 
a luminosity weighted age $\leq$ 6 Gyr, and 10\% $\leq$3 Gyr. 

Simulations show that ETGs may have formed through 
subsequent accretion or merging episodes that leave their
signatures in a younger stellar population. 
The ETGs morphological fine structures as well as their kinematical 
sub-components almost certainly formed as the result of either
accretion \citep[e.g.][]{Barnes02, Bournaud05, Naab06} or interactions 
episodes sometimes with gas-rich neighbors \citep{vangorkom97}. 
Some ETGs are also known to have detectable
amounts of cold gas \citep[see e.g.][]{Morganti06}.  
Rejuvenation episodes may then be common among ETGs. However, it is 
still unclear how significant and frequent they have been during the 
Hubble time.   

In this context,  although no morphological peculiarities are shown by
both NGC~3457 and NGC~3522, we investigated the galaxies SEDs (see Figure \ref{SED})
since FUV and NUV  bands are a sensitive probe  
of young stellar populations \citep[e.g.][]{Bianchi07, Bianchi09}.
In addition to {\it GALEX} and SDSS data
we also included  2MASS J, H, Ks magnitudes taken from the Extended Source Catalog \citep{Jarrett03}.
 Figure \ref{fits}  shows the measured SEDs from FUV to IR and the best-fit models obtained for
the two ellipticals. 
The top panels show SED $\chi^2$ fit results imposing the values from Table 1 (column 6) for the foreground extinction,
in the bottom panels both age and the value of the foreground E(B-V) were treated as free parameters. 
In the first case, NGC 3457 has a derived  age close to the oldest of the range 
found for spirals
and it is about three times more massive than NGC 3522.  
The fit with the models of passive evolution 
reproduces well the optical-IR SED of both galaxies, but a small FUV excess is seen in both objects.
This may indicate a different SFH than simply passive evolution.
The much younger ages in the second case are an effect of the higher E(B-V).
Allowing for a higher  foreground E(B-V) component, the whole wavelength range is reproduced
quite well. However such a higher E(B-V) with respect to previous estimates is hard to explain, and different SFHs will be explored
in a subsequent work, together with spectroscopy, to remove the degeneracy age-extinction.

The derived masses,  (between 2 and 4 $\times$ 10$^9$ M$_{\odot}$ for NGC 3457 and 1-2 $\times$ 10$^9$ M$_{\odot}$
for NGC 3522) are comparable to the least massive spirals in LGG 225
and  to the ellipticals in our LG \citep[NGC 205, M32,][]{Mateo98}. The mass of the most massive spirals in LGG 225 is comparable to M33, 
 therefore LGG 225 group may be considered an analog of our LG except for lacking of the brightest spirals. 

\begin{figure*} 
\psfig{figure=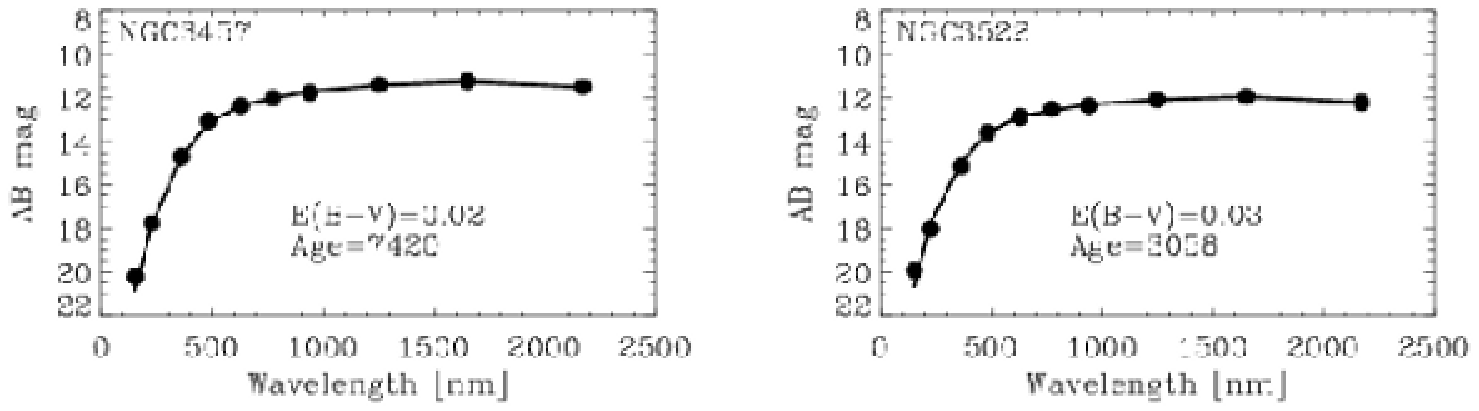,width=18cm}
\vspace{-0.4cm}
\psfig{figure=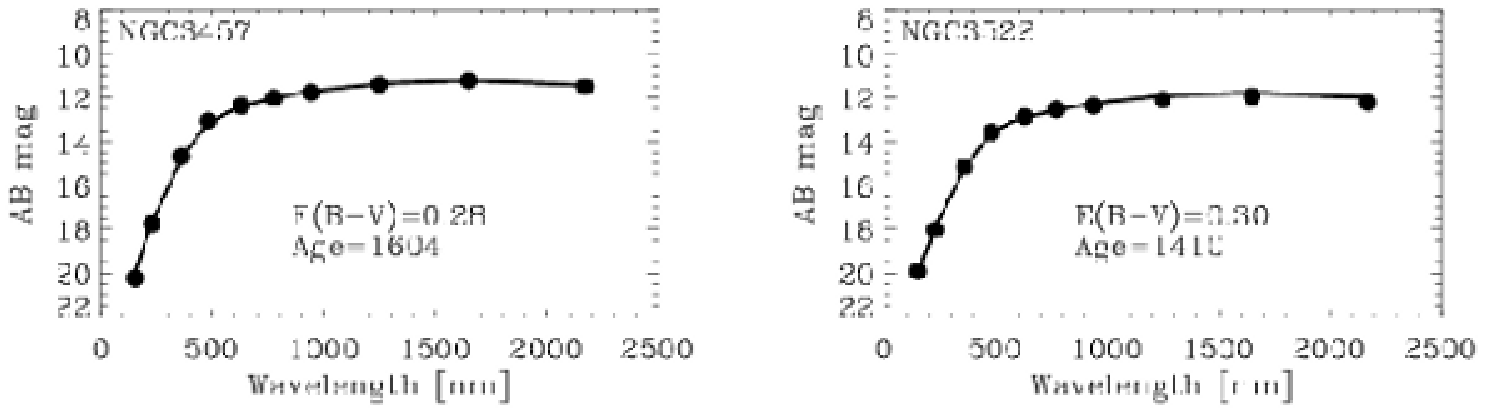,width=18cm}
 \caption{FUV to NIR SEDs from {\it GALEX}, SDSS, 2MASS (dots) of the two ellipticals in LGG 225 with the best-fit models (lines)
obtained assuming only foreground extinction (top) or  considering a possible additional extinction component (bottom).
Ages are in Myrs.}
\label{fits}
\end{figure*}

\begin{table}
\caption{Input parameters of the adopted SFH and spectro-photometric parameters \citep{Silva98}.}
\begin{tabular}{llllll}
\hline\hline
 \multicolumn{1}{l}{Type}  & \multicolumn{1}{l}{$\nu$} & \multicolumn{1}{l}{t$_{inf}$} & \multicolumn{1}{l}{t$_{esc}$}
  &\multicolumn{1}{l}{f$_{mc}$} & \multicolumn{1}{l}{IMF mass}  \\
	&  [Gyr$^{-1}$] & [Gyr] & [Myr] & [M $_\odot$] &\multicolumn{1}{l}{range}  \\ 
\hline
Elliptical  & 2 & 0.1 &100 & 0   & 0.15 - 120 \\
Spiral      & 0.6 & 4 & 18 & 0.5  & 0.1 - 100 \\
Irregular   & 0.8 & 9 & 20 & 0.5 & 0.2 - 100 \\
\hline
\end{tabular}
\label{che} 
\end{table}


\begin{table*}
\centering
\caption{Kinematical and dynamical properties of the group sample$^a$.}
\label{Dyn}
\scriptsize
\begin{tabular}{lcllllllllllll}
\hline\hline\noalign{\smallskip}
Group & Center  &  V$_{group}$  & Velocity  & D  & Harmonic &Virial  & Projected & Crossing& Group\\
 name & of mass &               & dispersion & & radius & mass &  mass &time$\times$H$_0$&Lumin. \\
   & RA [deg] ~ Dec &[km/s] &[km/s]& [Mpc] & [Mpc] & [10$^{13}$ M$_{\odot}$] &  [10$^{13}$ M$_{\odot}$] &   & [10$^{11}$ L$_{\odot}$]\\
\hline
  LGG 93$^{NUV}$ & 50.7510 -52.32828 &  950$\pm$90 & 200$\pm$61 & 12.66 & 0.45$\pm$0.04 & 1.96$\pm$0.44 & 7.6 $\pm$5.2& 0.11$\pm$0.10 & 1.94$\pm$0.03 \\
  LGG 93$^{B}$ & 50.7847 -52.32504   &  966$\pm$85 & 190$\pm$58 & 12.87 & 0.47$\pm$0.04 & 1.87$\pm$0.45 & 7.2 $\pm$5.4& 0.12$\pm$0.12 & 0.11$\pm$0.01 \\ 
  LGG 127$^{NUV}$& 76.5414 -34.27129 & 1000$\pm$65 & 198$\pm$46 & 13.33 & 0.52$\pm$0.03 & 2.22$\pm$0.33 & 26.2$\pm$16.3 & 0.45$\pm$0.17 & 4.28$\pm$0.07\\
  LGG 127$^{B}$ & 76.6130 -35.80034  & 1050$\pm$58 & 175$\pm$40 & 14.00 & 0.34$\pm$0.03 & 1.14$\pm$0.33 & 27.9$\pm$10.6 & 0.45$\pm$0.17 & 0.50$\pm$0.03\\ 
  LGG 225$^{NUV}$ &164.0016 17.44647 & 1104$\pm$27 & 105$\pm$19 & 14.72 & 0.11$\pm$0.02 & 0.14$\pm$0.26 & 3.1 $\pm$1.3&0.35$\pm$0.07 & 2.53$\pm$0.03\\
  LGG225$^{B}$ & 164.1968 17.74061   & 1125$\pm$30 & 118$\pm$21 & 15.00 & 0.29$\pm$0.03 & 0.45$\pm$0.27 & 4.1 $\pm$2.1 & 0.41$\pm$0.08 & 0.20$\pm$0.04\\
\hline
\end{tabular}

 $^a$Each quantity is computed weighting for NUV and B magnitudes, as marked on the top of the name groups.
 \end{table*}

\subsection{Group kinematic and luminosity--weighted dynamical properties}

A characterization of the group evolutionary phase requires
a kinematical and dynamical analysis. 
 We derived the kinematical and dynamical properties of our three groups, following the
luminosity-weighted approach described in \citet{Firth06}. This approach allow us 
to describe in an homogeneous way the properties of our groups and
to compare them with that of nearby groups.   

\citet{Firth06} analysis considered six nearby groups 
having  a central dominant galaxy and broad density range. 
These groups have both spiral and early-type members
as dominant galaxies.
In particular LGG 263 group is dominated by the Antennae (NGC~4038/4039), the
on-going merger of two spirals. The groups is mainly populated by late-type galaxies with a
scanty presence of lenticulars (ESO572-23, NGC~4024) and elliptical (NGC 4033) galaxies.
This kind of galaxy population is similar to our LGG~225 group showing 
 interacting (NGC 3447/NGC~3447a) 
and/or distorted galaxies (NGC~3443, NGC~3454, NGC~3507) and a
scanty population of early-type (NGC~3457 and NGC~3552).

The results of the kinematical  and dynamical analysis  are summarized in Table~\ref{Dyn}. 
All mass-related quantities are obtained by luminosity-weighting the contribution 
from each member galaxy. According to \citet{Firth06}, after luminosity weighting,
missing dwarf galaxies will not significantly alter the group velocity dispersion, 
virial mass estimates or crossing times.  The dynamical calculations are based on 
the formulae given in  \citet[][(their Table 6)]{Firth06}. Each galaxy is
weighted by its relative luminosity evaluated from the NUV-band magnitude converted to
relative luminosity. For comparison with the \citet{Firth06} groups we repeat the
dynamical analysis using  also the B-band total magnitudes given Table~1. 
As shown in Table~\ref{Dyn}, the B-band and the NUV-band dynamical analysis 
provide comparable results within the errors. 

The coordinate of the center of mass in column 2 of Table~\ref{Dyn} are
obtained by averaging the NUV  and B \ww coordinates of the group members. 
The NUV and B \ww mean velocity and  velocity dispersion are given in columns 3 
and 4 of Table \ref{Dyn} respectively. 
Fig.~\ref{helio} shows the distribution of the mean heliocentric radial velocity, with
overlapped the \ww mean velocity and the velocity dispersion  of the three groups
and of the LG. The mean velocity of the
three groups does not differ significantly while the velocity dispersion of LGG 93 and LGG
127,   which are comparable to that of the LG, are two times higher than that of LGG 225.  

In order to obtain a  measure of the compactness of the three groups, 
we have computed the harmonic mean radius (column 6 of Table \ref{Dyn}) using
the projected separations r$_{ij}$ between the i-th and j-th group member:
Figure~\ref{dist} shows the relative positions of the groups members in each group
on a linear spatial scale.   The NUV luminosity-weighted harmonic radius,
centered on the group center of mass is also drawn. 
For comparison we show in the same Figure (right panel)  the spatial distribution 
of galaxies in the three groups.

\begin{figure}[!ht]
\includegraphics[height=12.5cm,width=7.5cm]{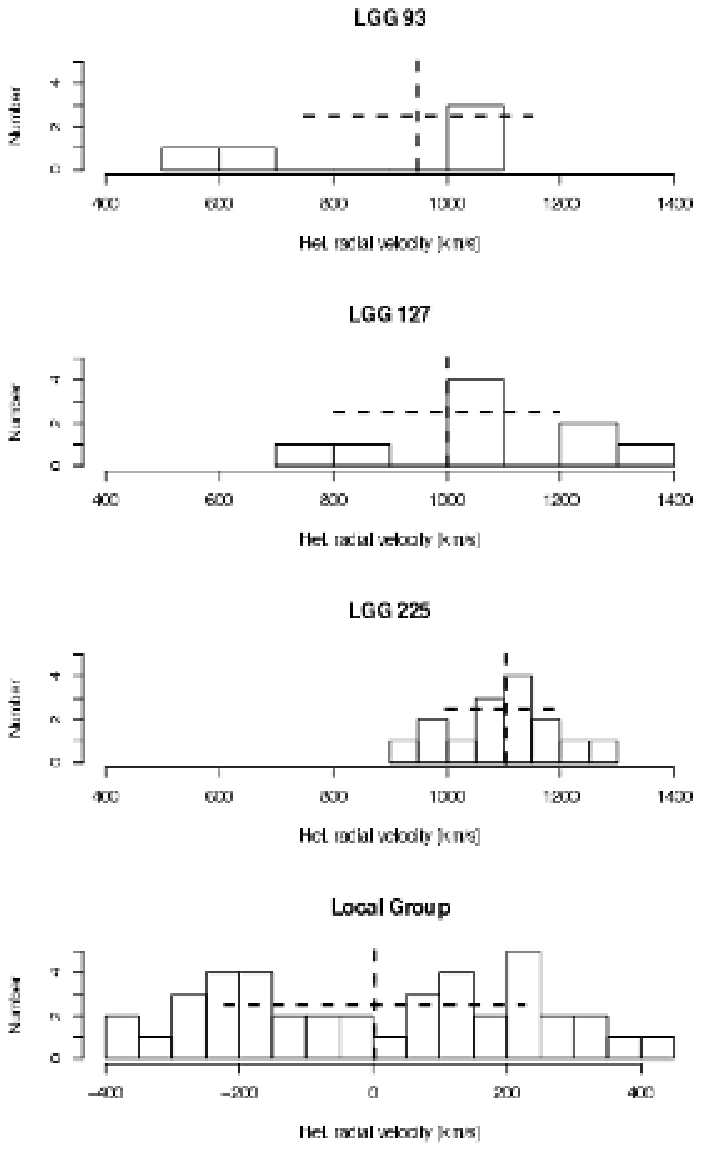}
 \caption{From top to bottom: histograms of the heliocentric radial velocity of the three groups
   and for comparison of the LG. 
 The NUV \ww   mean  velocity (vertical dashed lines) and the velocity dispersion 
 (horizontal dashed line)  shows the approximate dynamical boundaries of the three 
 groups along the radial velocity axis. For the LG, the mean  velocity and  the velocity dispersion 
 are not \ww.
}
\label{helio}
\end{figure}

\begin{figure*}[!h]
\includegraphics[height=6.5cm]{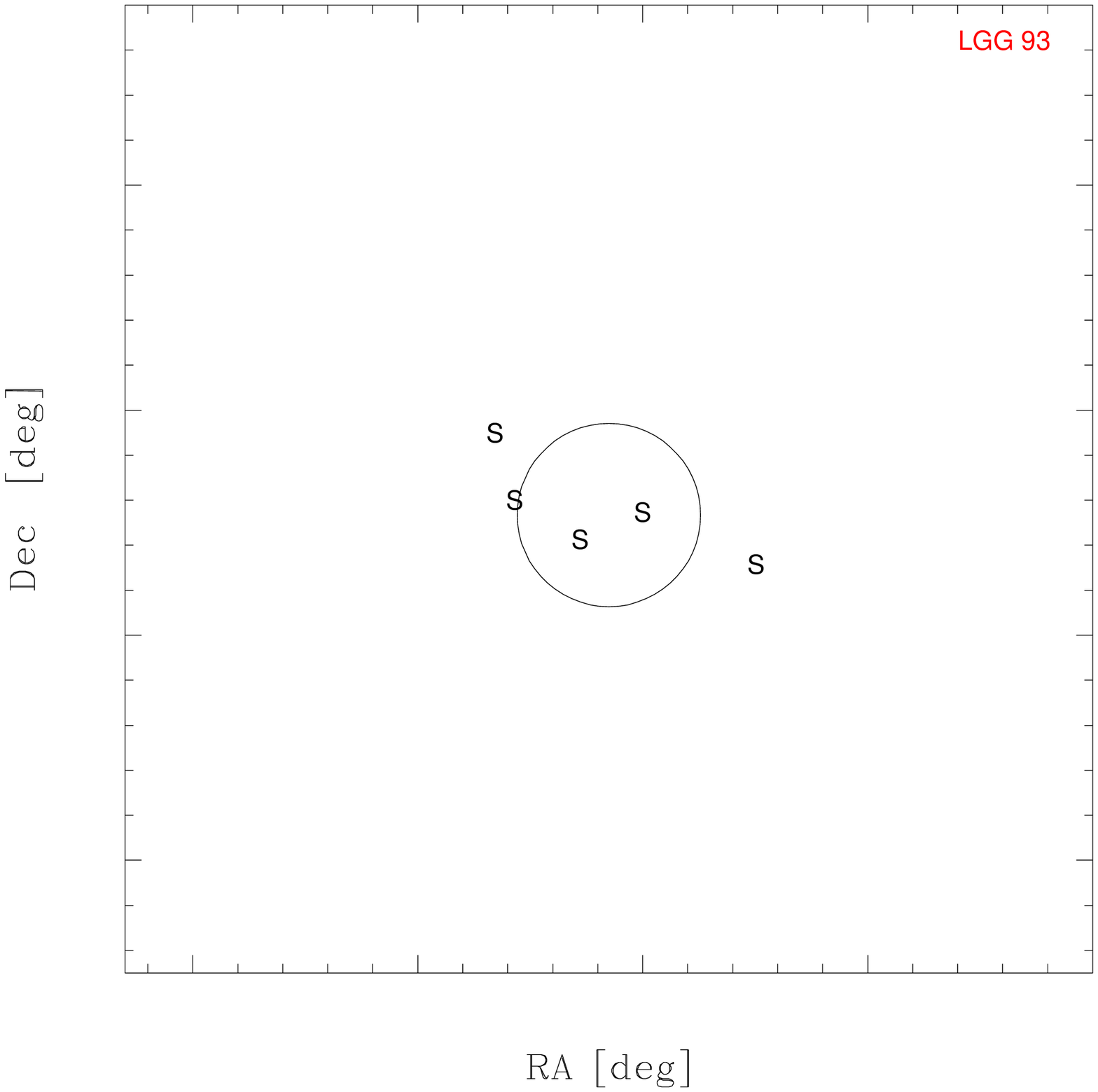}  \includegraphics[height=6.5cm]{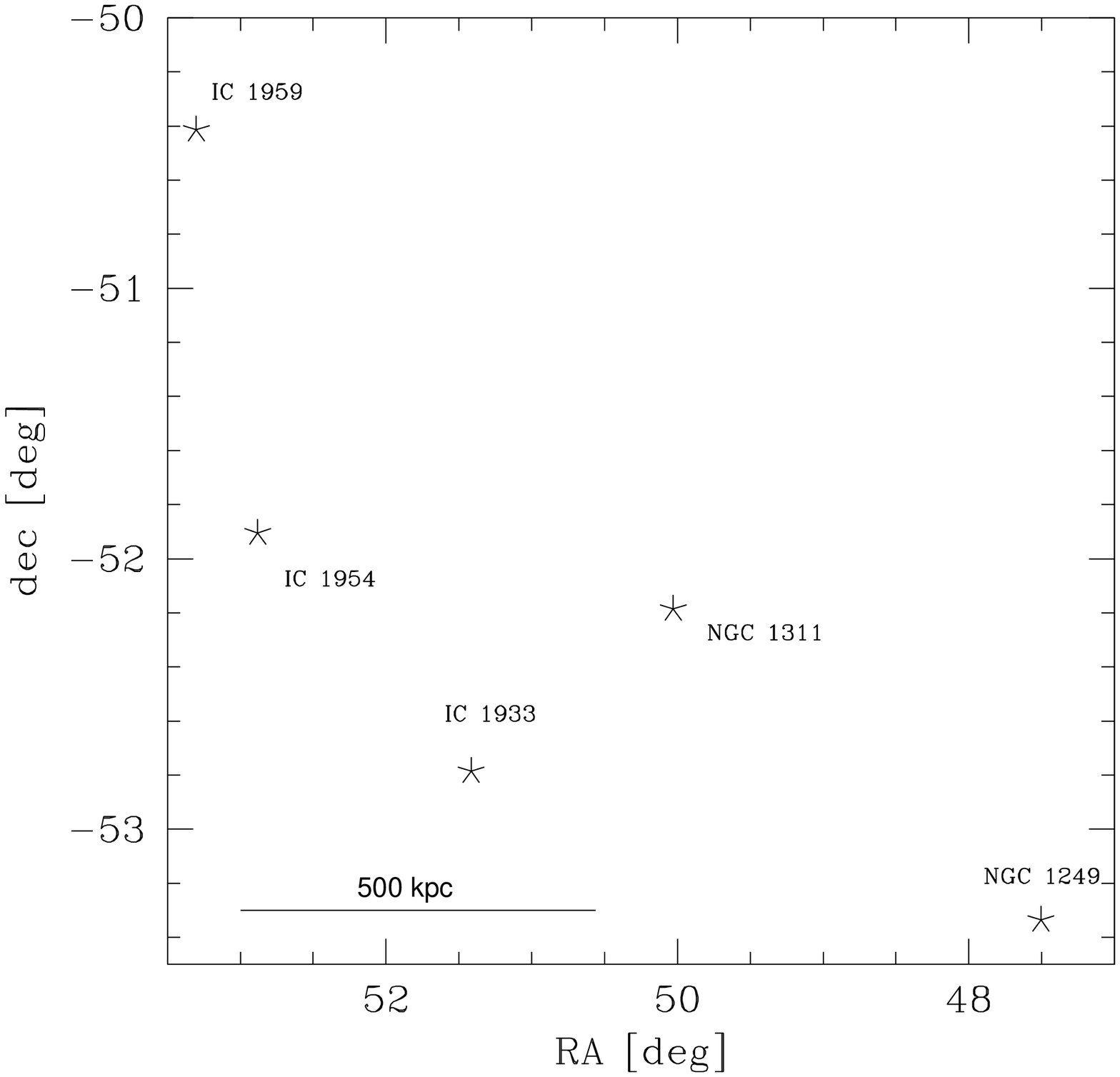}  \\
\includegraphics[height=6.5cm]{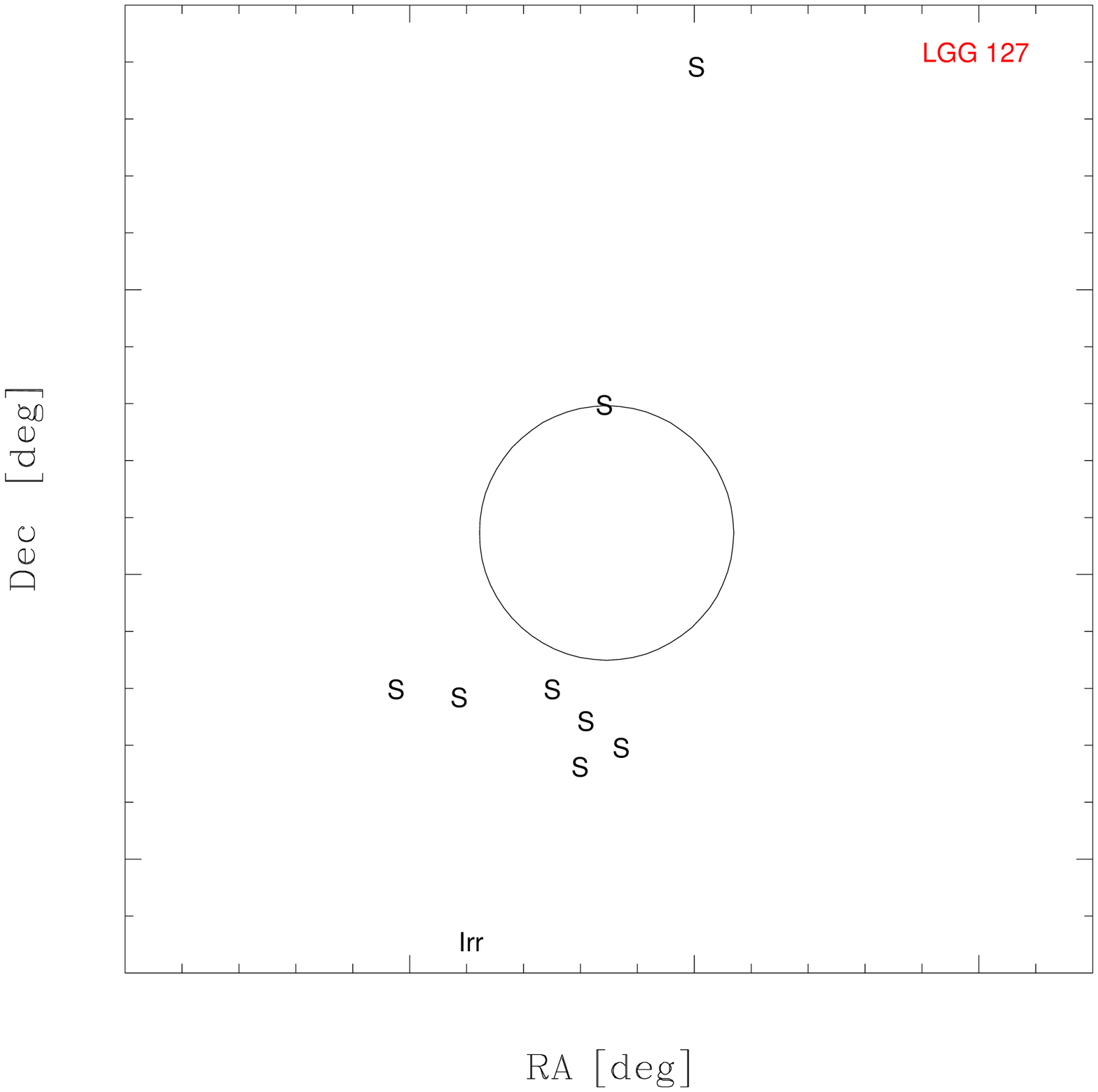} \includegraphics[height=6.5cm]{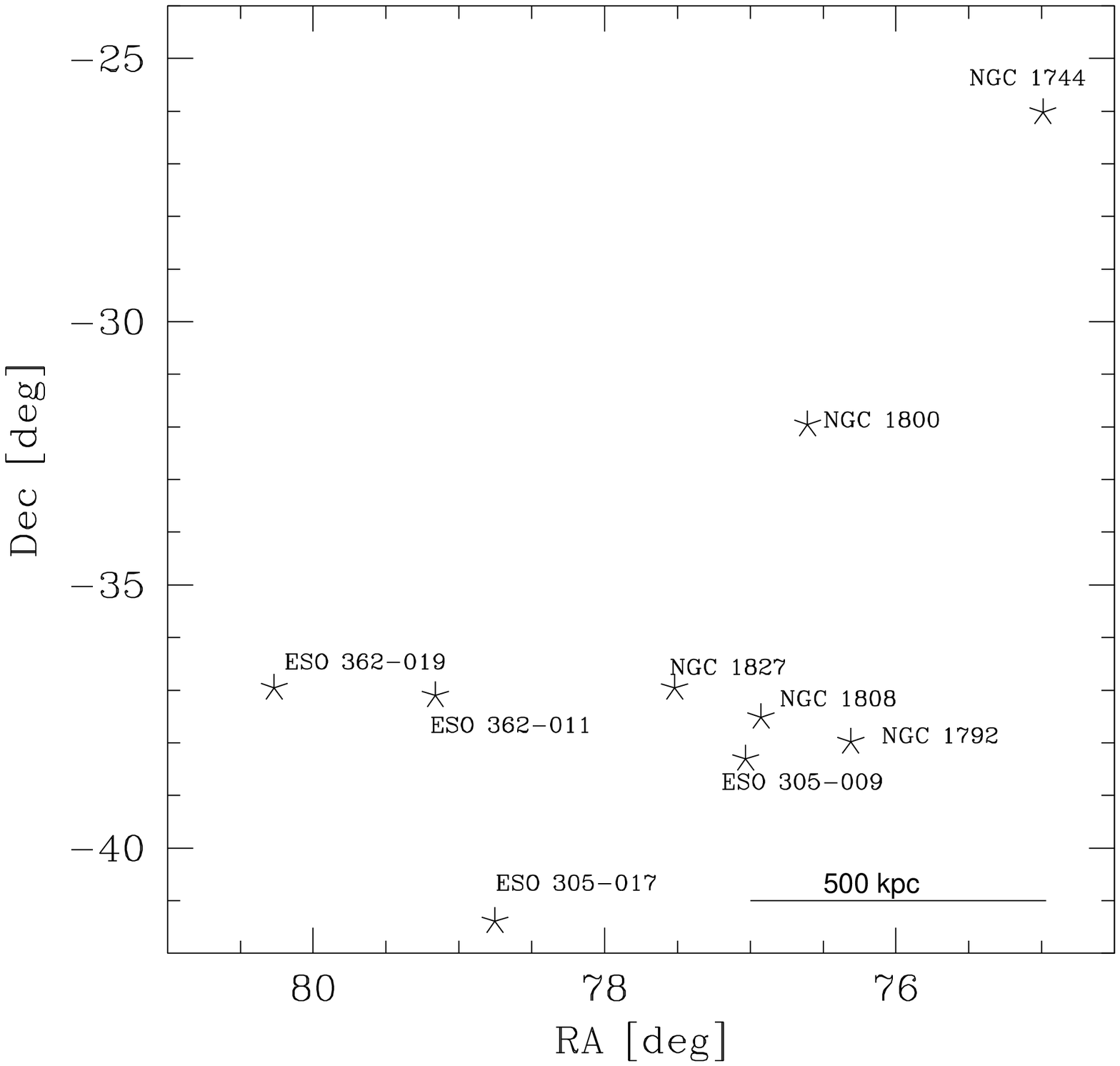} \\
\includegraphics[height=6.5cm]{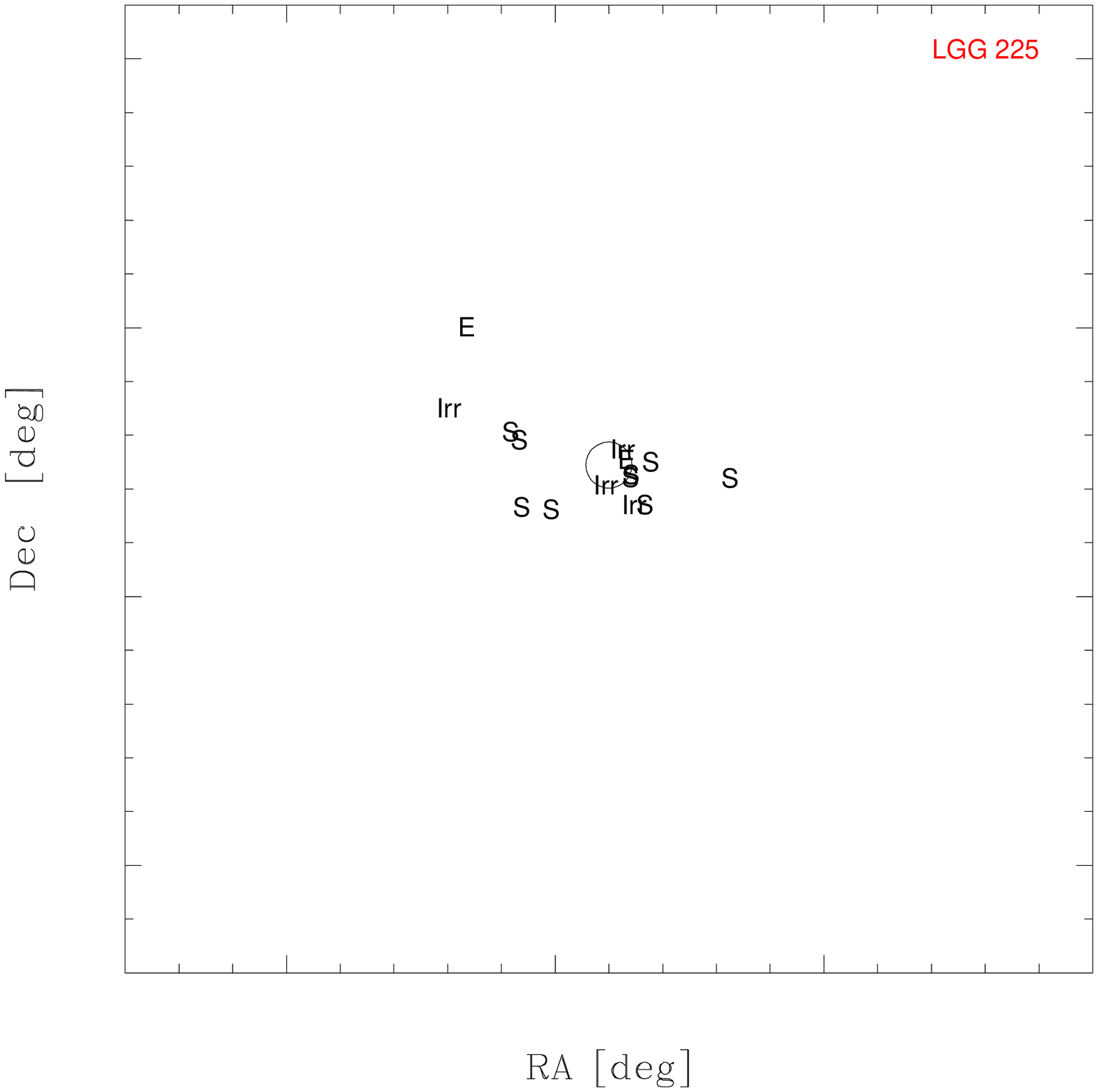} \includegraphics[height=6.5cm]{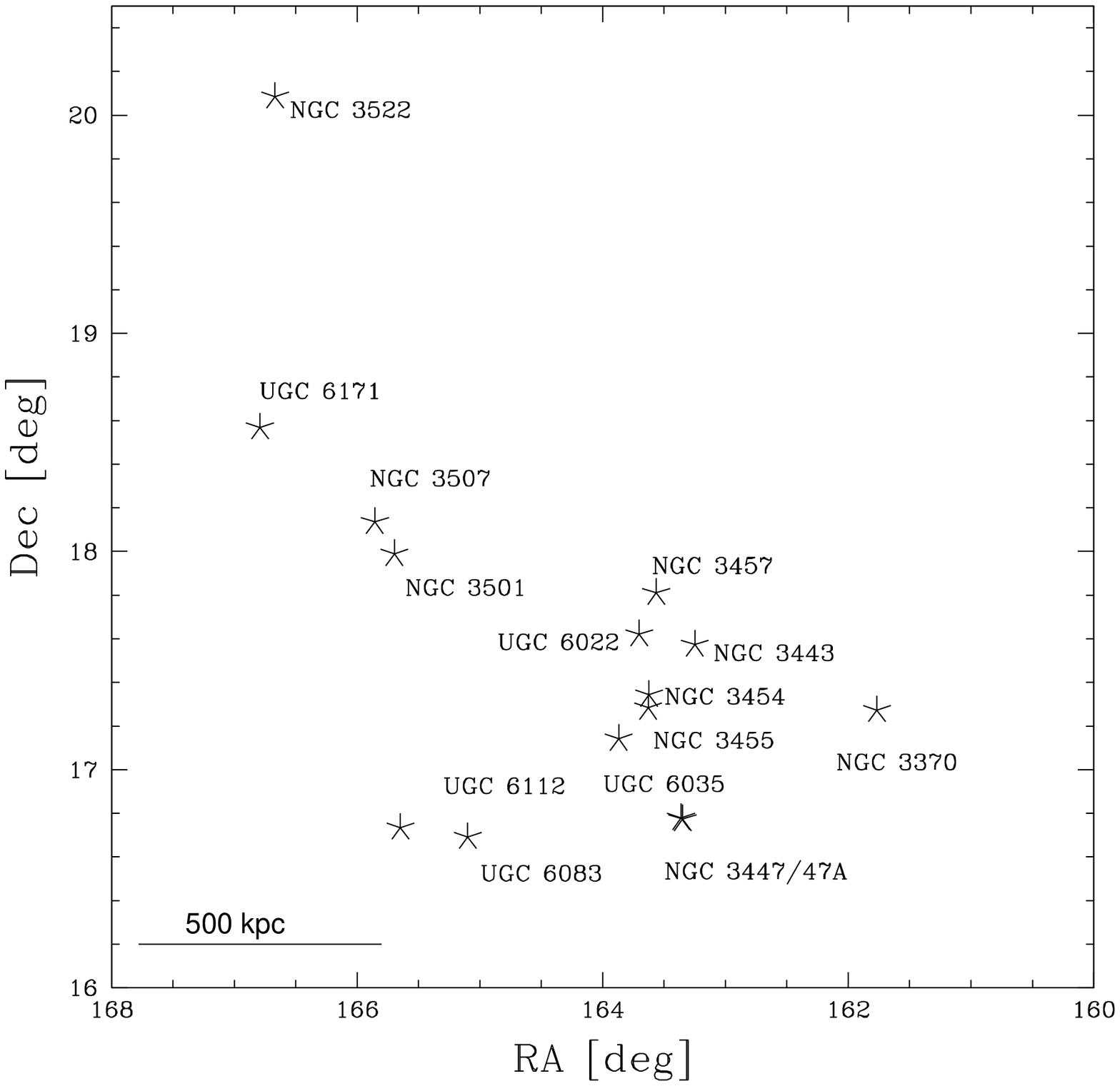} \\
 \caption{Left: the position of the galaxies in the three LGs in a    
   square approximately of 4.5 x 4.5 Mpc$^2$. The circle represents the NUV 
  luminosity-weighted
  harmonic radius centred on the  group center  of mass. 
  Right: The distribution of the galaxies on the sky for the three groups.  
}
  \label{dist}
 \end{figure*}

Columns 7 and 8 of Table \ref{Dyn} give the estimates of the virial mass and of the
projected mass for the three groups. The differences between the virial and the projected
mass are quite high, with the projected mass higher than the virial mass. This difference
is expected in systems where individual galaxies are close in projection.  
 \cite{Hei85} found that the virial mass underestimates the mass
while the moment of the projected mass gives more accurate values. Another reason for this
difference could be that the groups are not virialized. The crossing time is usually
compared to the Hubble time to determine whether the groups are virialized
\citep{Ferguson90}.   
The crossing time  is smaller than the
Hubble time, suggesting that the groups  are gravitationally bound systems. 
However, following \citet{Firth06}, the crossing times of LGG 127 and LGG 225 
exceed 0.2 Hubble times, suggesting that they could be yet unvirialized.

Assuming Gaussian distributions for all input measurements, errors in Table \ref{Dyn} 
are computed  using Montecarlo simulations. Galaxy coordinates were considered accurate.
We generated a set of 10000 groups having the respective  number of members 
of the three groups and a Gaussian distribution of respective mean velocity and 
dispersion given in Table \ref{Dyn} to derive a set of Gaussian output distributions.

Summarizing, the kinematical and dynamical analysis suggest that our three groups
can be considered loose although gravitationally bound groups. As  NGC~4038 and NGC~4697,
studied by  \citet{Firth06}, LGG~127 and LGG~225  are likely in a pre-virial collapse phase and 
probably still undergoing a dynamical relaxation. The luminosity--weighted
virial mass estimate may be therefore questionable for these groups. 
LGG~93 luminosity--weighted
analysis suggest that the group is likely virialized with crossing time and
harmonic radius similar to that of NGC~5084 and Dorado in the \citet{Firth06} sample.

\section {Discussion and conclusions}

We analyzed the  UV photometric properties of three nearby groups  
dominated by late-type galaxies,  LGG~93, LGG~127 and LGG~225.
The  UV data-set  was integrated by SDSS optical
photometry in the case of LGG~225.

FUV and NUV morphologies are very similar in most galaxies indicating that very
young populations dominate their emission.  
 UV and optical classifications are typically consistent. Bars are  easily
 recognizable in the UV bands,  although  both NGC~1311 and IC~1959,
 indicated as barred galaxies in the RC3 classification,
do not show a bar in our UV images.

Some LGG~225 members display unambiguously interacting 
(NGC 3447/NGC~3447A, NGC 3454/NGC~3455)  and distorted morphologies 
(UGC 6035) particularly evident in the UV imaging.  Furthermore, 
UV images reveal young stellar populations in extreme outskirts of these galaxies,  
extending much farther out than in the optical images, as found
in about 30\% of spiral galaxies by \citet{Thilker07,  Bianchi09}.

A possible signature of past interaction events is the presence of
a bar structure, seen in LGG~127's, NGC~1744 and ESO305-009 
which show in addition open arms, and multiple arms, 
as  NGC~1808  in the same group. 
Numerical simulations \citep[e.g.][]{Noguchi87} show that a bar
structure and open arms may develop in a disk galaxy after few galactic 
rotations, close to the passage of a companion. 

(FUV-NUV) colors  suggest  very recent episodes
of star formation. There is evidence that such episodes are triggered
by  on-going interaction as in the case of NGC~3447A or 
 NGC~3454/NGC~3455.  The (FUV-NUV) color profile 
of neither M~31 or M~33  suggests the presence of such young components, visible
  in the color profile of M~51 (Bianchi et al. 2005).
  
We derived ages and masses by fitting  the observed SEDs of LGG 225 galaxy members, 
that extends from far-UV to NIR, with populations synthesis models. 
The UV bands proved to be crucial in  disentangling evolution and extinction
effects.

Almost all galaxies in LGG~93 and LGG~127 
are detected or have upper limit fluxes at 60$\mu$m and 100$\mu$m in the 
IRAS faint source catalogue \citep{Moshir90}.
Only two galaxies in LGG~225 are detected by IRAS.  
Their cold dust masses are estimated in the range  
$10^3$ - $10^5$ M$_{\odot}$ with a mean ratio $M_{dust}/L_B$=-4.65$\pm$0.92
typical of late-type galaxies \citep{Bettoni03}. 
All galaxy members of LGG93, eight out of nine of LGG 127 and only seven out of fifteen
of LGG 225
are detected in J, H, Ks bands in the 2MASS survey,
NGC 3447 is not detected. 
The average color of the detected objects, of  $\langle (B-Ks) \rangle$=3.02$\pm$0.65
is bluer than the typical (B-Ks) color of a 10-15 Gyr old disk  \citep[$\approx$3.5, 
see][]{Mazzei92} but in agreement with our GRASIL model predictions for 3-4 Gyr spirals
of 2.98$\leq$B-Ks$\leq$3.54, confirming our composite population SED analysis. 

The Spitzer MIPS f$_{24\mu m}$ $\sim$1.7 mJy \citep{Temi09} of the elliptical NGC 3522
is compatible with that expected from our best fit shown in Fig. \ref{Mod}. 
No Spitzer public data are yet available for LGG 225.

We investigated the group luminosity-weighted dynamics. 
The comparison with \citet{Firth06} groups suggests that ours are physically 
bound loose groups. The crossing times and harmonic radii  indicate that LGG~127 
and LGG~225 are likely in a pre-virial phase, not dissimilar from other 
nearby groups like NGC~4038, NGC~4697 groups, which, at odd with ours, are
 dominated by a central galaxy. LGG~93 is likely virialized.
The unvirialized phase of LGG~127 and in particular of LGG~225 could 
explain the active star formation   in this latter group and the indication
of interactions in LGG~127. 
 \citet{Plionis06} computed the median dynamical characteristics of the \citet{Eke04}
group sample. For groups with $z\leq0.03$ and 4-30 galaxy members they found velocity dispersion
of 157$\pm$35 km~s$^{-1}$  and
virial mass  of  $6.2 \times 10^{12}$  ($h^{-1}_{72}$ M$_{\odot}$) 
comparable, within errors, to our groups. In particular their median crossing time is 1.3$\times$10$^9$ yr
comparable to our value for LGG~93, the likely virialized group (see Table~6). 
 The stellar masses of the three groups estimated from the multiband photometry range from a few 10$^9$  
to 10$^{10}$ M$_{\odot}$, although  for two groups only three photometric bands are available. 
 Dynamical masses are higher (see Table 6), 
comparable to the LG mass 
estimated to be between 1.8 and 
5.3 $\times$10$^{12}$ M$_{\odot}$ \citep{Li08}. However for the two groups, LGG 127 and LGG 225 yet unvirialized,
dynamical mass estimates are questionable. The most massive galaxy in LGG 225 is comparable to M33  in the LG,
no very massive spirals are found.
  
Dynamical analyses of  large sets of poor groups 
\citep[e.g.][and references therein]{Tovmassian09}  
suggest that groups in the local universe are a family of cosmic 
structures  presently at various stages of their virialization
processes. The morphology-density relation found at high density
regimes \citep{Dressler80} is at work also at the group scale
\citep{Postman84}. As  galaxy-galaxy 
interactions and merging events proceed, the host group dynamically 
evolves and the fraction of early-type galaxies should appear
 high in dynamically advanced, high velocity dispersion groups 
\citep[e.g.][]{Tovma04, Agu07}. 

Recent studies suggest that the galaxy evolution in
low density environments, like loose groups, is 
in delay with respect to high density ones \citep[e.g.]
[and references therein]{Clemens09}.  
Groups offer then the possibility to investigate 
the main mechanisms that drive galaxies
towards the passive evolution we measure
in dense structures.
 We are extending our multi-wavelength analysis to a large 
number of groups with different characteristics. In particular a  group sample  
ranging from LGA  to groups containing 
an increasing fraction of early-type galaxies.

 \begin{acknowledgements}
A.M. acknowledges the support of the Italian Scientists and 
Scholars in North America Foundation (ISSNAF) through  an ISSNAF fellowship 
in Space Physics and Engineering, sponsored
by Thales Alenia Space. 
We want to thank the referee for useful comments.
{\it GALEX} is a NASA Small Explorer, launched in April
2003. 
This work is based on {\it GALEX} data from GI program 121 and 
archival data. We gratefully acknowledge NASA's support for construction, operation
and science analysis of the GALEX mission, developed in cooperation with the Centre National
d'Etudes Spatiales of France and the Korean ministry of Science and Tecnology. 
Some of the data presented in this paper were obtained from the 
Multimission Archive at the Space Telescope Science Institute (MAST). 
STScI is operated by the Association of Universities for Research in Astronomy, 
Inc., under NASA contract NAS5-26555. Support for MAST for non-HST data is 
provided by the NASA Office of Space Science via grant NAG5-7584 and by other 
grants and contracts.
We acknowledge the usage of the HyperLeda database (http://leda.univ-lyon1.fr).
 \end{acknowledgements}

{\it Facilities:} {\it GALEX}, Sloan

 \bibliographystyle{aa}
 \bibliography{13216}

\end{document}